\documentstyle[sprocl,psfig]{article}

\def\kov{$\check{\rm C}$erenkov }
\def\hz{\,{\rm Hz}}

\def\mhz{\,{\rm MHz}}

\def\mev{\,{\rm MeV}}
\def\gev{\,{\rm GeV}}

\def\nsec{\,{\rm ns}}
\def\dmsq{\Delta m^{2}~}
\def\musec{\mu s}
\def\numu{\nu_{\mu}}
\def\numubar{\bar\nu_{\mu}}
\def\nue{\nu_e}
\def\nuebar{\bar\nu_e}
\def\nutau{\nu_{\tau}}

\def\pimue{\pi \rightarrow \mu \rightarrow e}

\def\sinsqtheta{\sin^2 2 \theta~}
\def\mutoe{\numu \rightarrow \nue}

\def\pizero{\pi^0~}

\def\gtwid{\mathrel{\raise.3ex\hbox{$>$\kern-.75em\lower1ex\hbox{$\sim$}}}}
\def\ltwid{\mathrel{\raise.3ex\hbox{$<$\kern-.75em\lower1ex\hbox{$\sim$}}}}
\newcommand{\text}[1]{\mbox{#1}}

\begin{document}


\begin{titlepage}

\bigskip
\bigskip
\title{\Large A letter of intent for an experiment to measure 
$\mutoe$ oscillations and $\numu$ disappearance at the Fermilab
Booster:  
BooNE \\  \today
}
\bigskip
\author{E. Church, I. Stancu, G. J. VanDalen}

\smallskip

\address{University of California, Riverside, CA 92521}

\smallskip

\author{R. A. Johnson}

\smallskip

\address{University of Cincinnati, Cincinnati, OH 45221}

\smallskip

\author{J. M. Conrad,{\footnotesize *} J. Formaggio, T. Ochs, M. H.
Shaevitz, E. G. Stern,\\
B. Tamminga}

\smallskip

\address{Columbia University, Nevis Labs, Irvington, NY 10533}

\smallskip

\author{D. Smith}

\address{Embry Riddle Aeronautical University, Prescott, AZ 86301}

\smallskip

\author{G. T. Garvey, W. C. Louis,{\footnotesize *} G. B. Mills, V. Sandberg, R. Tayloe, 
D. H. White}

\smallskip

\address{Los Alamos National Laboratory, Los Alamos, NM 87545}

\smallskip

\author{H. J. Kim, R. Imlay, W. Metcalf, N. Wadia}

\smallskip

\address{Louisiana State University, Baton Rouge, LA 70803} 

\smallskip

\author{Z. D. Greenwood}

\smallskip

\address{Louisiana Tech University, Ruston, LA 71272}

\author{~~~\hrulefill~~}
\address{\footnotesize * Co-spokespersons: J. Conrad and W. C.
 Louis} 

 
\maketitle

\end{titlepage}

\vfill\eject

\tableofcontents

\vfill\eject

\section{Executive Summary}

{\it The MiniBooNE experiment will be capable of observing both
$\nu_\mu \rightarrow \nu_e$ appearance and $\nu_\mu$ disappearance.
In addition, the experiment will be able to measure $\Delta m^2$ and
$\sin^22\theta$ and search for CP violation in the lepton sector.
By using the phototubes and electronics from LSND, the detector cost
is estimated to be \$1.6M and we estimate that the neutrino beam line
would  cost \$3M.} \\

This Letter of Intent describes a search for neutrino oscillations motivated
by the LSND observation, which has been interpreted as $\bar \nu_\mu
\rightarrow \bar \nu_e$, and by the atmospheric neutrino deficit which may
result from $\nu_\mu$ oscillations. The BooNE (Booster Neutrino Experiment)
program will have two phases. The first phase, MiniBooNE, is a single
detector experiment designed to:
 
\begin{itemize}
\item  Obtain $\sim 400$ events per Snowmass-year ($1\times 10^7$ s) if the
LSND signal is due to $\nu _\mu \rightarrow \nu _e$ oscillations.
 
\item  Extend the search for $\nu _\mu \rightarrow \nu _e$ oscillations
approximately one order of magnitude in $\Delta m^2$  beyond what has been
studied previously if no signal is observed.
 
\item  Search for $\nu _\mu $ disappearance, to address the atmospheric
neutrino deficit, through the suppression of  the expected 50,000 $\nu _\mu
C\rightarrow \mu +X$ events per Snowmass-year.
 
\item  Test CP-violation in the lepton sector if oscillations are observed
by running with separate $\nu _\mu $ and $\bar{\nu}_\mu $ beams.
\end{itemize}
 
The second phase of the experiment introduces a second detector, with the
goals of:
 
\begin{itemize}
\item  Accurately measuring the $\Delta m^2$ and $\sin ^22\theta $
parameters of observed oscillations.
 
\item  Determining the CP violation parameters in the lepton
sector.
\end{itemize}

The MiniBooNE experiment (phase 1) would begin taking data in 2001.
By using phototubes and electronics from the LSND experiment, the
MiniBooNE Detector is relatively inexpensive, \$1.6 M, and 
able to be constructed on a short time scale.   The detector
would consist of
a double-wall cylindrical tank 
which is 11 m in diameter and 11 m high. The inner tank
would be covered on the inside by 1220 8-inch phototubes
(10\% coverage) and filled 
with 600~t of mineral oil, resulting in a 400 t fiducial volume.
The volume between the tanks would
be filled with scintillator oil to 
serve as a veto shield for identifying particles both entering and leaving 
the detector. The detector would be located 1000 m
from a 
neutrino source.

The neutrino beam constructed using the 8 GeV proton Booster at FNAL
would service both phases of the experiment.
The neutrino
beam line would consist of a target followed by a focusing system
and a $\sim$30 m long pion decay volume. 
The low energy, high intensity and 1 $\mu$s time-structure of 
a neutrino beam produced from the booster beam 
are ideal for this experiment.   
The sensitivities discussed in this Letter of Intent 
assume the experiment receives 5 Hz of protons in one Snowmass-year.
The cost of this beam line is expected to be \$3 M.

This Booster experiment is compatible with the Fermilab collider
or the fixed-target MI programs.  The FNAL Booster
is capable of running at 15 Hz ($5 \times 10^{12}$ 
protons per pulse),  or 30 Booster batches per 2 s Main Injector
Cycle.   The antiproton stacking requires only 6 Booster batches at
the start of the Main Injector cycle.   In principle, this means 
the BooNE beam line could receive 12 Hz, well above the expectation 
on which our sensitivities are based.

The BooNE experiments represent an opportunity to resolve 
two outstanding neutrino oscillation questions on a short-time scale.
Within the upcoming five years, no existing or approved experiments 
will be able to address conclusively the LSND signal region.   
Also, there are 
no accelerator-based experiments within this time scale that can 
prove conclusively that oscillations are
the source of the atmospheric neutrino deficit.
Thus BooNE represents an important and unique addition to
the Fermilab program.

A formal proposal for this experiment will be submitted to the Fermilab
Physics Advisory Committee in the autumn of 1997.

\section{Introduction}

{\it The MiniBooNE experiment is motivated by the evidence for neutrino
oscillations from the LSND and atmospheric neutrino experiments.
The detector will be similar to that in the LSND experiment
and will be located 1000m from a neutrino beam line fed by the 8 GeV proton
Booster.} \\

\subsection{Motivation}

This experiment is motivated by two important pieces of evidence for neutrino
oscillations.    The first is the observation of events by the 
LSND collaboration that are consistent with 
$\bar \nu_\mu \rightarrow \bar \nu_e$ oscillations.
The second is the observed deficit of atmospheric neutrinos 
which may be attributed to $\nu_\mu$ disappearance through 
oscillations.   Here we briefly review these results and  
the expectation for what MiniBooNE can contribute.   Chapter 3 
provides further details on these results and other experimental 
evidence for neutrino 
oscillations.

\begin{figure}
\centerline{
\psfig{figure=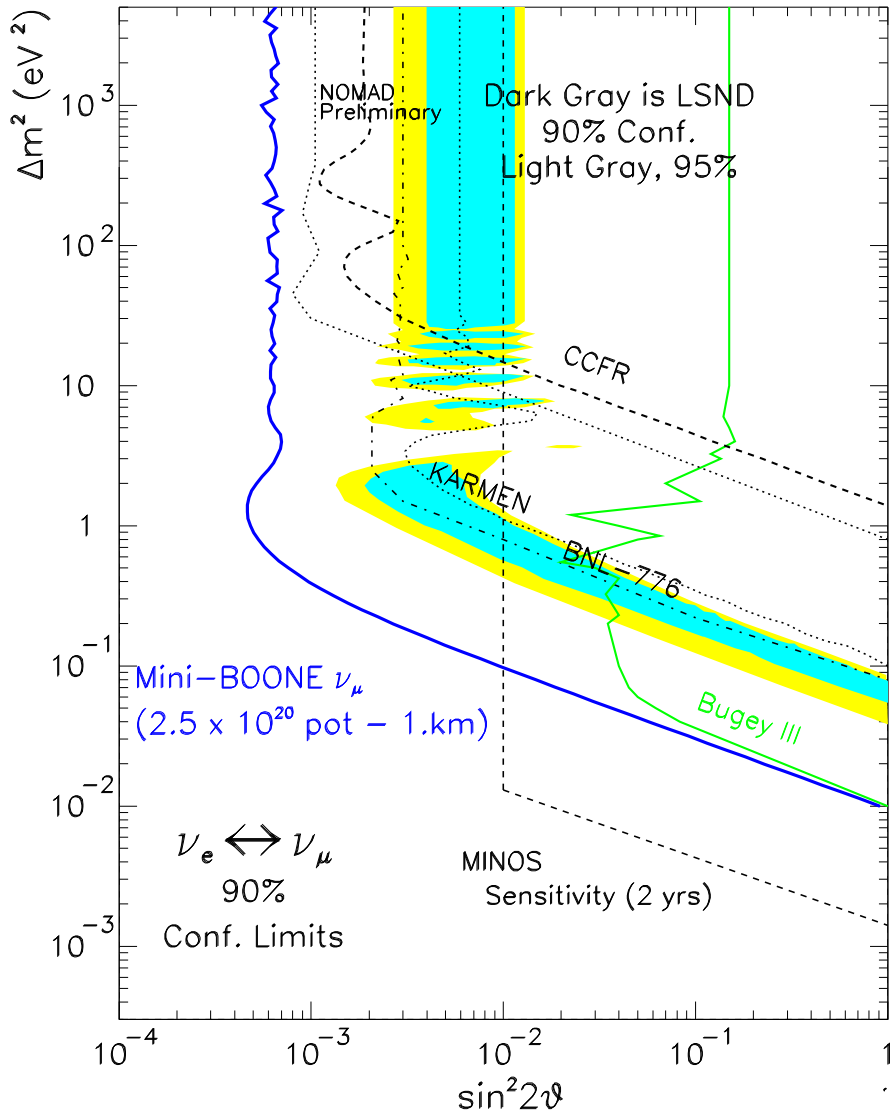,bbllx=100bp,bblly=100bp,bburx=600bp,bbury=600bp,height=5.5in,width=5.5in,clip=T}
}
\caption{
90\% C.L. limit expected for MiniBooNE for $\nu_{\mu} \rightarrow \nu_e$
appearance after one year of running, including 10\% systematic error,
if LSND signal is not observed (solid line).  
Summary of results from past experiments and 
expectations for the future MINOS experiment
are also shown.}
\label{fig:mueapp}
\end{figure}

The LSND experiment at Los Alamos has reported evidence\cite{bigpaper2}
for $\bar \nu_{\mu} \rightarrow \bar \nu_e$ oscillations with an
oscillation probability of $\sim 0.3\%$.  
The allowed values of $\Delta m^2$ and $\sin^2 2\theta$
corresponding to this oscillation probability are indicated in Fig.
\ref{fig:mueapp} by the grey region.
Previous oscillation searches have not seen oscillations in
the LSND allowed region for $\Delta m^2>4$ eV$^2$, as shown in 
Fig. \ref{fig:mueapp}.    This isolates the most favored region 
at low $\Delta m^2$.
LSND also is able to search for $\nu_{\mu}
\rightarrow \nu_e$ oscillations using $\pi^+$ that decay in flight in the
beam stop. This decay-in-flight oscillation search has different
backgrounds and systematics than the decay-at-rest search, and the
presence of an excess that is consistent with the decay-at-rest search
provides additional evidence that the LSND results are due to
neutrino oscillations.

\begin{figure}
\centerline{
\psfig{figure=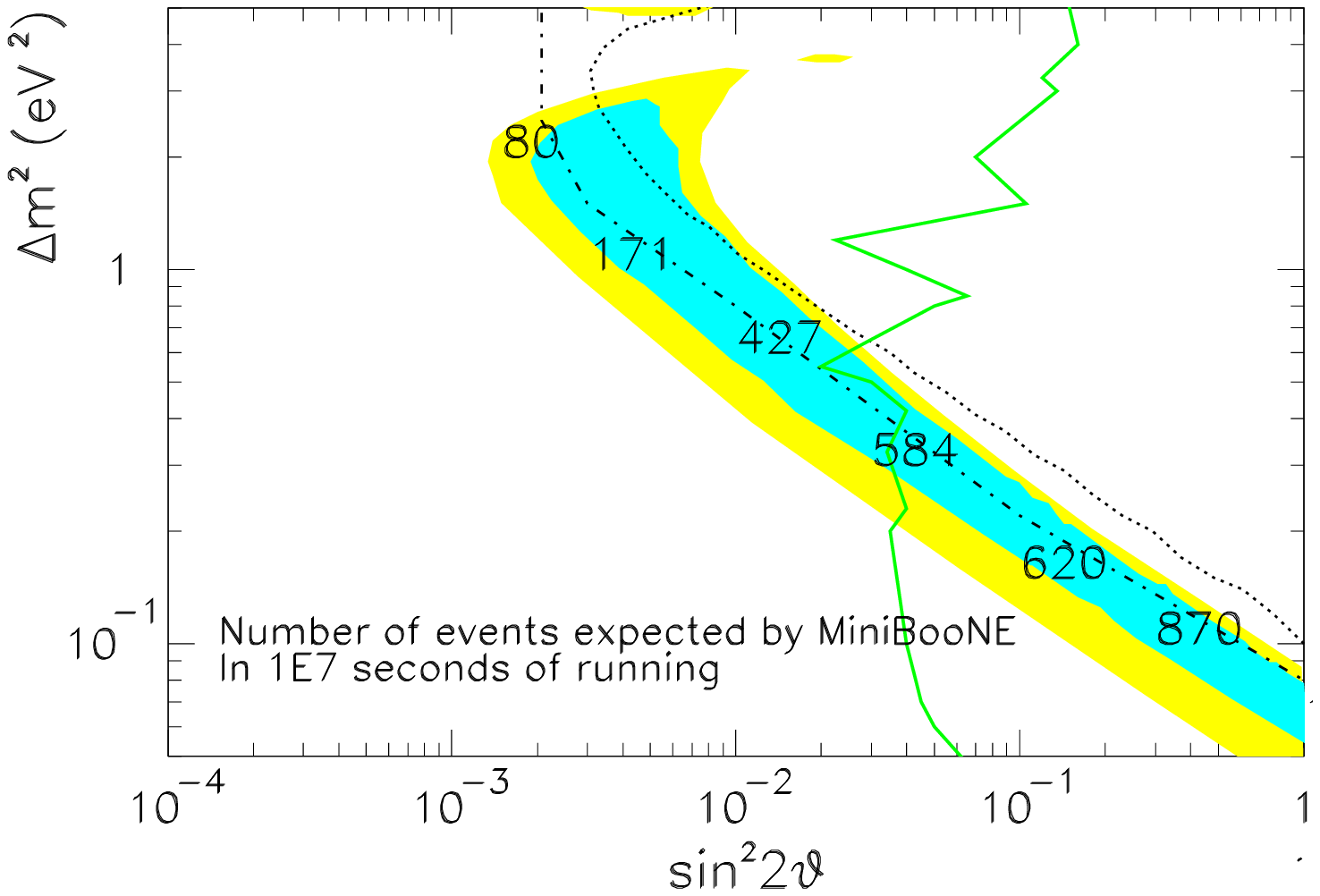,bbllx=50bp,bblly=400bp,bburx=600bp,bbury=650bp,height=3.5in,width=5.5in,clip=T}
}
\caption{If LSND signal is observed, this plot shows the
number of events expected in 1 Snowmass-year of running for 
MiniBooNE for the low $\Delta m^2$ favored region for LSND (shaded).   Lines
indicate regions excluded by past experiments (see Fig. 1).   }
\label{fig:nevts}
\end{figure}

If the LSND signal is due to neutrino oscillations, MiniBooNE expects 
between 100 and 400 events per Snowmass-year, depending on the 
$\Delta m^2$ and $\sin^2 2\theta$ of the oscillation, outside of 
the regions ruled out by previous experiments.   The expectations 
are shown in Fig. ~\ref{fig:nevts} (solid line).   The MiniBooNE 
systematics are significantly different to the LSND 
experiment.   Thus MiniBooNE will be able to  verify or disprove the
LSND result.   The full BooNE two-detector system will then accurately
measure the oscillation parameters.    

If the LSND signal is not observed by 
MiniBooNE, then the expected sensitivity is shown in Fig. \ref{fig:mueapp}.
This experiment extends approximately an order of magnitude in $\Delta
m^2$ beyond previous limits.   

The second important hint for neutrino oscillations comes from 
experiments which 
indicate a 
deficit of muon neutrinos from cosmic ray
production in the atmosphere.   
The Kamioka and IMB experiments\cite{kamioka1,imb1} 
determined the ratio of $\nu_\mu/\nu_e$ to be only about 60\% of the 
theoretically expected
ratio for neutrino energies below $\sim$1 GeV,  
independent of the visible
energy of the charged lepton and the projected zenith angle of the
atmospheric neutrinos. Interpreting the shortfall 
as arising from oscillation of muon neutrinos  
requires a large mixing angle ($\sin^22\theta \sim 0.5$)
and a $\Delta m^2 >10^{-3}$ eV$^2$.   
The atmospheric problem can be attributed to either
$\numu \rightarrow \nue$  
or $\numu \rightarrow \nutau$ oscillations.  Because the 
Bugey result, shown in Fig. ~\ref{fig:mueapp}, excludes the 
atmospheric neutrino deficit region, $\numu \rightarrow \nutau$
oscillations are considered to be the likely candidate.

Upper limits on the possible $\Delta m^2$ range come from 
previous accelerator-based experiments and the zenith angle 
dependence of the atmospheric neutrino deficit.
The CDHS search for $\nu_\mu$ disappearance indicates
$\Delta m^2 < 0.4$ eV$^2$. 
The Kamioka group has observed a zenith angle dependence of the high 
energy (greater than 1 GeV) atmospheric neutrino sample\cite{atmoskam}
which indicates that  $\dmsq << 0.5$ eV$^2$ (Kamioka prefers a $\Delta m^2 
\sim 10^{-2}$ eV$^2$), although the uncertainties are large.
However a recent publication from the IMB collaboration\cite{imb} reports no
zenith angle dependence.  Also, preliminary data from
the Super Kamiokanda collaboration shows
a zenith angle distribution that is consistent with being
flat and, in any case, with less angular dependence (see section 3.4).

\begin{figure}
\centerline{
\psfig{figure=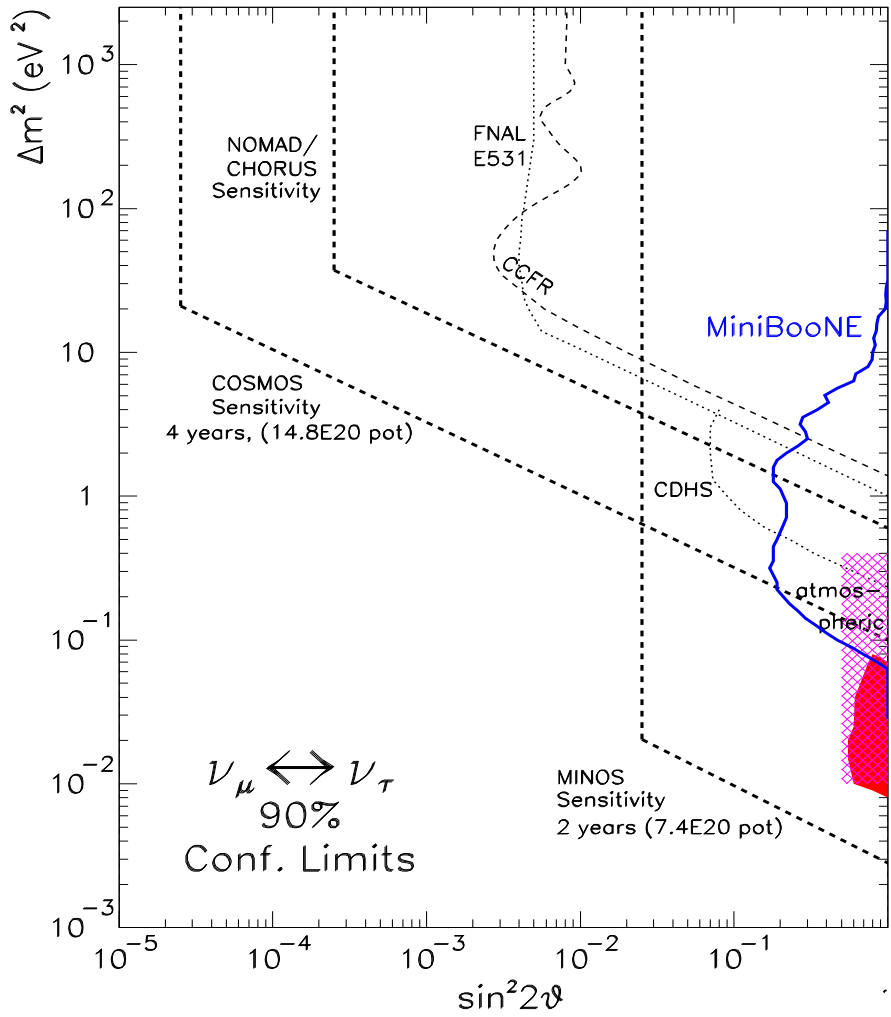,bbllx=100bp,bblly=150bp,bburx=600bp,bbury=700bp,height=5.5in,width=5.5in,clip=T}
}
\caption{Summary of results from past experiments (narrow, dashed and dotted), 
future approved experiments (wide, dashed) and 
90\% C.L. limit expected for MiniBooNE (solid) for $\nu_{\mu}$ 
disappearance after one year of running at 1 km.   
Solid region indicates the favored region for the atmospheric neutrino
deficit from the Kamioka experiment.  
A result from Kamiokanda
indicating no zenith angle dependence extends the favored region to
higher $\Delta m^2$ as indicated by the hatched region.}
\label{fig:mudisap}
\end{figure}

Figure \ref{fig:mudisap} shows an overview of past experiments 
(narrow dashed and dotted lines) 
and expectations for future approved experiments (wide dashed lines)
for $\nu_\mu \rightarrow \nu_\tau$ searches. 
In light of the changing situation concerning the 
zenith angle dependence, Fig. \ref{fig:mudisap}
shows the allowed region if there is 
a zenith angle dependence (solid) and if there is no dependence (hatched). 
In the higher $\Delta m^2$ scenario, MiniBooNE (solid line)
can address the atmospheric neutrino oscillation question
by searching for $\nu_\mu$ disappearance.
MiniBooNE  is sensitive to variations in the 
flux with energy that are consistent with
oscillations.   Statistical and 10\% systematic errors were
included in this determination.    Issues relating to the precision of the
$\nu_\mu$ disappearance search in MiniBooNE are discussed in section 9.4.

The case where the $\Delta m^2$ values 
from atmospheric neutrino experiments are compatible
with LSND provides a useful example of a three-generation mixing
formalism which may be applied to explain the present pieces of evidence
for neutrino oscillations.
If $\Delta m^2$ for $\nu_\mu \rightarrow \nu_e$ from LSND is 
approximately the same as $\Delta m^2$ from $\nu_\mu \rightarrow \nu_\tau$
in the atmospheric case, then three-generation
mixing models\cite{fuller} with only one dominant mass value for the mass
eigenstates, {\it i.e.}, $m_3\gg m_1\approx m_2$, apply.   
In this case, the LSND result is 
explained via $\nu_\mu \rightarrow \nu_3 \rightarrow \nu_e$ while 
the atmospheric result is explained through 
$\nu_\mu \rightarrow \nu_3 \rightarrow \nu_\tau$.  This and other 
examples of three generation mixing models are discussed in 
section 3.5.

MiniBooNE is an experiment which can and should be done on a short
time scale.   Construction of MiniBooNE can be completed by 2001,
as discussed in sections 5 and 6.   Thus the data-taking and analysis 
presented above is expected to be completed by 2003.   
Within the next five years there are no other 
accelerator experiments
that can address fully the oscillation issues described above.   The KARMEN
experiment, which is presently running at the ISIS facility with a 
segmented liquid scintillator detector,  
may have sufficient statistics to verify
the LSND signal with similar systematics to LSND if $\Delta m^2$ is
greater than 1 eV$^2$; however, 
KARMEN will have difficulty proving that the signals are due to neutrino
oscillations because of similar systematics.\cite{Klaus}   In the low $\Delta m^2$
region, KARMEN will not have sufficient statistics to verify fully the 
LSND result.\cite{Klaus}      
Super Kamiokanda will greatly refine the atmospheric neutrino 
deficit measurement using past techniques.   While this is quite important,
an experiment such as MiniBooNE, with very different systematics 
is required to verify that the deficit is due to neutrino
oscillations.   Finally, the MINOS experiment is not scheduled to
begin running until 2003 and requires two years to obtain the
sensitivity
indicated on Figs. ~\ref{fig:mueapp} and ~\ref{fig:mudisap}.
Thus, MiniBooNE fills an important gap within the global neutrino
program.    It should be noted that there are competitive proposals
in preparation for submission to CERN.   Therefore, it is important
that Fermilab seize the initiative in pursuing accelerator-based 
oscillation physics within the next decade.

\subsection{Neutrino Beam and Detector}

Discussion with FNAL staff and management indicate that the
Booster is capable of providing an additional 5 pulses per second at
$\sim 5 \times 10^{12}$ protons per 1 $\mu$ sec pulse at 8 GeV
beyond the requirements for antiproton stacking.
As 2 GeV pion production is copious from an 8 GeV beam, we propose 
building a focusing system capable of producing 
a parallel beam of pions centered
on 2 GeV/c.
This beam has a relatively short decay length of 30 m, so that
the fraction of $\nu_e$ in the beam from the $\pimue$ decay chain is kept at 
a low level, as 
these $\nue$ are a basic background for the appearance measurement.
The focusing system will be capable of operation in either 
positive or negative polarity. The positive polarity yields a higher
neutrino flux, although the $\nue$ background from kaon decays is lower
for the negative polarity.
The duty factor of the Booster beam with single turn extraction makes
cosmic ray background manageable and makes the data acquisition problem 
much simpler than for LSND.   The decay region is designed to reduce 
the $\nue$ contribution from kaon and muon decays.    Particle production
in the beam line is monitored using similar systems to those in the
NuTeV experiment.   The beam design is described in Chapter 5.

We propose building a 600 ton detector located at 1000 m
from the neutrino source. The detector
will be capable of measuring the $\nu_e$ and
$\numu$ energy spectra though quasi 
elastic scattering as described in Chapter 7, and the event energy
distribution in the detector will allow the determination
of the neutrino oscillation parameters.
The detector is very similar to that in the LSND experiment, 
allowing transfer of analysis expertise, particularly in the area of 
particle identification.

\section{Status of Neutrino Oscillation Experiments}

{\it  
Evidence for neutrino oscillations come from the solar neutrino
experiments, the atmospheric neutrino experiments, and the LSND 
experiment. These results can be interpreted within three-generation
neutrino mixing models.   Although future experiments are planned,
MiniBooNE fulfills a unique niche in addressing the present evidence 
for neutrino oscillations.}
\\

\subsection{Neutrino Oscillation Formalism}

If neutrinos have mass, it is likely that the interaction responsible for
mass will have eigenstates which are different from the weak eigenstates
that are associated with weak decays. 
In this model, the weak eigenstates
are mixtures of the mass eigenstates and lepton number is not strictly
conserved. A pure flavor (weak) eigenstate born through a weak decay will
oscillate into other flavors as the state propagates in space. This
oscillation is due to the fact that each of the mass eigenstate components
propagates with a different phase if the masses are different, $\Delta
m^2=\left| m_2^2-m_1^2\right| >0$.
The most general form for 3-component oscillations
is 
\[
\left( 
\begin{array}{l}
\nu _e \\ 
\nu _\mu \\ 
\nu _\tau
\end{array}
\right) =\left( 
\begin{array}{lll}
U_{e1} & U_{e2} & U_{e3} \\ 
U_{\mu 1} & U_{\mu 2} & U_{\mu 3} \\ 
U_{\tau 1} & U_{\tau 2} & U_{\tau 3}
\end{array}
\right) \left( 
\begin{array}{l}
\nu _1 \\ 
\nu _2 \\ 
\nu _3
\end{array}
\right) 
\]
This formalism is analogous to the quark sector, where strong 
and weak eigenstates are not identical and the resultant mixing is described 
conventionally by a unitary mixing matrix.    The oscillation
probability is then:
\begin{equation}
{\rm Prob}\left( \nu _\alpha \rightarrow \nu _\beta \right) =\delta
_{\alpha \beta }-4\sum\limits_{j>\,i}U_{\alpha \,i}U_{\beta \,i}U^*_{\alpha
\,\,j}U^*_{\beta \,\,j}\sin ^2\left( \frac{1.27\;\Delta m_{i\,j}^2\left(
{\rm
eV^2}\right) \,L\left({\rm km}\right) }{E_\nu \left({\rm GeV}\right) }%
\right)  \label{3-gen osc}
\end{equation}
where $\Delta m_{i\,j}^2=\left| m_i^2-m_j^2\right| $ .
Note that there are three different 
$\Delta m^2$ (although only two are independent)
and three different mixing angles.
The oscillation probability also depends upon the length, L, from
the source and neutrino energy, E$_\nu$.

Although in general there will be mixing among all three flavors
of neutrinos, two-generation mixing is often assumed for simplicity.
If the the mass scales are quite different ($m_3 >> m_2  >> m_1$ for
example), then the 
oscillation phenomena tend to decouple and the two-generation 
mixing model is a good approximation in limited regions.
In this case, each transition can be described 
by a two-generation mixing equation:
\begin{equation}
\label{eq:P}
P  =  \sinsqtheta  \sin^2(1.27 \dmsq L/E_\nu)
\end{equation}
where $\theta$ is the mixing angle.
However, it is possible that experimental results
interpreted within the two-generation 
mixing formalism may indicate very different $\dmsq$ 
scales with quite different apparent strengths for the same oscillation.
This is because, as is evident from equation \ref{3-gen osc},
multiple terms involving different mixing strengths and $\Delta m^2$ 
values contribute to the transition probability for $\nu_\alpha \rightarrow
\nu_\beta$.

\subsection{LSND Results}

\begin{figure}
\centerline{\psfig{figure=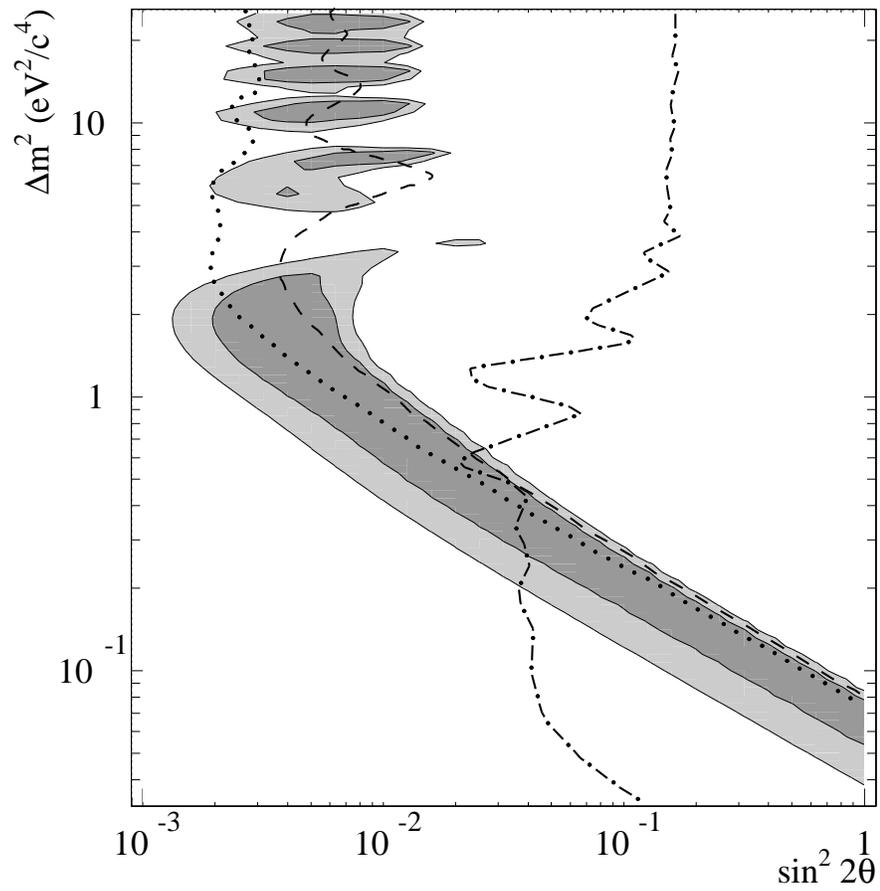,width=4.7in,silent=}}
\caption{Allowed regions in the $\dmsq$ vs $\sinsqtheta$ parameter
space from the LSND $\bar \nu_{\mu} \rightarrow \bar \nu_e$
appearance experiment. Also shown are $90\%$
C.L. limits from KARMEN at ISIS (dashed curve), E776 at BNL (dotted curve),
and the Bugey reactor experiment (dot-dashed curve).}
\label{fig:loglik}
\end{figure}

\begin{figure}
\centerline{\psfig{figure=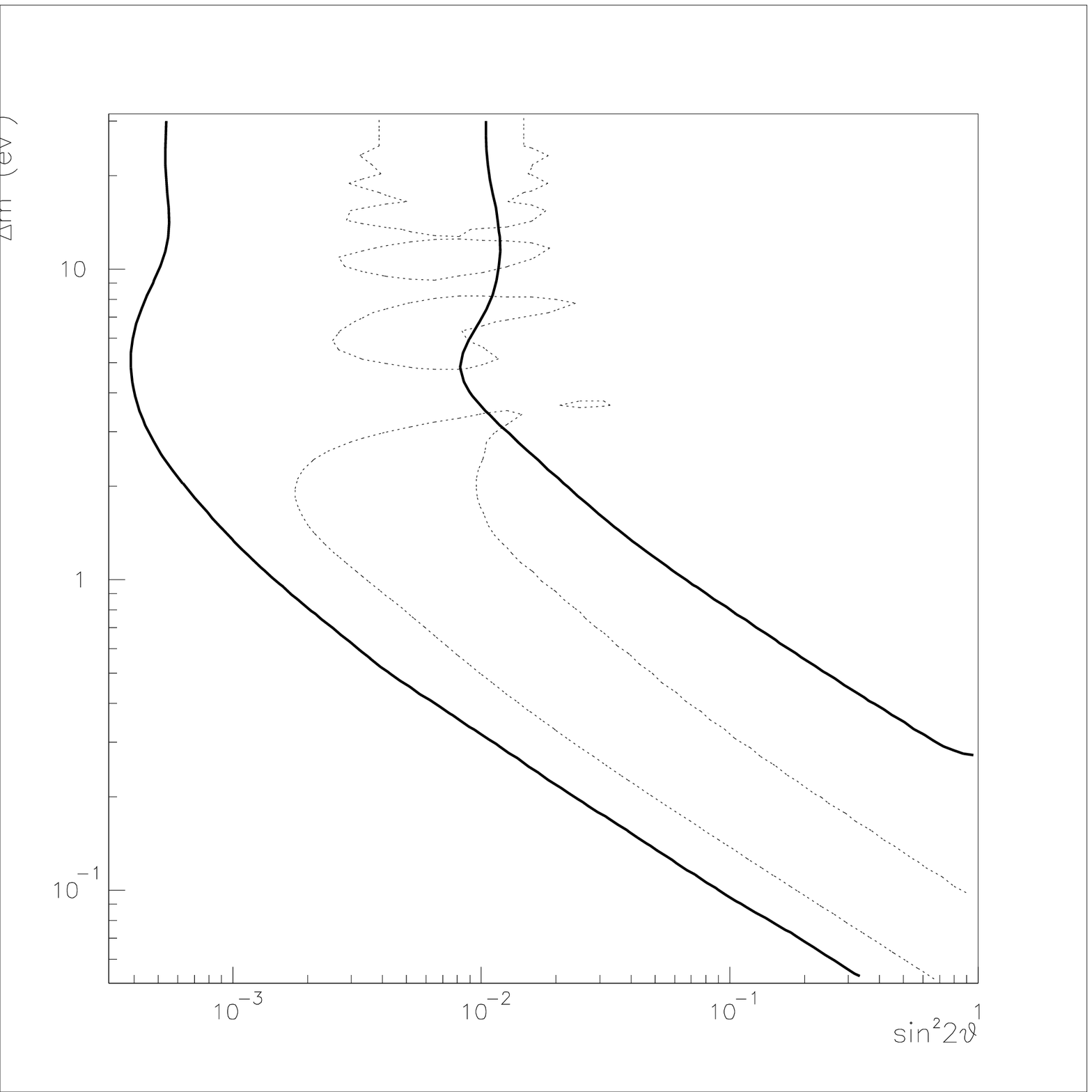,width=4.7in,silent=}}
\caption{Allowed regions in the $\dmsq$ vs $\sinsqtheta$ parameter
space from the LSND $\nu_{\mu} \rightarrow \nu_e$ (solid curve)
and $\bar \nu_{\mu} \rightarrow \bar \nu_e$ (dashed curve)
appearance experiments.}
\label{fig:loglik2}
\end{figure}

The LSND accelerator 
experiment uses a detector composed of 167 tons of dilute liquid
scintillator placed 30 m from the beam stop of the LAMPF 800 MeV
proton beam.\cite{bigpaper2}
Neutrinos are produced from stopped 
$\pi ^{+} \rightarrow \mu^+ \nu_\mu$ and $\mu^+ \rightarrow e^+
 \nu_e \bar \nu_\mu$ decays.
At the detector the $\bar \nu_e$ are detected by
$\bar \nu_e p\rightarrow e^{+}n$, and the neutron is then
captured producing a 2.2 MeV gamma. 
The experiment has reported evidence\cite{bigpaper2} for 
$\bar \nu_\mu \rightarrow \bar \nu_e$
oscillations by
observing 
$51.0_{-19.5}^{+20.2}\pm 8.0$ $\bar \nu_e$ events above
background.
This corresponds to an oscillation probability of Prob$\left( 
\bar \nu_\mu \rightarrow \bar \nu_e\right) = (0.31\pm 0.11\pm
0.05)\%$ and with an oscillation
parameter allowed region shown in Fig. \ref{fig:loglik}.   
Due to the low 800 MeV
proton energy of the LAMPF beam, the neutrino backgrounds are quite
small and well understood. The largest background is from $\mu^-$
decay at rest in the beam stop, which is suppressed by a 
factor of $7 \times 10^{-4}$ relative to $\mu^+$ decay at rest. The
suppression results from the following three factors: the ratio of
$\pi^-$ to $\pi^+$ (0.12) times the probability that the $\pi^-$
decays in flight (0.05) times the probability that the $\mu^-$
decays at rest (0.12).

The LSND signal is strengthened by a complementary
$\nu_\mu \rightarrow \nu_e$ oscillation search,\cite{DIF} which has
completely different systematics and backgrounds than the
$\bar \nu_\mu \rightarrow \bar \nu_e$ oscillation search. 
The neutrinos for this search come from pion decay in flight and
are of higher energies than those produced by stopped muons.
Thus it is interesting that the 
decay in flight analysis does show a signal, although of 
lesser significance, which indicates the same favored regions of 
$\Delta m^2$ and $\sin^2 2\theta$ (see Fig. \ref{fig:loglik2})
as the decay at rest analysis.
A global analysis of the decay-in-flight and decay-at-rest data
is in progress.

\subsection{Solar Neutrino Experiments}

Since the first observation of $\nue$ interactions in a Cl target in
the Homestake mine by 
Davis and collaborators,\cite{davis} three additional experiments have
measured solar neutrino interactions and have determined that there
are fewer neutrinos from the sun than are expected from the Standard
Solar Model.
There have been 
two experiments using Ga as target, GALLEX and SAGE, 
and the experiment at Kamioka in which neutrinos 
were scattered from electrons in water.
The Ga experiments\cite{gallium} have the lowest energy threshold, while 
Kamioka\cite{kamioka} is limited by the $\sim 7\mev$ 
detection energy threshold for electrons.
Hata and Langacker published\cite{hata} a thorough 
analysis of these data two years ago.
They considered experimental errors in detail as well as possible variations in 
the Standard Solar Model which is used to predict the flux of neutrinos 
that is expected from the sun in the absence of neutrino oscillations.
Each of the experiments is sensitive to different parts of the neutrino 
spectrum. These sensitivities are shown in Table \ref{tab:reactions} using 
the Standard Solar Model of Bahcall and Pinsonneault\cite{bahcall}.
The experimental results 
are shown in Table \ref{tab:ssn}, where it is clearly seen that the
measured solar neutrino flux is below the prediction of the Standard
Solar Model.

\begin{table}[t]

\caption{Sensitivities of the three types of solar neutrino
experiments on the expected reactions 
in the Standard Solar Model.}

\label{tab:reactions}

\vspace{0.4cm}

\begin{center}

\begin{tabular}{|c|c|c|c|}
\hline 
~~& Kamioka & Homestake & GALLEX/SAGE \\
\hline
pp & & & 0.538 \\
$^7$Be I & & & 0.009 \\
$^7$Be II & & 0.150 & 0.264 \\
$^8$Be & 1 & 0.775 & 0.105 \\
pep & & 0.025 & 0.024 \\
$^13$N & & 0.013 & 0.023 \\
$^15$O & & 0.038 & 0.037 \\
\hline
\end{tabular}
\end{center}
\end{table}

\begin{table}[t]

\caption{The measured rate from the three types of solar neutrino experiments 
compared to the Standard Solar Model.}

\label{tab:ssn}
\vspace{0.4cm}
\begin{center}
\begin{tabular}{|c|c|}
\hline 
Experiment & Rate \\
\hline
GALLEX/SAGE & $0.62 \pm 0.10$ \\
Homestake & $0.29 \pm 0.03$ \\
Kamioka & $0.51 \pm 0.07$ \\
\hline
\end{tabular}
\end{center}
\end{table}

\begin{figure}
\centerline{\psfig{figure=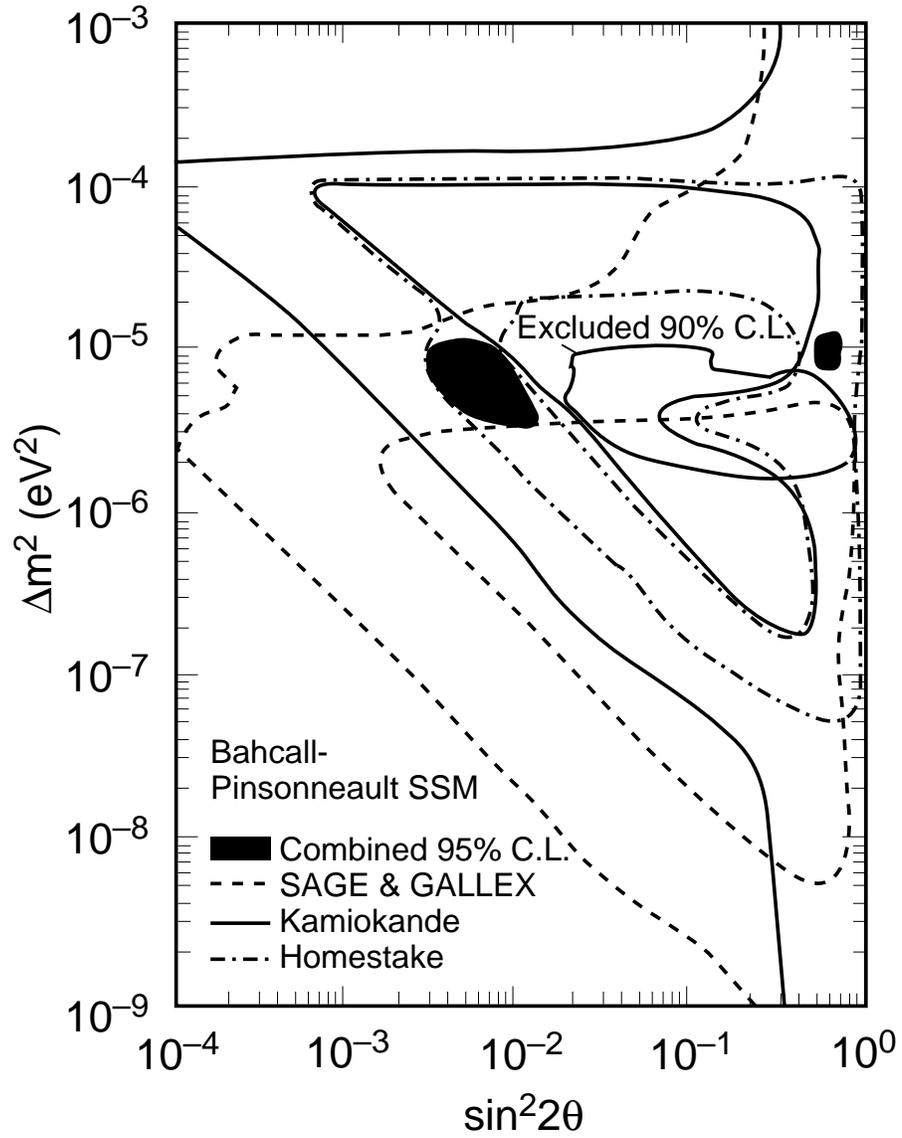,width=4.7in,silent=}}
\caption{Allowed regions in the $\dmsq$ vs $\sinsqtheta$ parameter
space from the four solar neutrino experiments.} 
\label{fig:solar}
\end{figure}

Hata and Langacker conclude that 
the experimental data cannot be explained by variations in solar physics 
and that neutrino oscillations are strongly favored.
Also, resonant transformation of $\nue$ to other flavors through the MSW 
effect is preferred, leading to the allowed region in the $\dmsq$ -  
$\sinsqtheta$ parameter space shown in Fig. \ref{fig:solar}.
Although we have stressed that three generations of neutrinos 
must be considered in general, that 
need not apply to the solar neutrino discussion as long as the masses 
$m_1$, $m_2$, $m_3$ are described by $m_3
>> m_2, m_1$.
As the small mixing solution is favored, a value for $\theta_{12}$ of about 
$3 \times 10^{-2}$ is implied.
It is worth emphasizing that these data have been gathered over an extended 
period of time and that many systematic checks have been performed.
Furthermore, the solar model is very much constrained 
by the solar luminosity, particularly 
in the case of the Gallium reaction.
The Kamioka result should be confirmed by Super Kamioka 
in the near future with significantly better statistics.
Also, the neutrino oscillation hypothesis will be tested 
by the SNO experiment, which will measure both charged and neutral current 
solar neutrino interactions. 
Overall, the solar neutrino experimental observation appears firm,
although uncertainties in solar dynamics are still a cause for concern.

\subsection{Atmospheric Neutrino Experiments}

Cosmic rays interacting in the upper atmosphere produce pions that 
decay into muons and neutrinos.
In approximate terms, this decay chain produces two muon neutrinos for each 
electron neutrino.
Charged-current reactions from these neutrinos 
have been observed in a number of detectors.
The sensitivity of these detectors to electrons and muons varies over the 
observed energy range, and so the experiments depend on a Monte Carlo 
simulation to determine the relative efficiencies.
For example, electron events are mostly contained in the detector, while muon 
events have longer range and escape the detector at the higher energies.
The experiments report the observed ratio of 
muon to electron events 
divided by the ratio of events calculated in a Monte Carlo simulation.
The experiments also model the absolute neutrino flux.
A recent summary of the experimental situation is discussed by Gaisser and 
Goodman\cite{g&g} and shown in Table \ref{tab:atmo}. Several experiments
measure the ratio of observed events to expected events to be less than one.
For Kamioka the data is divided into low and high energy samples.
The low energy events have a muon (with energy typically
less than 1 GeV) that is contained in the detector.
Muons are identified in two ways by both Kamioka and IMB.
The first way involves identification of the \kov ring, which is significantly 
different for electrons and muons.
(There is also a significant difference between the efficiency for detection
of electrons and muons, but this is presumably in the detector simulation.)
The second method uses the fact that muons that stop in H$_2$O usually decay, 
allowing the observation of decay Michel electrons to facilitate identification 
of the muon.
The ratio is consistent using both methods of identification.

\begin{table}[t]
\caption{Ratio of $\numu$ to $\nue$ observed events divided by the ratio
of expected events for the different atmospheric oscillation experiments.}
\label{tab:atmo}
\vspace{0.4cm}
\begin{center}
\begin{tabular}{|c|c|c|}
\hline 
Experiment & Exposure & Flavor Ratio \\
\hline
~~& kT - year & $\mu$ / e \\
IMB1 & 3.8 & $0.68 \pm 0.08$ \\
Kamioka Ring & 7.7 & $0.60 \pm 0.06$ \\
Kamioka Decay &  & $0.69 \pm 0.06$ \\
IMB - 3 Ring & 7.7 & $0.54 \pm 0.05$ \\
IMB - 3 Decay &  & $0.64 \pm 0.07$ \\
Frejus Contained & 2.0 & $0.87 \pm 0.13$ \\
Soudan & 1.0 & $0.64 \pm 0.19$ \\
NUSEX & 0.5 & $0.99 \pm 0.29$ \\
\hline
\end{tabular}

\end{center}

\end{table}

For the fully contained events the muon energy is sufficiently low that
the cross section is over predicted 
by the Fermi gas model\cite{fermi_gas} used by all 
experiments. 
This is a valid concern but seems unlikely to explain the large 
and persistent effect
observed.
Moreover, this problem should not afflict the second sample of partially 
contained events for which the energy is typically $\sim 5 \gev$.
It is our view, and that of the experimenters, that the 
$\nu_{\mu}/ \nu_e$ ratio is suppressed, 
although some systematic effects are still to be understood.

\begin{figure}
\centerline{\psfig{figure=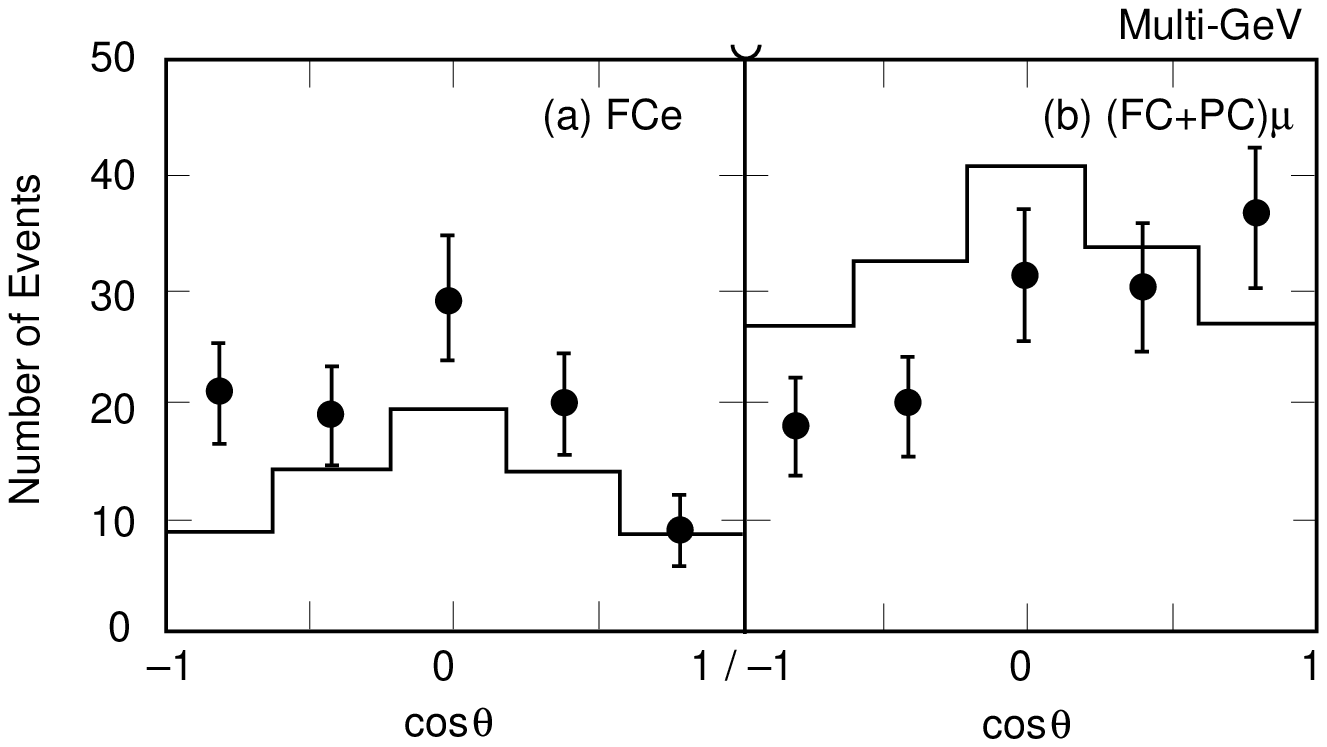,width=4.7in,silent=}}
\caption{Zenith-angle distributions from the Kamioka experiment
for (a) the electron-like events
and (b) the muon-like events. The circles with error bars show the
data and the histogram the MC simulation (without oscillations).}
\label{fig:k1}
\end{figure}

\begin{figure}
\centerline{\psfig{figure=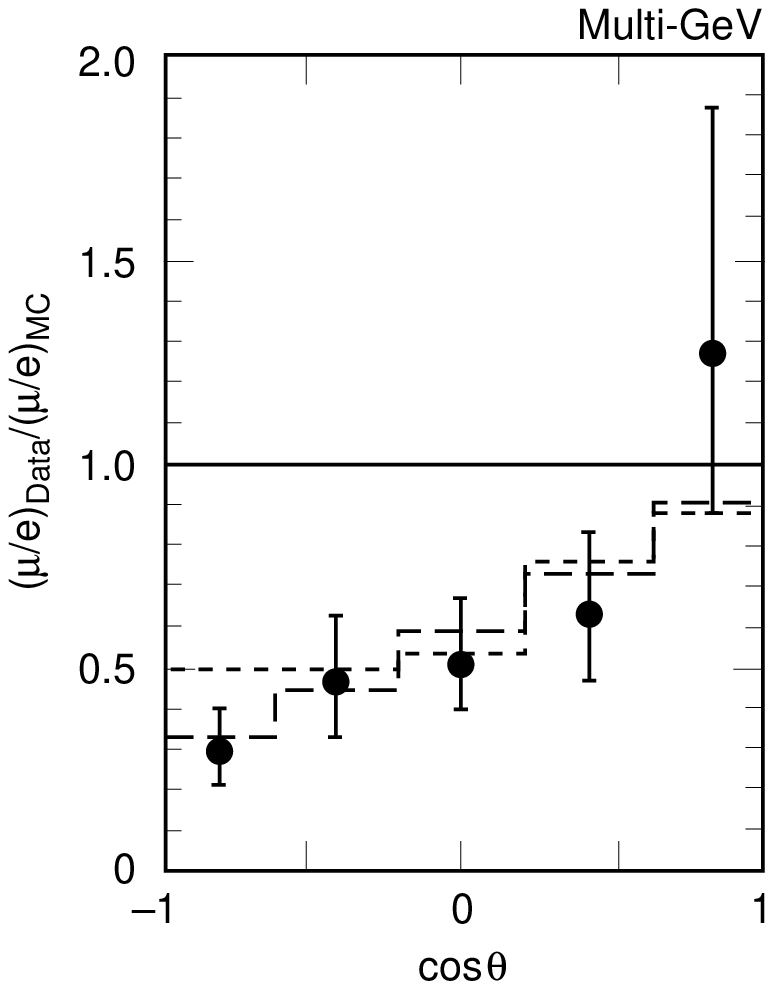,width=4.7in,silent=}}
\caption{Zenith-angle distribution from the Kamioka experiment
of $(\mu/e)_{data}/(\mu/e)_{MC}$.
The circles with error bars show the data. Also shown are the
expectations from the MC simulations with neutrino oscillations for
parameter sets $(\Delta m^2,\sin^22\theta)$ corresponding to the
best-fit values to the multi-GeV data for $\nu_\mu \rightarrow \nu_e$
(($1.8 \times 10^{-2}$ eV$^2$, 1.0), dashes) and $\nu_\mu \rightarrow \nu_\tau$
(($1.6 \times 10^{-2}$ eV$^2$, 1.0), dots) oscillations.}
\label{fig:k2}
\end{figure}

The Kamioka group has reported a zenith angle dependence of the 
apparent atmospheric neutrino deficit\cite{atmoskam},
based on their examination of higher energy atmospheric neutrino events
(visible energy greater than 1.3 GeV and average energy equal to 6 GeV).
In this instance the observed $\nu_\mu/\nu_e$ 
ratio was $0.57 \pm 0.08 \pm 0.07$,
consistent with the earlier observations, but with a strong dependence
of the ratio on the zenith angle of the projected neutrino direction. The
ratio for these high energy neutrinos coming from directly overhead
(zenith angle of about $0^0$) was reported as $1.3 \pm 0.4$. Thus, the high
energy muon neutrinos coming from large distances (zenith angle greater
than $90^0$) evidenced large depletion, while high energy muon neutrinos
coming from overhead showed no such loss. 
The $\nu_{\mu}$ and $\nu_e$ distributions are shown in Fig. \ref{fig:k1},
and the ratio of these two distributions is shown in Fig. \ref{fig:k2}.
The probability of $\nu_{\mu}$
disappearance is given by equation~\ref{eq:P}, which depends
upon $L$, the distance from the neutrino's origin in km and  $E_\nu$, the
neutrino energy in GeV. 
The fact that
little disappearance effect is observed for a zenith angle of $\sim 0^0$ means
that
$$\sin^2(1.27 \Delta m^2 L/E_\nu) \sim 0$$ or
$$L/E_\nu << 1/1.27 \Delta m^2.$$
With $L \sim 30$ and $E_\nu \sim 6$, one finds that $\Delta m^2 << 0.1$.
This small value for $\Delta m^2$ has greatly influenced a number of 
subsequent proposals using large detectors at 
hundreds of kilometers from the neutrino source to investigate
the phenomena associated with the atmospheric observations.

\begin{figure}
\centerline{\psfig{figure=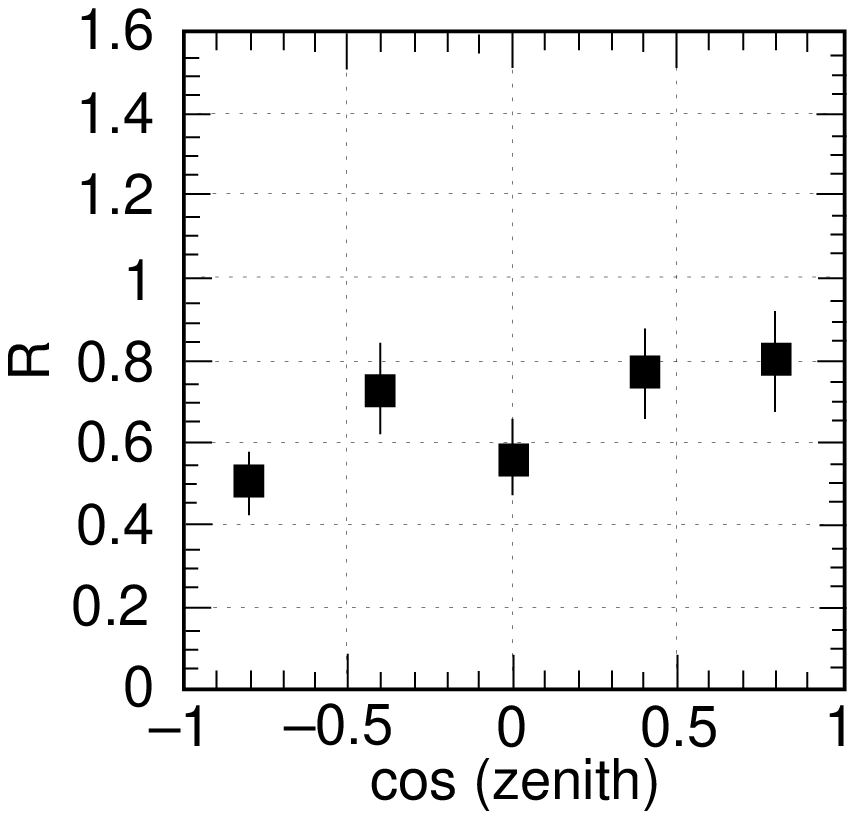,width=4.7in,silent=}}
\caption{The Super Kamioka preliminary zenith angle distribution for the
sub-GeV contained events.}
\label{fig:superk_low}
\end{figure}

\begin{figure}
\centerline{\psfig{figure=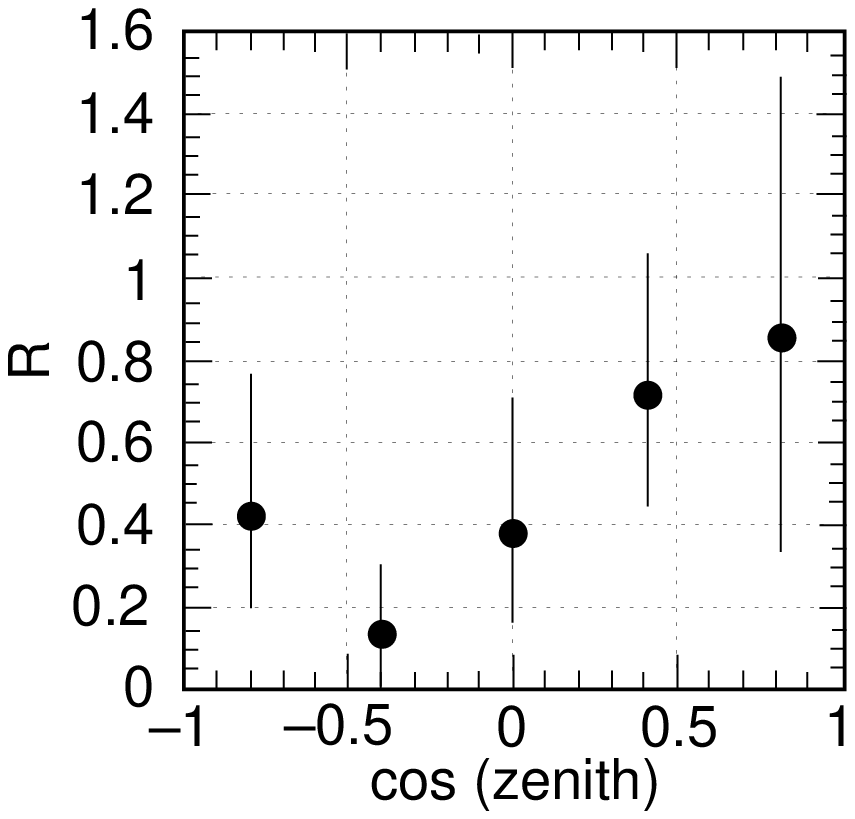,width=4.7in,silent=}}
\caption{The Super Kamioka preliminary zenith angle distribution for the
multi-GeV contained events.}
\label{fig:superk_high}
\end{figure}

The significance of the reported dependence is not large, and 
it appears, moreover, that the observed zenith angle dependence reported
by Kamioka may not be a real effect. 
A preprint from the IMB collaboration\cite{imb}
reports no such dependence.   Also, as shown in Figs. \ref{fig:superk_low}
and \ref{fig:superk_high},
the Super
Kamioka collaboration sees a zenith angle dependence that is consistent
with being gently sloped or flat. This new result
changes the range of values for $\Delta m^2$. 
Averaged over the neutrino energy, $<\sin^2(1.27 \Delta m^2 L/E_\nu)>\sim 0.5$.
If the zenith angle distribution is indeed flat, then
$$L/E_\nu > \pi/4(1.27 \Delta m^2) = 0.62/\Delta m^2$$
which leads to $\Delta m^2 > 0.15$ eV$^2$.
This value of $\Delta m^2$ is compatible with the LSND observation.
Thus, it appears that 
there may be a common value of $\Delta m^2$ for $\nu_\mu$ 
disappearance and for $\nu_\mu \rightarrow \nu_e$
oscillations (see section 3.5).

In this proposal we assume that the evidence for 
a discrepancy in the ratio of $\nu_\mu$ events to $\nu_e$ events over that 
expected is significant enough to receive serious attention.
However, we also assume that the value of $\dmsq$ that is deduced from the 
Kamioka zenith angle distribution may be taken with caution.
The atmospheric neutrino problem makes the
$\nu_\mu$ disappearance experiment an important part of the BooNE proposal. 

\subsection{Theoretical Interpretation of the Data}

It is difficult to make a fit to the experimental evidence described above
with the general form given by Eq. \ref{3-gen osc}.
Therefore, simplifications must be made, resulting in various models
with different assumptions. For example, the ``maximal
mixing'' model\cite{max-mix} has a mixing matrix given by 
\[
U_{ij}=\frac 1{\sqrt{3}}\left( 
\begin{array}{lll}
\omega _1 & \omega _2 & \omega _3 \\ 
\omega _1 & \omega _2 & \omega _3 \\ 
\omega _1 & \omega _2 & \omega _3
\end{array}
\right) {\rm ~~where~~ } \omega _i=\left( 
\begin{array}{l}
Complex\;Cube \\ 
Roots\;of\;Unity
\end{array}
\right) 
\]
With $\Delta m_{12}^2<10^{-11}$eV$^2$ and $\Delta m_{23}^2\approx 10^{-2}$eV$%
^2$, this model can accommodate the solar, atmospheric and reactor
measurements but not the LSND signal.

Various models with two of the masses being ``almost degenerate'' can
produce interesting oscillation patterns. In these models, the mixing matrix
has large mixing between the degenerate partners and would be given
approximately by
\[
|U_{\alpha i}|=\left(
\begin{array}{lll}
0.99 & 0.03 & 0.03 \\
0.03 & 0.71 & 0.71 \\
0.03 & 0.71 & 0.71
\end{array}
\right)
\]
An example of such a model is one with $m_1\approx m_2\ll m_3$ which leads
to two $\Delta m^2$ scales given by a small $\Delta m_{12}^2$ and a larger $%
\Delta m_{13}^2\approx \Delta m_{23}^2$. In this model, each oscillation
channel ($\nu _\alpha \rightarrow \nu _\beta $) can be treated using the
two-generation formalism with $\sin ^22\theta \approx 4\left| U_{\alpha
\,3}\right| ^2\left| U_{\beta \,3}\right| ^2$and the appropriate $\Delta m^2$%
. With effectively only two mass scales, it would seem hard to explain the
three $\Delta m^2$ scales associated with solar ($\Delta m^2\approx 10^{-5}$
eV$^2$), atmospheric ($\Delta m^2\approx 10^{-2}$ eV$^2$), and LSND ($\Delta
m^2\approx 10^{-1}$ eV$^2$) experiments. Cardall and Fuller\cite{fuller}
have suggested that the atmospheric and LSND $\Delta m^2$ values could be
similar if one discounts the zenith angle dependence; the common value would
be in the range $0.1<\Delta m_{LSND,\;atmos}^2<0.5$ eV$^2$. The solar
oscillation signal is accommodated by having $\Delta m_{12}^2\approx 10^{-5}$
eV$^2$. With the above mixing matrix and mass hierarchy, the solar,
atmospheric and LSND data can all be explained by oscillations through the
various mass eigenstates.
 
\noindent $
\begin{array}{lllclll}
\text{LSND:} & \nu _\mu  & \rightarrow  & \left(
\begin{array}{l}
\nu _1 \\
\nu _3
\end{array}
\right) \text{ or }\left(
\begin{array}{l}
\nu _2 \\
\nu _3
\end{array}
\right)  & \rightarrow  & \nu _e &
\begin{array}{l}
\Delta m^2\approx 0.1-0.5\text{ eV}^2 \\
\sin ^22\theta \approx 5\times 10^{-3}
\end{array}
\\
\text{Atmospheric:} & \nu _\mu  & \rightarrow  & \left(
\begin{array}{l}
\nu _2 \\
\nu _3
\end{array}
\right)  & \rightarrow  & \nu _\tau  &
\begin{array}{l}
\Delta m^2\approx 0.1-0.5\text{ eV}^2 \\
\sin ^22\theta \approx 1
\end{array}
\\
\text{Solar:} & \nu _e & \rightarrow  & \left(
\begin{array}{l}
\nu _1 \\
\nu _2
\end{array}
\right)  & \rightarrow  & \nu _\mu  &
\begin{array}{l}
\Delta m^2\approx 10^{-5}\text{ eV}^2 \\
\sin ^22\theta \approx 3\times 10^{-3}
\end{array}
\end{array}
$
 
\noindent 
The MiniBooNE experiment has the sensitivity to test for both $\nu
_\mu \rightarrow \nu _e$ and $\nu _\mu \rightarrow \nu _\tau $ oscillations
in the $\Delta m^2=0.1-0.5$ eV$^2$ mass region at the above mixing levels
and, thus, offers an opportunity to explore this possible inclusive
scenario.

\subsection{MiniBooNE and Future Experiments}

The goal of MiniBooNE is to address the LSND result and the
atmospheric neutrino deficit questions.   
Two other future accelerator-based experiments are aimed at 
addressing these results; however,
neither are likely to have definitive results within the next five
years.  These experiments are 
KARMEN and MINOS. 

The KARMEN experiment\cite{karmen} 
is of similar design to LSND. It has a total mass of about 56 tons
and is located about 17.5 m from the 200 $\mu$A proton source,
compared to the 167 ton mass of LSND, which is located about 30 m from
the 1000 $\mu$A LAMPF source. The KARMEN experiment has completed an
important upgrade of their veto detector shielding which will reduce the
cosmic ray background by a factor of $\sim 40$ from previous running.
However, despite greatly reducing their
chief background with the new veto shield, KARMEN's event sample will be
less than the LSND sample due to the accelerator intensity. 
As a result, KARMEN may able
to verify the LSND signal over the $\Delta m^2 > 1$ eV$^2$ range,
except perhaps for the full 6 eV$^2$ region, after 2-3 years of data
taking, but the signal will be less significant than 
the LSND signal.  MiniBooNE will cover the full LSND range,
particularly the $\Delta m^2 < 1$ eV$^2$ region with an
expectation of $\sim 5\sigma$ for a signal.
Furthermore,
because the neutrino source and experimental signal are so similar to
that of LSND, KARMEN will not provide the systematic check required.
One wants to verify the LSND oscillation signal in a new
energy region where the backgrounds and systematic are
completely different.   

The MINOS experiment, scheduled to begin taking data in 2003, has
sensitivity over only part of the LSND signal range. For this experiment, $%
\nu _\mu \rightarrow \nu _\tau $ oscillations can be detected through three
``disappearance'' measurements: the absolute $CC$ rates, the $NC/CC$ ratio,
and the $E_\nu $ distribution in the two detectors. With two years of data,
the experiment will cover the $\nu _\mu \rightarrow \nu _\tau $ region with $%
\sin ^22\theta >2\times 10^{-2}$ and $\Delta m^2>10^{-2}-10^{-3}$ eV$^2$ as
shown in Fig. \ref{fig:mudisap}. MINOS was specifically designed to address
the atmospheric neutrino deficit and has limited sensitivity for $\sin
^22\theta <0.02$. If the mixing is above about $0.1$, MINOS will be able to
determine the $\nu _\mu \rightarrow \nu _\tau $ oscillation parameters from
the $E_\nu $ dependence of the observed effects. For $\nu _\mu \rightarrow
\nu _e$ oscillations in the LSND signal region, the MINOS experiment will
see an effect if $\sin ^22\theta >\sim 0.01$ but will only be able to
determine oscillation parameters if $\sin ^22\theta >\sim 0.1$. The $\sin
^22\theta >0.05$ region of the LSND signal has already been ruled out by the
Bugey reactor experiment,\cite{bugey} which means that the MINOS experiment
will not have sensitivity to address the LSND signal region in any detail.
Thus, it is unlikely that MINOS will present definitive results
on either the LSND or atmospheric neutrino problem before the latter half of
the next decade.

The ICARUS Experiment, which has not yet been approved to run at CERN,
may have comparable sensitivity to MiniBooNE.
This experiment plans to expose 600 ton liquid Ar TPC modules
located in the Gran Sasso tunnel to high energy neutrinos from 
CERN at a distance of 732 km. The proposal has the potential 
of high sensitivity and a reach down to $\Delta m^2 \sim 10^{-3}$ eV$^2$.
The required large scale liquid Ar technology has been developed
by the ICARUS collaboration. In addition, there are also
plans to place another 600 ton liquid Ar TPC module in the
Jura at a distance of 17 km from the existing CERN-SPS wide band
neutrino beam. The range in $L/E_\nu$ is 0.17-3.4 km/GeV and
7.3-146 km/GeV for the Jura and Gran Sasso detectors, respectively.
The $L/E_\nu$ for BooNE is 0.3-5.0 km/GeV, covering what we believe is the most
important range in $\Delta m^2$. Note that ICARUS in the Jura location
has a sensitivity to $\nu_\mu \rightarrow \nu_e$ oscillations that
is comparable to MiniBooNE.


\section {MiniBooNE Capabilities and Issues}

{\it The MiniBooNE experiment is designed to make use of a high intensity
neutrino beam from the 8 GeV Booster along with a large, mineral-oil-based
detector. The beam has low }$\nu _e${\it \ contamination and the detector
has high efficiency and precision particle identification ability down to
low energies. The detection and particle identification techniques are
derived from the LSND experiment and are, thus, well understood. This
section outlines the MiniBooNE capabilities and issues with subsequent
sections providing the detailed descriptions and simulation results.}

As shown in the previous sections, there is a need for experiments to probe $%
\nu _\mu \rightarrow \nu _e$ oscillations in the $0.01-1.0$ eV$^2$ mass
region with mixings down to $\sin ^22\theta \approx 10^{-3}-10^{-2}$. For $%
\Delta m^2=0.1$ eV$^2$, an experiment needs an $L/E$ value between 2 to 4.
Since the rate from a neutrino source falls as $1/L^2$, the most cost
effective way to probe this region is with the smallest L for the available
E$_\nu $ value. A neutrino beam from the 8 GeV Fermilab Booster is almost
optimal for this region using an $L$ value of $\sim 1$ km combined with $%
0.15<E_\nu <1.0$ GeV. In addition, a sensitive search for $\nu _\mu
\rightarrow \nu _e$ oscillations requires low intrinsic $\nu _e$ background
in the beam. A Booster $\nu $ beam would have low $\nu _e$ background event
rate $(\nu _e/\nu _\mu \approx 10^{-3})$ since the K production source is
reduced with the low primary proton energy and a short decay pipe can be
used, thus, minimizing the $\mu $ decay source.

A low-energy Fermilab $\nu $ experiment is possible due to the very high
proton fluxes available from the Booster. Combining the high proton flux
with a high efficiency horn focused secondary beam will provide over 100,000 
$\nu _\mu $ events/kt-yr at 1 km from the source. The high intensity and
rapid cycling of the Booster does make important requirements for the beam
design. There will need to be significant shielding to meet radiation safety
requirements. The beam elements including the high-current horn will need to
be reliable at cycle rates of $\approx 5$ Hz. A new underground enclosure
must be constructed to house and provide access to the beam, the 30 m decay
pipe and the dump. The neutrino beam will be directed horizontally at
7 m below the ground level, thereby minimizing any surface radiation.
In order to make the experimental costs low, the
enclosure needs to be made using conventional 
construction techniques and existing
shielding materials where possible.

The proposed experiment would start with a single detector with the goal of
probing the LSND mass region and establishing definitive indications of
neutrino oscillations. If a positive signal is observed, this first stage
would be followed up using a two detector experiment in the same neutrino
beam. For the initial single detector experiment, MiniBooNE, accurate $\nu $
flux and background calculations will be needed. Modern simulation tools can
accurately model beam transport and scraping but need to be augmented and
checked using direct measurements. A primary ingredient for the simulation
is the particle production spectrum from the 8 GeV proton interactions in
the thick production target. Data does exist from Argonne and KEK but will
need to be supplemented by measurements taken with the actual neutrino beam.
Position and profile monitors similar to those used by NuTeV and BNL 776 in
the primary beam, decay pipe, and post-dump region will provide important
constraints on the beam simulation (see section 5.6). 
It is also possible to measure the
momentum spectrum by allowing a tiny part of the secondary beam to pass
through the dump into a small spectrometer. Analysis methods have been
developed in previous neutrino experiments to use the measured neutrino
spectrum in the detector to determine the secondary particle fractions. For
example, this technique has been successfully used to fix the charged $\pi /K
$ fraction in the NuTeV experiment and, thus, fix the $\nu _e$ background
from $K^{+}/K^{-}$ decay.

The MiniBooNE experiment needs a detector with a large fiducial mass and
good particle identification for neutrino events in the $0.15<E_\nu <2.0$
GeV energy region. At these low energies, a totally active detector is
necessary. A detector based on a large volume of dilute, mineral-oil-based
scintillator is both cost effective and very powerful for particle
identification using the techniques developed for the LSND experiment.
Mineral oil has several advantages over distilled water as a detection
medium: a) more \v Cerenkov light, b) no purification requirements, c) shorter
radiation length, d) less $\mu ^{-}$ capture probability, and e) the ability
to form a dilute scintillator mixture for better particle identification. 
Many of the critical detector components are available from the LSND
experiment including the 1220 eight-inch photomultiplier tubes (PMTs) with
readout and data acquisition system. The dilute scintillator will be
contained in a double wall tank 11 m in diameter by 11 m high leading to a
fiducial volume corresponding to $\sim 400$ tons. The 70 cm region between
the two tanks will be filled with regular liquid scintillator and
instrumented with PMTs as a veto. In order to minimize costs, the tank will
use standard commercial oil/water tank technology and safety standards and
be partially buried. 

Particle mis-identification is an important limitation for the $\nu _\mu
\rightarrow \nu _e$ oscillation measurement. Using the techniques developed
and tested in the LSND experiment, the mis-identification of $\nu _\mu $
events as $\nu _e$ events can be reduced to the $\approx 2\times 10^{-3}$
level while keeping the $\nu _e$ and $\nu _\mu $ efficiency above 50\% . The
identification techniques are based on the spatial and time correlation of
the detected \v Cerenkov and scintillation light by the PMTs lining the walls
of the detector.  A further strength of the experiment is the ability to
measure these backgrounds from the preponderance of events which are
identified correctly. Muon neutrino events are identified by observing an
exiting $\mu $ or a decay electron with the correct time and position
correlation with the $\mu $ track. For $\nu _\mu $ events, about 8\% of the
outgoing $\mu ^{-}$s are captured before decay and must be identified by
the spatial and time signature of the \v Cerenkov and scintillation light. In
this type of detector, a $\mu $ has a more focused \v Cerenkov ring and
relatively more scintillation light than an $e$ or $\gamma $ interaction.
(Scintillation light can be isolated due to its much broader time
distribution.) The signature for a  $\nu _e$ event is a diffuse \v Cerenkov
ring with relatively low scintillation light. This signature can also be
satisfied by $\nu _\mu N\rightarrow \nu _\mu \pi ^0X$ events where the $%
\gamma $s from the $\pi ^0$ decay are identified as electrons. The cross
section and E$_\nu $ threshold for the $\pi ^0$ production reduces the rate
substantially for this process with respect to $\nu _\mu N\rightarrow \mu X$
scattering but a rejection factor of 100 is still needed to reduce this
background to the $10^{-3}$ level. This rejection is available for the
MiniBooNE detector by detecting the second $\gamma $ or by detecting late
scintillation light from an energetic recoil proton or outgoing muon.
Further reductions are also possible by minimizing the high energy component
of the $\nu _\mu $ beam where $\pi ^0$ production is largest. The high
energy component can be reduced by moving the detector off the beam axis and
by adding dispersion in the beam optics. 

The 8 GeV Booster $\nu $ beam using a focusing horn system and a 30 m decay
pipe will provide $\sim 50,000\left( 10,000\right) $ identified $\nu _\mu (%
\overline{\nu }_\mu )$ events/yr in the 400 ton MiniBooNE detector. For the $%
\nu _\mu \rightarrow \nu _e$ oscillation measurement, the beam-related and
mis-identification backgrounds can be held to less than $\sim 0.5\%$ and be
known  with a systematic uncertainty of $10\%$. If oscillations exist at
the LSND level, MiniBooNE should see several hundred anomalous $\nu
_e$ events over a beam-related (mis-identification) background of 100 (150)
events, establishing the signal at the $>5\;\sigma $ level.  If no
oscillation signal is observed, the experiment will exclude $\nu _\mu
\rightarrow \nu _e$ oscillations with $\sin ^22\theta >6\times 10^{-4}$ for
large $\Delta m^2$ and $\Delta m^2>0.01$ eV$^2$ for $\sin ^22\theta =1$ as
shown in Fig.~\ref{fig:mueapp}.

\section{Neutrino Beam}

{\it The neutrino beam will be fed by the 8 GeV proton Booster
operating at a rate of 5 Hz. The secondary pions will be focused by a
double-horn system and then decay in a 30 m long decay volume.} \\

\subsection{General Considerations}

Several criteria were used to design the neutrino beam.   The first 
is to maximize the low energy flux, which provides the sensitivity to
low $\Delta m^2$.  Lower energy neutrinos also result in fewer 
interactions producing $\pi^0$s which may be misidentified in the 
detector as a $\nu_e$ event.
The second criteria is to maintain a small 
($<3 \times 10^{-3}$)
ratio of $\nue$ ($\nuebar$) to $\numu$ ($\numubar$)
while still obtaining high statistics.   The $\nu_e$ beam background
results mainly from muon and kaon decays.

The solution is a low energy beam made by focusing pions and kaons 
produced in a target that may be as much as two interaction lengths
long.  For this letter of intent we have designed a beam using
focusing  elements similar
to a design used at Brookhaven\cite{bnlbeam}.
Practical issues limit the length of this space to perhaps a third of a pion 
decay length or less depending on the $\nue$ contamination
requirement, and here we have chosen 30 m.

\subsection{Booster}

The Fermilab booster synchrotron (Booster) accelerates protons to 8 $\gev/c$
at a peak repetition rate of 15$\hz$.  A fraction of these pulses will be 
injected into the Main Injector, and 
we assume that 5$\hz$ of the remaining pulses will 
be available for the neutrino beam.
Single turn extraction of the Booster beam will provide an output beam pulse 
1$\musec$ long, which makes rejection of cosmic ray background straightforward
and provides a timing signal suitable for data acquisition.
The RF frequency 
at full energy is 50$\mhz$, which gives a 20$\nsec$ substructure for the
extracted beam and which is nearly ideal for the kind of neutrino detector 
envisaged for this proposal.
This time separation facilitates event reconstruction, and the 1.5 ns 
$\sigma$ of the
RF bucket is an excellent addition to particle identification.
We propose to operate as an extension to the  transfer line that has 
been constructed for Main Injector operation at
MI 10.  

\subsection{Focusing System}

Focusing is typically managed in two elements; one acting as a collection lens
close to the target and one to make the momentum acceptance broad.
The object is to transform as much pion and kaon flux as possible into a 
parallel beam and into an acceptable decay space.

At this time, the relatively old fashioned horn-focused
technology has provided 
a satisfactory solution, and we present it here as an existence 
proof rather than a final solution.
A horn design from Brookhaven National Laboratory\cite{bnlbeam}
is used for the results presented in this letter of intent.
This focusing system was designed to optimize the
neutrino flux for 28 GeV protons on target and   
without concern for the $\nue$ background.
Although with this design we do meet the criteria of high flux and 
low backgrounds, our ongoing studies indicate that 
simple modifications will produce improvements.  

Alternatives to the horn design are also being pursued.
In particular, a Helical Quad focusing system with permanent
magnets is under consideration.

\subsection{Design Issues Related to $\nu_e$ Backgrounds}

A principal concern  is that the $\nue$
impurity in the beam be low enough that this background to the oscillation
signal be sufficiently small.   This intrinsic $\nu_e$ background
results mainly from muon and kaon decays.    In order to reduce this 
background we plan to:
\begin{itemize}
\item Use a relatively short, 30 m decay pipe, which 
optimizes the pion decays while
minimizing the muon decay background.
\item Run in negative and positive focusing polarities.   
The positive polarity provides a very high $\nu_\mu$ flux.   The
negative focusing causes the background from charged kaons to be lower
because $K^-$ production is smaller than $K^+$ production
at a proton energy of 8 $\gev$, providing a cleaner sample and an
opportunity to check our systematic understanding of kaon production
in the positive mode.
\item Use a relatively small radius decay tunnel.   This minimizes the
backgrounds from  $K_L$, which 
have a significant branching ratio into $\nuebar$ and $\nue$ 
and are not focused.
\end{itemize}

We have simulated the beam with the GEANT 3.21/FLUKA transport code
to verify that the 
achievement of low $\nue$ contamination is straightforward 
as long as the proton energy is low, and these results are 
presented in the following sections.
It is clear that with the geometry described above the experiment is 
feasible and backgrounds can be reduced to an acceptable level.
It should be noted that 
in the calculations presented below a narrow decay pipe has not
been implemented, 
so the background estimates are somewhat conservative.

There is one improvement that we have considered which has the promise 
of improving the beam in a substantial way.
If between the first focusing element and the second a bending magnet is 
inserted with approximately 25 degrees of bend, 
then the secondary beam will be 
deflected so that neutral kaons will not penetrate to the decay region, 
thus lowering the $\nue$ contamination from this source.  
A magnet with 2T and 1.5 m long is adequate, and lowers the neutral
kaon $\nue$ contamination by a factor of ten for a 30 meter decay region. 
This option also lowers 
the high energy component of the beam that penetrates the decay 
region can also be restricted, and hence the high energy flux in the $\numu$ 
beam will be much reduced.
This is of considerable advantage because the neutral current $\pizero$ 
production that is a potential background to 
the oscillation signal will also be reduced considerably.
The general point that is raised by this suggestion is that careful detailed 
design of this beam is likely to improve the background situation described 
in this proposal, making the experiment significantly better and more 
sensitive.  This detailed design will be pursued in the near future.

\subsection{Neutrino Fluxes}

\begin{figure}
\centerline{\psfig{figure=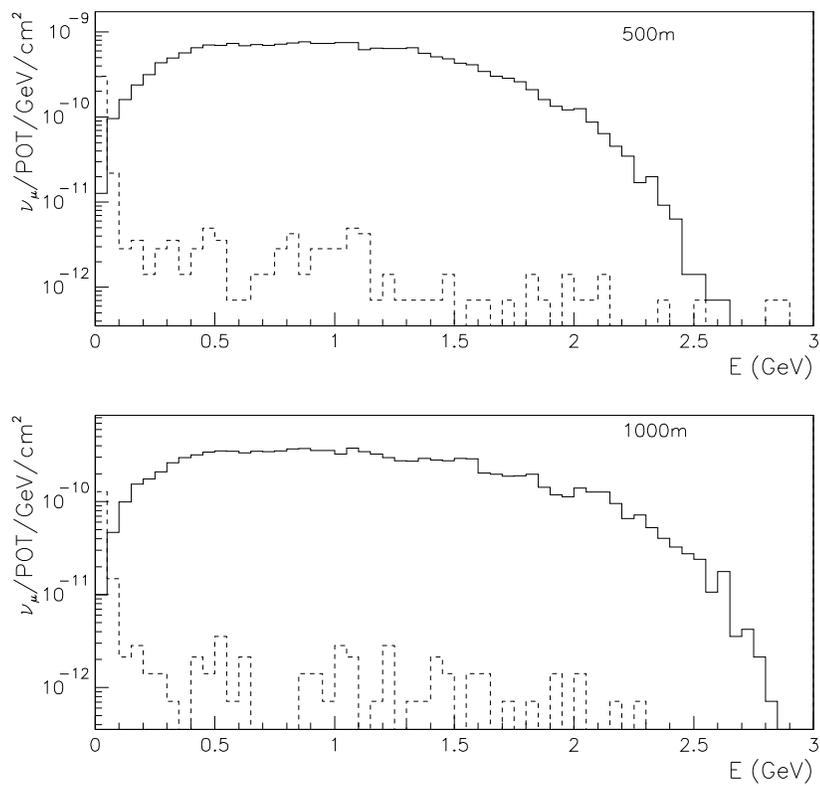,width=4.7in,silent=}}
\caption{Flux of $\nu_\mu$ (solid histogram) and $\nu_e$ 
(dashed histogram) from a 30m decay length 
beam line at 500 m and 1000 m 
from the target (at the position of the beam dump).}
\label{fig:pos_flux}
\end{figure}

Fig. \ref{fig:pos_flux} shows the $\nu_\mu$ flux (solid histogram)
from a 30m decay length 
beam line at 500 m and 1000 m from the target.
There are three sources of $\nue$ (or $\nuebar$) in a decay-in-flight neutrino 
beam.
Charged kaon decay lengths are sufficiently short (one tenth that of pions) 
that they decay dominantly in the vicinity of the target, and about
5\% of the charged kaon decay rate leads to $\nu_e$ from the branching mode 
$K^+ \rightarrow \pizero e^+ \nu_e$.
Secondly, since virtually all of the pions that decay produce a muon, this muon
may also decay occasionally into $\nu_e$.
Although this fraction is small, it is still comparable with the $\nu_e$ 
background produced 
from charged kaons.
A simple guideline is that the ratio of $\nu_e$ to $\nu_\mu$ is proportional to 
the fraction of the pions that decay or to the length of the decay region.
A third contribution to the $\nu_e$ flux in the beam comes from $K_L$ decay.
At the energy of the Booster, 
$K^+$ are produced much more frequently than $K^-$, and
the $K_L$ are produced roughly half as frequently as $K^+$.
The $\nu_e$ flux (dashed histogram) from the 30m decay 
length beam line at 500 m and 1000 m 
from the target is also shown in Fig. \ref{fig:pos_flux}.
Data will be taken with both positive and negative focused beam.
A 30 m decay region is a good compromise that maintains high flux 
while
keeping the contribution from the $\pimue$ decay chain and
from kaon decay at an acceptable
 level.
Finally, the transverse shape of the decay region should be made 
somewhat narrow to keep the
 contribution from $K_L$ at a minimum.

\subsection{Beam Monitoring}

The primary beam position, profile, and intensity will need to be
monitored to maintain the highest possible flux and to constrain
beam simulations.  
The harsh environment of the booster beam precludes the
continuous use of standard beam line swics and sems.  BPMs will be
used to measure beam position, and a beam current toroid can be used
to measure beam intensity (BPMs can also measure intensity, but the
toroid is necessary for an initial calibration).  Two X-Y BPM pairs
are needed, one close to the target and one farther back to measure
the targeting angle and position.  A wire secondary emission monitor,
to measure beam profiles, will be moved into the beam periodically and
then retracted to prevent radiation damage to the secondary emission
efficiency.

The secondary (pion and Kaon) beam profiles and intensities will be
monitored with a large swic/ion-chamber similar to that used by NuTeV
in their NW2 enclosure.  The exact positioning of this device depends
on the design of the focusing horn and decay pipe.  Another swic will
be positioned downstream of the hadron beam dump to measure the
profiles and intensities of the muon beam (again, similar to the
current NuTeV experiment NW4 enclosure).

Most of the necessary readout electronics for the beam monitors are
already available at Fermilab.  We will use the standard swic scanners
for swic and wire sem readout.  The BPMs and beam current toroid can
use electronics similar (and perhaps identical) to those already in
use in the transfer gallery (copies of this electronics would have to
be fabricated).\cite{Drucker}

\subsection{Neutrino Flux Angular and $r$ Dependence}

\begin{figure}
\centerline{\psfig{figure=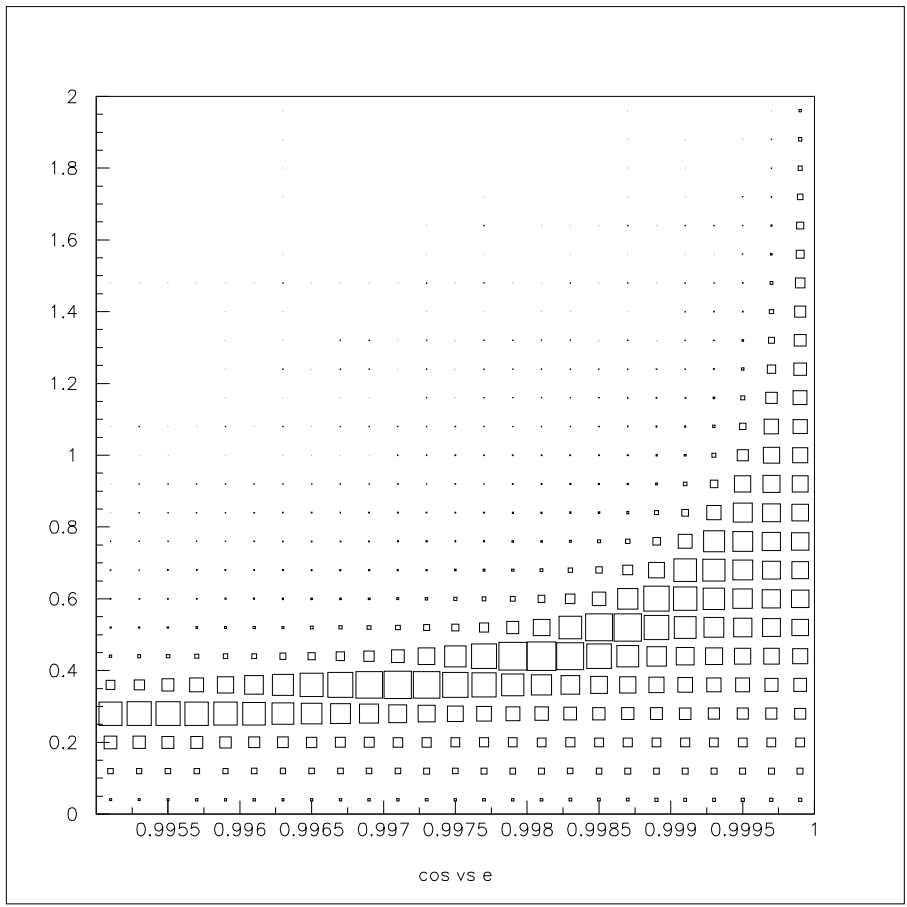,width=4.7in,silent=}}
\caption{The neutrino flux as a function of neutrino
energy and $\cos \theta$, where $\theta$ is the angle of the
neutrino relative to the incident proton direction.}
\label{fig:eang}
\end{figure}

Fig. \ref{fig:eang} shows the neutrino flux as a function of neutrino
energy and $\cos \theta$, where $\theta$ is the angle of the
neutrino relative to the incident proton direction. Although the
total neutrino flux peaks at $\theta = 0$, for a given neutrino energy
the maximum neutrino flux
varies as a function of $\theta$. For example, for $E_\nu = 0.6$ GeV
the neutrino flux peaks at $\cos \theta = 0.999$ and for 
$E_\nu = 0.4$ GeV the peak flux occurs at $\cos \theta = 0.998$, or 
$\theta=63$ mrad.
Therefore, we may plan to offset the detector from the forward
direction in order to decrease the higher energy neutrino flux and
increase the lower energy neutrino flux. Our present design has the
center of the detector tank at $\theta = 0.008$ or $\cos \theta =
0.99997$.   In the final design, we may further offset the detector.
This can be accomplished by repositioning the detector on site or 
by bending the beam downward.   

\begin{figure}
\centerline{
\psfig{figure=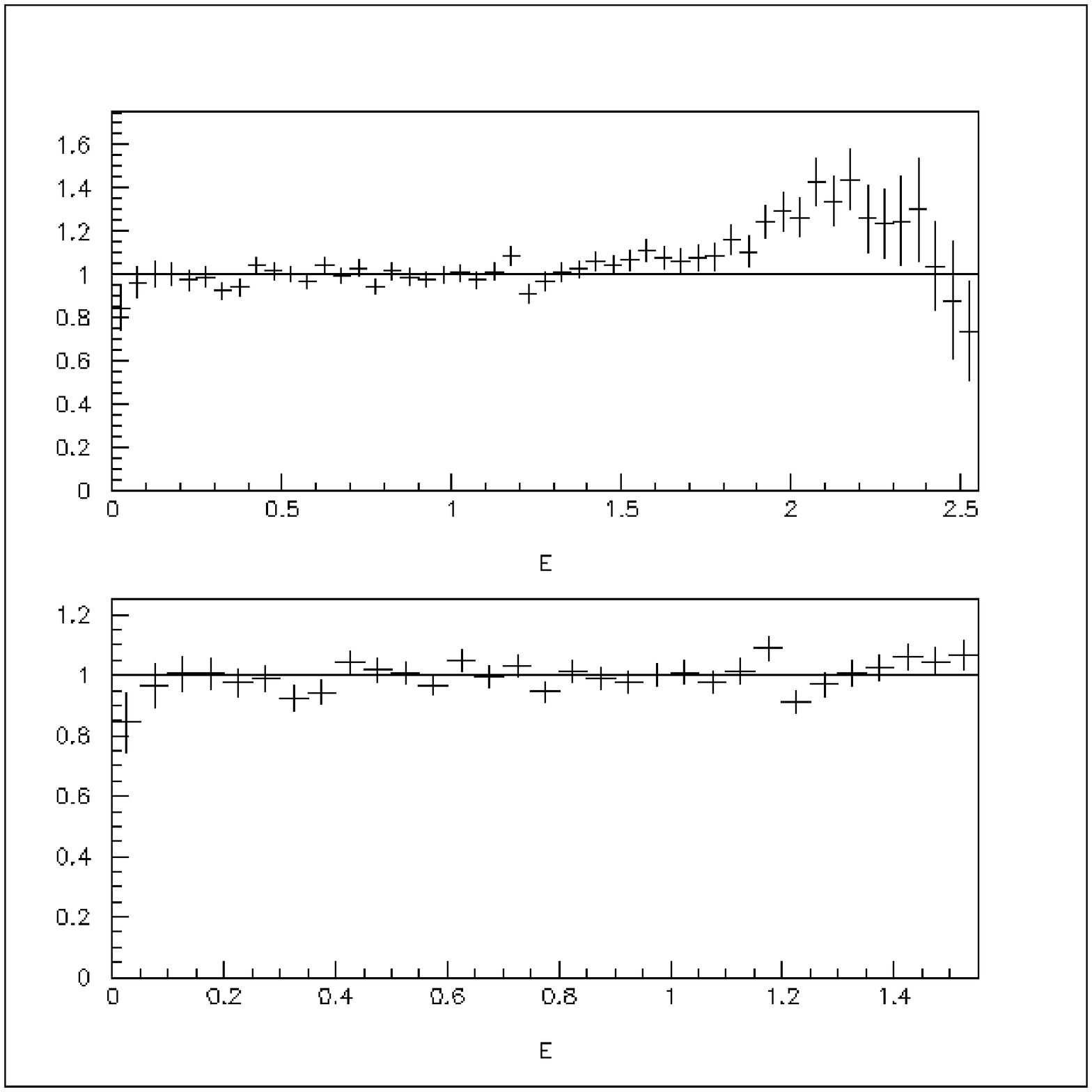,bbllx=0bp,bblly=0bp,bburx=600bp,bbury=800bp,height=5.5in,width=6.in,angle=90,clip=T}
}
\caption{The ratio of the neutrino flux passing
through the BooNE detector at 1000 m
compared to 500 m as a function of neutrino energy in GeV. The ratio
is normalized such that it equals one for a pure $1/r^2$ dependence.
The upper figure shows the full energy range, while the lower figure enlarges
the region of interest.}
\label{fig:rdep}
\end{figure}

In the second phase of BooNE, when interactions in 
two detectors are compared to
determine the oscillation parameters, it is important that the 
flux show a simple $1/r^2$ spatial dependence.  The beam line has been 
designed with this in mind.
The neutrino flux, as shown in Fig. \ref{fig:pos_flux} at 500 and 1000 m,
has a $1/r^2$ spatial dependence to very good
approximation. 
Fig. \ref{fig:rdep} shows the ratio of the neutrino flux passing
through the BooNE detector at 1000 m
compared to 500 m as a function of neutrino energy. 
Although some deviation is
observed between 1.5 and 2.5 GeV  (Fig.
\ref{fig:rdep}a), no deviation
from a pure $1/r^2$ dependence is observed for neutrino energies
less than 1.5 GeV, which is the range that 
will be used in the analysis (Fig. 
\ref{fig:rdep}b).
Averaged over the entire 
neutrino energy range,
there are about 3.91 (instead of 4.00)
times as many neutrinos passing through the
detector at 500 m compared to 1000 m.

\subsection{Beam Construction Issues}

\begin{figure}
\centerline{
\psfig{figure=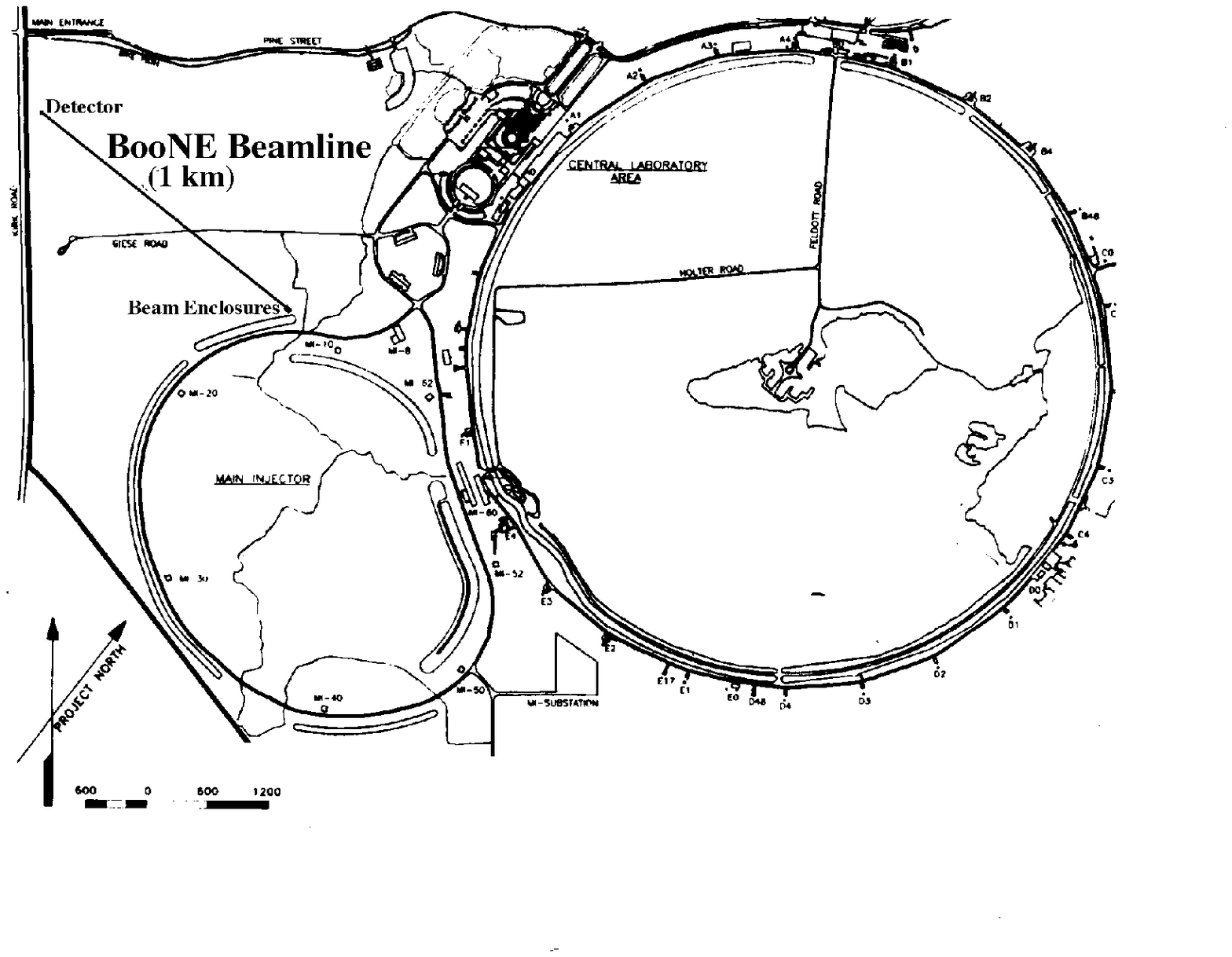,bbllx=0bp,bblly=100bp,bburx=300bp,bbury=400bp,height=5.5in,width=5.5in,clip=T}
}
\caption{Southwest region of the Fermilab site.  The BooNE beam line
originates at MI10.   Targeting and Dump Halls are located north 
of the MI cooling ponds.   The detector is located at the northwest 
edge of the schematic between Giese, Kirk and Pine roads. The beam line is 
1 km in length}
\label{fig:layout}
\end{figure}

A schematic of the site  
of the BooNE beam line is shown in Fig.~\ref{fig:layout}.
The beam line begins at MI10.   The target and dump enclosures are
positioned beyond the Main Injector cooling ponds.   The beam direction
points northwest.   The detector is located 1 km from the dump
enclosure.   

The beam will originate at enclosure MI10.
A transport pipe will be installed to bring
the beam to the targeting enclosure.   The length of the transport
pipe will depend upon the best position of the enclosure relative to
other structures, duct-work and ponds.   
For cost estimation purposes, we assume 
100 m.   Beam transport will use permanent magnets.

\begin{figure}
\psfig{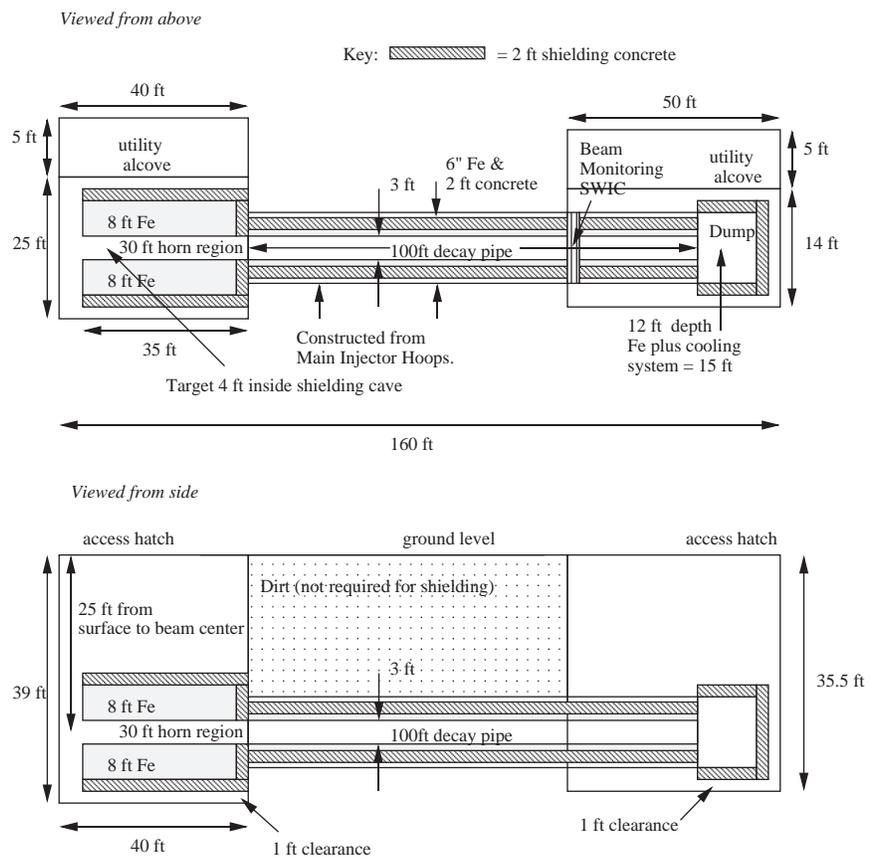}
\caption{Schematic of enclosures with estimated shielding.}
\label{fig:enc}
\end{figure}

Schematics of the
beam targeting enclosure, meson decay pipe and dump enclosure are shown in 
Fig. \ref{fig:enc}.   The targeting and dump enclosures are
designed to permit visual inspection of the caves, if required for
radiation safety monitoring.   The decay pipe will be surrounded by
a structure made from Main Injector Hoops, which are 8 ft $\times$ 10 ft.

Shielding represents a significant cost because of the intensity of the beam.
We have worked with Don Cossairt, Associate Head, Fermilab Radiation 
Protection Group, to develop a preliminary estimate for the required
shielding.   The shielding shown in Fig. \ref{fig:enc} 
represents a conservative estimate and was
determined using energy and intensity scaling
of existing approved beam lines.  Scaling can be used for 
approximating overall cost, but {\it will not} be used to
determine actual shielding 
requirements in the final design.  The CASIM program will be run
during Summer, 1997, in order to accurately determine the shielding needed.   

\section{Detector}

{\it The MiniBooNE detector will consist of an 11.0 m diameter 
double-wall tank
that is filled with 600 tons of mineral oil and covered on the
inside with 1220 eight-inch phototubes (10\% coverage). The outer
volume of the tank will serve as a veto for outgoing and incoming particles.} \\

\begin{figure}
\centerline{\psfig{figure=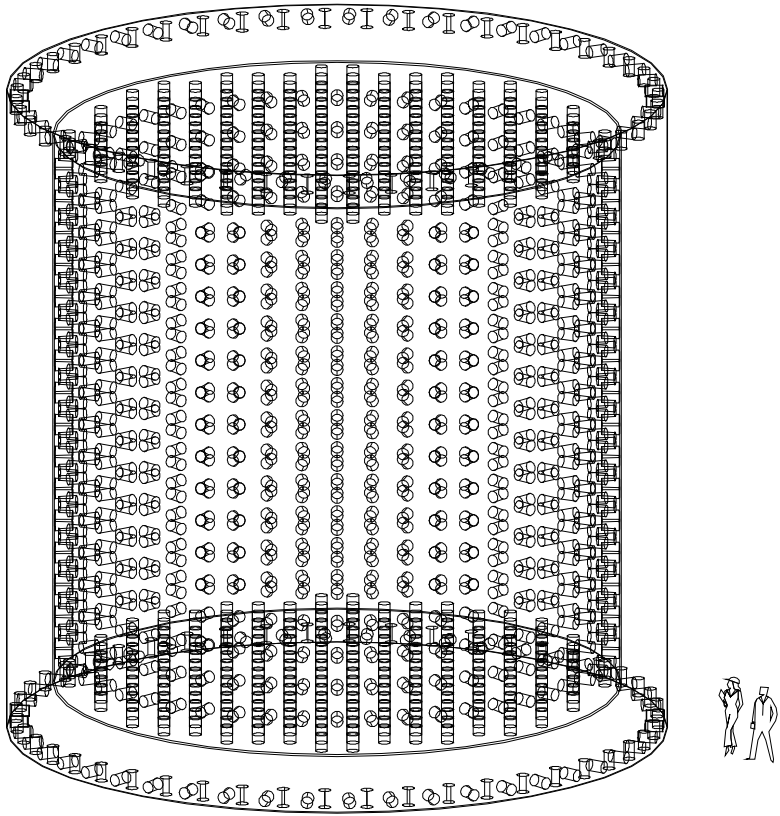,width=4.7in,silent=}}
\caption{Schematic of the proposed detector.}
\label{fig:detector}
\end{figure}

The BooNE experiment consists of
a 600 ton imaging \v Cerenkov detector.
The detector will be placed 
1000 m from the neutrino source, such
that the energy distributions of $\nue C \rightarrow e^- X$
and $\nu_{\mu} C \rightarrow \mu^- X$ quasi-elastic events
in the detector determine $\dmsq$. 
Characteristics of the detector are shown in Table \ref{tab:detector}.
A schematic of the proposed detector is shown in Fig. \ref{fig:detector}.
The detector is a double-wall cylindrical tank, 11 m in diameter and
11 m high.
The outer volume serves as a veto shield for uncontained
events and is filled with a high light-output liquid scintillator.
The inner (main detector) volume is a right cylinder 9.6 m in diameter
and is filled with mineral oil plus a small concentration
of b-PBD scintillator such that 90\% of the light generated by
electrons in the tank is \v Cerenkov light and 10\% is scintillation light.
A total of 1220 eight-inch photomultiplier tubes (PMTs) cover 10\% 
of the area of the inner tank.

The outer (veto) volume is equally important for identifying
uncontained events and for vetoing cosmic rays.   The scintillator oil
in this region will be 
viewed by 292 PMTs distributed along the 
circumference of the two endcaps.
The PMTs in the top ring, for example, are arranged so that half are 
sensitive to scintillator light from the cylindrical jacket and half to 
light from the top disk of veto fluid. 
The veto volume is separated from the main detector volume by a 2.5 cm
steel wall.
This wall provides optical isolation between the veto and central detector,
and it helps reduce the background from $\pizero$ production.
Neutrino induced 
$\pizero$ production is a serious background in this type of detector because of the
similar behavior of gammas and electrons (see Chapter 5 for details). 
If only one of the gammas from a $\pizero$ decay converts in the main detector, 
that $\pizero$ will look like a $\nue$ appearance candidate.
The 2.5 cm wall will convert about 2/3 of the escaping gammas into showers 
detectable by the veto shell.

In this chapter, details of the detector design are explained.
Issues related to the construction are considered and  
the electronics and data acquisition are described.
The PMTs, electronics, and data acquisition will be the same as those used
for the LSND experiment.

\begin{table}[t]
\caption{The characteristics of the BooNE detector.}
\label{tab:detector}
\vspace{0.4cm}
\begin{center}
\begin{tabular}{|c|c|c|}
\hline 
&Detector&Veto\\
\hline
Volume&695 m$^3$&400 m$^3$\\
Mass&591 t&340 t\\
PMTs&1220(10\%)&292 \\
Fiducial Volume&449 m$^3$&\\
Fiducial Mass&382 t&\\
\hline
\end{tabular}
\end{center}
\end{table}

\subsection{Detector Construction Issues}

\begin{figure}
\psfig{figure=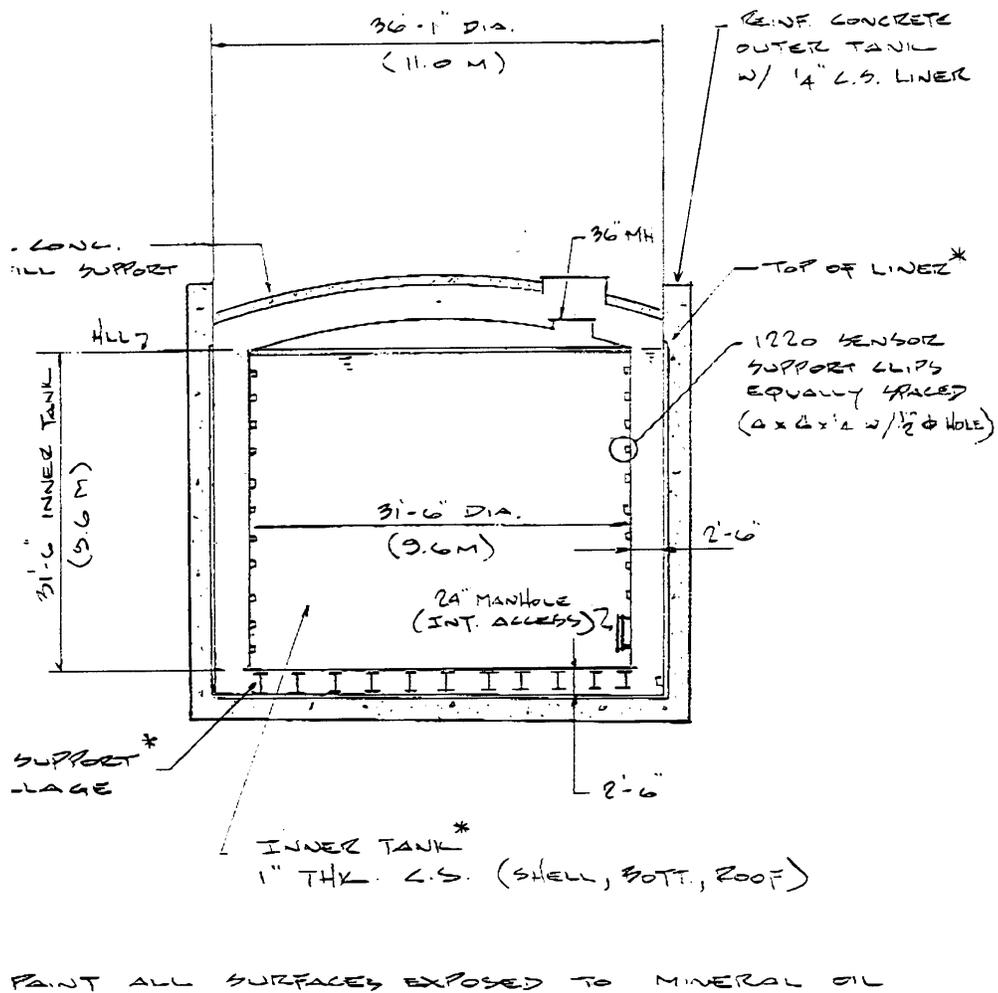,bbllx=100bp,bblly=100bp,bburx=600bp,bbury=700bp,height=6.in,width=5.5in,clip=T}
\caption{Schematic of the detector for cost estimates.}
\label{fig:tanks}
\end{figure}

Several designs for the two-tank structure are under
consideration, with one example shown in Figure~\ref{fig:tanks}. 
The tanks will be constructed from carbon steel.
All portions of the detector exposed to oil will be painted with
black epoxy.   In the design shown, 
phototube cables exit the tanks through the manholes 
at the top.  In the inner tank, the photoubes will be mounted on the 
top, bottom and walls of the tank.   Each region between 
supporting beams at the bottom will be instrumented with 
multiple phototubes for vetoing cosmic rays.   These phototubes are 
positioned by accessing the region between the tanks through the lower
manhole.  Veto phototubes will also be located 
around the circumference between the two tanks 
at the top facing downward.   Solid
scintillator will be placed across the top of the inner tank to
provide the upper veto.  The tanks are surrounded by a concrete
structure which provides support.  

The detector will be placed partially below ground level.
If the above design is used, in order to provide 
protection from weather and dirt while phototubes are installed,
an inflatable temporary dome will be placed over the 
top of the tank.   This will be connected to a temporary 
portakamp where supplies can be stored.    After installation
of the phototubes, the inner and outer tanks will be simultaneously
filled with oil.   The dome and portakamp will be removed.  
Finally, a dirt embankment forming a hill over the 
tank will be installed to provide shielding for the 
above-ground-level portion of the detector.
Electronics will be located in a trailer near to the detector.

In this letter of intent, 
the cost estimates for the construction of the tank system
are from discussions with 
estimators at the Chicago Bridge and Iron Company of
Plainfield, Illinois and are for the tank design shown
in Fig.~\ref{fig:tanks}.~\cite{Ned}

\subsection{Electronics and DAQ}

The front-end electronics and the data acquisition system for BooNE must
record the time of occurrence and the charge in each photomultiplier
tube pulse, define a trigger for event selection, and provide for
efficient event building.  The system must provide time recording
resolution of $\sim 0.2$
 ns and charge resolution of $\sim 0.2$ pe. The system must
span time scales of millisecond durations to allow for the correlations
of prompt events with signals from subsequent decays.  The dynamic range
of the charge digitizer must be adequate to meet the needs of the event
fitter and the energies that are expected from the neutrino scattering
processes.  Finally, the system must be low cost and fabricated with
appropriate technologies.  Since neutrino experiments tend to run for
long periods, the system must operate reliably with low maintenance.  An
updated version of the LSND architecture can meet these requirements.
In the following paragraphs the relevant features of the LSND
electronics and data acquisition system are reviewed and the updates and
modifications for the BooNE experiment are described.

\begin{figure}
\centerline{\psfig{figure=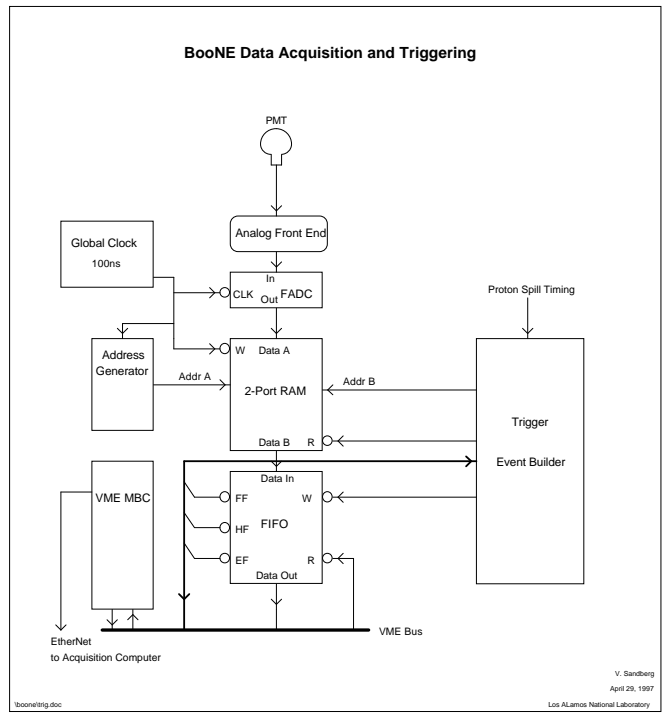,width=4.7in,silent=}}
\caption{The proposed BooNE PMT channel data acquisition.}
\label{fig:daq1}
\end{figure}

The LSND data acquisition architecture consists of a fast analog front
end, a flash analog to digital conversion (FADC) stage, and digital
memories to store and sort the digitized data.  A block diagram of this
system as modified for BooNE is shown in Fig. \ref{fig:daq1}.  The system operates
synchronously with a GPS-referenced clock providing the 100ns time base.
This clock is scaled by a binary counter, the output of which is used as
the address to store the digitized PMT data.

\begin{figure}
\centerline{\psfig{figure=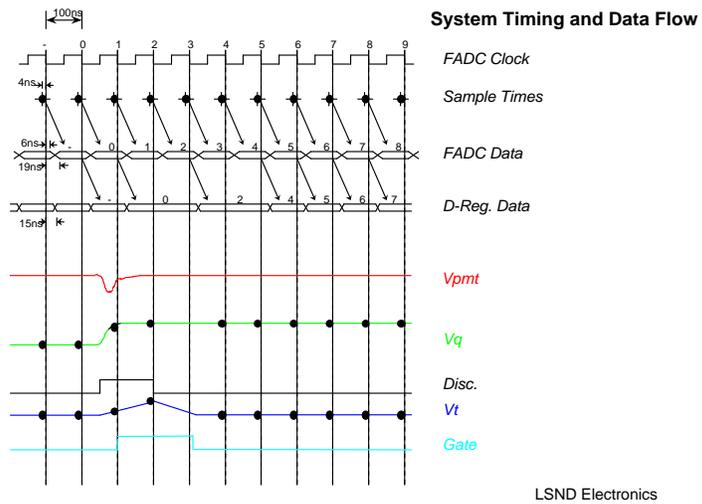,width=4.7in,silent=}}
\caption{PMT pulse acquisition and signal timing relationships.}
\label{fig:daq2}
\end{figure}

Each photomultiplier tube signal is connected to a channel that performs
a leading edge time interval determination (the interval being between
the PMT pulse and the following edge of the global clock) and an
integration of the pulses.  The signal acquisition and data timing are
shown in Fig. \ref{fig:daq2}.  
These signals are sampled every 100ns by FADCs and the
digitized data are written into dual ported memories addressed by the
global time base.  At each clock tick the charge and time analog
voltages are digitized by the FADCs and written into the dual ported
memory.  The address is then incremented and the cycle repeated.  This
part of the acquisition system is free running.  The memories are
necessarily finite in size and hold a finite time history (for LSND
204.8 $\mu$s with 100 ns resolution).  They are configured as
circular buffers, that is, the memory is overwritten after the clock has
cycled through the address space of the memory.  This provides for
temporary storage of the immediate past history for a time long enough
for a trigger decision to be made by a monoboard computer.  It also
allows for a pretrigger to determine the state of the detector prior to
the arrival of an event, that is, it can ``look backwards in time''.

\begin{figure}
\centerline{\psfig{figure=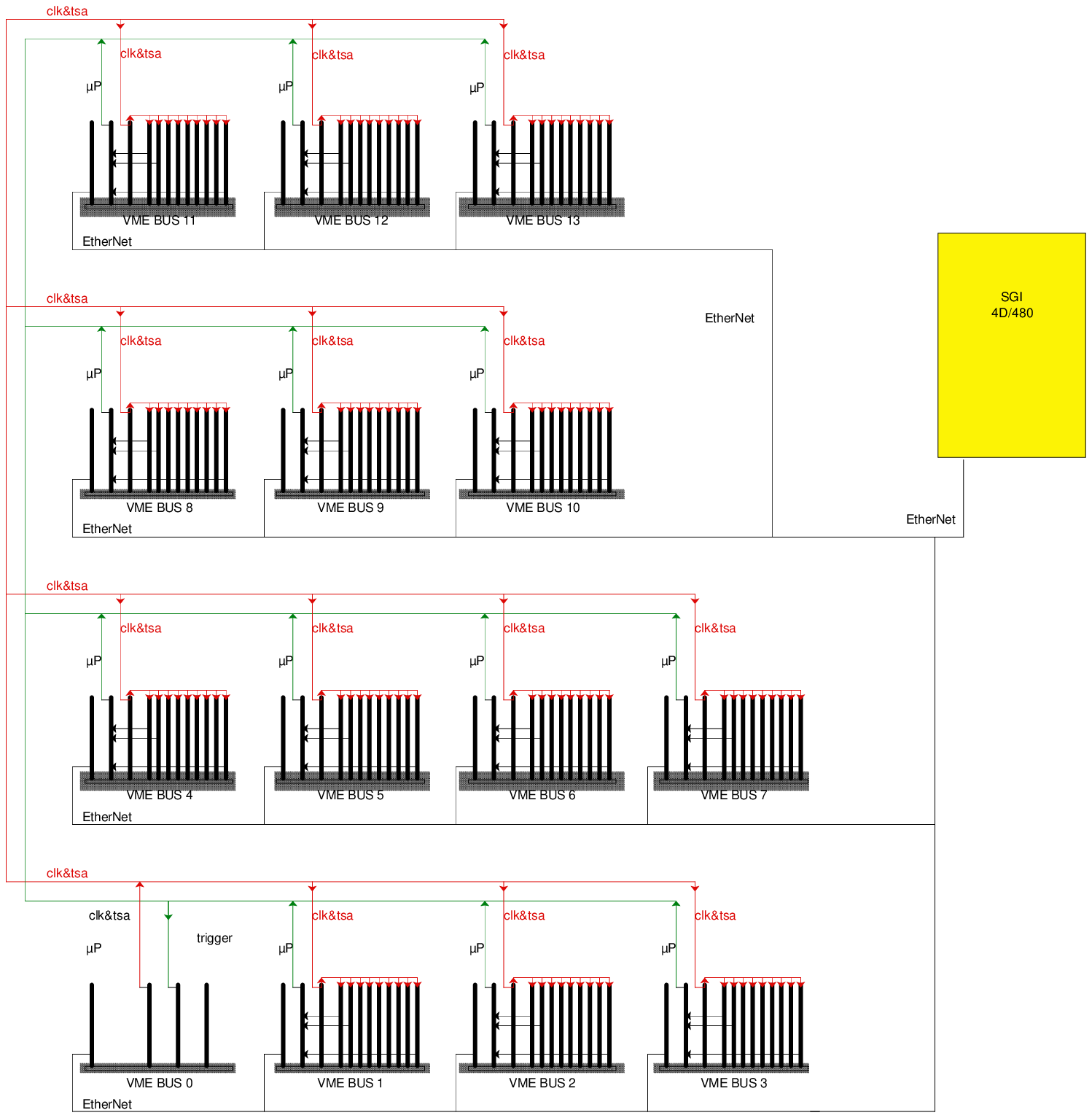,width=4.7in,silent=}}
\caption{System configuration of the individual VME QT crates and their
interconnections.}
\label{fig:daq3}
\end{figure}

Data are extracted from the front-end dual-ported memories and built
into events by the trigger system.  Each front-end leading-edge
discriminator provides a digital pulse to a global summation module,
which computes the digital sum of all of the tubes that fired in the
previous global-clock interval.  A bank of digital comparators signal
when the sum exceeds a preset value and the global-clock time when this
happens is loaded into a FIFO memory.  The trigger monoboard computer
(for LSND it was an MVME-167) continually reads this FIFO.  From the
resulting times series of multiplicity data the trigger determines the
candidate events and initiates an event build by broadcasting those
global-clock times when the particular event took place to the other
address port of the dual-ported memories that contain the PMT charge and
time data.  The system transfers the data at these ``event time stamps''
from their temporary location to FIFOs located adjacent to the
dual-ported memories, where the data resides until read by other
monoboard computers.  These computers (MVME-167) calibrate and compact
the data, assemble them into ethernet packets, and send them to the
experiment's analysis computer.  A system configuration diagram is shown
in Fig. \ref{fig:daq3}.

Modifications to this system for BooNE would include the design of the
appropriate trigger hardware and the modifications to the trigger
acquisition code.  The front-end electronics is being redesigned to make
use of faster, cheaper memory to allow a larger circular buffer for a
longer recorded history.  A revision of the digital architecture (all
components reside in VME crates) is being undertaken to simplify the
dedicated data and addressing lines and keep compatibility with the VME
standards. This aims to improve the reliability and reduce system costs.
A new, all digital time interpolator is in design to enhance the time
resolution to the 0.2 ns specification.  Finally, a 10-bit or 12-bit
FADC is being considered to meet the resolution-dynamic range
requirements on the charge measurement.  Alternatively, a dual slope
analog front end is being designed that has piecewise linear gains and
which matches the desired low pulse height sensitivity and large pulse
height dynamic range onto the 2V range of the FADCs.

The size of each PMT window is determined by the trigger computer's
software, as is the triggering decision.  It is straightforward to
modify the trigger code to configure the above system for running with
the beam time structure of the Fermilab booster.  For example, the
front-end electronics would be left free running so that the prehistory
could be acquired.  When the booster delivered a proton pulse to the
target, a timing pulse would be generated and sent to an unused ``PMT
input'' in the DAQ system.  This would allow a precise time determination
of when to begin considering events as neutrino candidates.  The front
end continues to record the activities in the detector, then after all
anticipated events are finished, the trigger would read all of the data
in the dual-ported memories.  (It is more likely that such a read would
begin sooner to remove any time overhead, but his is a detail in the
design.  The LSND system runs continuously and only reads that part of
memory that contains a window 500 ns wide around a candidate event's
time stamp.  The LSND system notes whether or not the accelerator has
spilled beam on the neutrino production target by recording the beam
status through an unused ``PMT channel'' as described above.  Beam off
events are handled identically to beam on events.  The trigger does not
know if the accelerator is off or on and as such is unbiased.  This also
allows a large sample of the background to be collected, which is of
great importance for LSND.  It is not as critical to the BooNE
experiment, which takes place at higher energies.  It may be relevant to
nu-p elastic scattering, however.)

\subsection{Trigger Operating Modes}

The data transfer rate on the VME backplane of a QT crate is set by the
monoboard computer (MBC).  For LSND a Motorola MVME167 was used and had
a maximum transfer rate of 20 MBytes/s, which was realized by one VME
bus cycle every 200 ns.  With this rate it required 13 $\mu$s to collect
one time-stamp record of a full loaded QT crate, consisting of 128 PMT
channels plus one header long word (260 bytes of data).  To collect the
six time-stamp records of an event molecule required 78 $\mu$s.

Conventional LSND DAQ operates with the trigger controlling the event
builds and only selected event molecules being read from QT memory into
MBC memory.  In this mode the VME back plane traffic is low.  In view of
the (relatively) low repetition rate of the booster, several alternative
modes of operation are possible.  A ``zero-threshold mode'' that records
the full detector history and a hybrid scheme that expands the history
window are possible, as is a reduced threshold mode that would provide
some data sparsification at the trigger level.  For example, the BooNE
DAQ could operate in several different scenarios:

\noindent 1. LSND-mode where by an initiating ``high-threshold trigger'' flags a
``low-threshold mode'' and selected events are built.  This has low back plane
traffic, but at the cost of a threshold for events.

\noindent 2. Zero threshold-mode where the front-end clock is enabled 
$\sim 50 \mu$s before
the proton spill begins, the time of the spill is recorded, and the
clock is allowed to run for (204.8-50.0) $\mu$s before being shut off and the
entire QT memory is read into MBC memory.  This mode collects the entire
time history of the detector from $\sim 50 \mu$s prior to the spill to 
$\sim 150 \mu$s
after the spill.  It requires between 13 ms to 26 ms of back plane time to
read the QT memory into the MBC.  Assuming a worst case of a repetition
rate of 15 spills per second (66 ms per spill), this mode allows for an
overhead of at least 40ms between spills for MBC processing and full
event building to the main acquisition computer.

\noindent 3. A hybrid scheme that takes advantage of the buffering capability of
the LSND front-end memory system.   This scheme leaves the front-end
clock running continuously and initiates QT-memory-to-FIFO transfers
$\sim 20 \mu$s before the spill.  
This QT-memory-to-FIFO link is maintained until
the FIFOs are half full, where upon the trigger interrupts this link and
operates now in the LSND-low-threshold mode.  The QT-FIFO readout is
begun by the MBC as soon as there is data in the FIFOs, which would
occur at the $\sim 20 \mu$s before spill time.  Notice that the data flow is
event driven with  throttling controls (N.B., the use of the half-full
flags on the QT FIFOs) and FIFO buffer capacity to allow for smooth
transfer between the no-threshold and low-threshold modes of operation.

\noindent 4.  A modified hybrid scheme which combines the timing with respect to
the beam spill with a low threshold (but not zero) to collect events
that had, e.g., greater than 4 PMTs on; rather than collect everything,
which would be mostly zeros and burden the data acquisition for no good
reason.

To implement these schemes on the LSND DAQ system requires the addition
of a hardware card to signal the proton spill timing and the writing of
new trigger software.

\section{Detector Simulation}

{\it The GEANT-based detector Monte Carlo package accurately 
simulates events in the detector and allows the determination
of the event reconstruction resolutions and the particle identification
efficiencies. All of these techniques have been developed and tested
at the LSND experiment.} \\ 

\subsection{GEANT Implementation}

The proposed BooNE detector was simulated using the GEANT
Monte Carlo package. A double wall cylindrical tank filled with
mineral oil was coded into the program with dimensions as described
in Section 6. An approximation of a PMT was modeled and multiple copies
were positioned as shown in Fig. \ref{fig:detector}.

The GEANT implementation (V3.21) simulates the production of
\v Cerenkov light for particles with velocities above threshold and
tracks this light through the media using the (wavelength-dependent)
optical properties of the mineral oil. This optical photon tracking
was also used to track scintillation photons that were generated
isotropically and in proportion to the amount of energy lost by
charged particles in the mineral oil.
These optical photons were tagged as \v Cerenkov or scintillation photons
so that the amount of scintillation light could be adjusted
in the reconstruction phase. If an optical photon intersected with the
photocathode surface of the PMT volume, it was considered detected
with an efficiency equal to the quantum efficiency of the PMT
(also wavelength-dependent). These PMT ``hits'' were written to an
ntuple along with the veto and input particle information.

In order to study the response of the proposed detector, first,
single particle ($\mu^{\pm},e^{\pm},\pi^0$) events were generated
isotropically over a range of energies (50-1200 MeV) within the
mineral oil volume of the inner tank.  Then, to better understand
the relative rates for the signal and background reactions,
multi-particle events of interest to the BooNE experiment were
simulated. The most important reactions studied were:
$\nu_{e} C \rightarrow e^- X$, $\nu_{\mu} C \rightarrow \mu^- X$, and
$\nu_{\mu} C \rightarrow \nu_{\mu} \pi^0 X$. The final states of
these reactions often include an energetic proton which was included in
the simulation. These events were generated within the fiducial volume of the
inner tank and were weighted by the estimated cross sections
and neutrino fluxes.

\subsection{Event Reconstruction}

Events are reconstructed using a fitting algorithm that is similar
to that developed for LSND. As in LSND, the chi square of the
position and angle fits ($\chi_r$ and $\chi_a$) are minimized
to determine the event position and direction. In addition,
these minimized chi squares, together with 
the fraction of PMTs with late hits ($\chi_t$), provide excellent
particle identification and low $\mu^{\pm} - e^{\pm}$ and
$\pi^0 - e^{\pm}$ misidentification. 

To determine the event position and
time (corresponding to the midpoint of the track), the position
chi square, $\chi_r$, is minimized. $\chi_r$ is defined as
$$\chi_r = \sum q_i\times (t_i-t_o-r_i/v)^2/Q ~ ,$$
where $q_i$ is the charge of hit phototube $i$, 
$t_i$ is the time of hit phototube i, $t_o$ is the fitted
event time, $r_i$ is the distance between the fitted position and
phototube i, v is the velocity of light in oil (20.4 cm/ns), $Q$ is the
total number of photoelectrons in the event and the
sum is over all hit phototubes. 
Similarly, the event direction
(for particles above \v Cerenkov threshold) is determined by minimizing
the angle chi square, $\chi_a$. $\chi_a$ is defined as
$$\chi_a = \sum q_i\times (\alpha_i-47^o)^2/Q ~ ,$$
where $\alpha_i$ is the angle between the fitted direction
and the line segment extending from the fitted position
to hit phototube i, $47^o$ is the \v Cerenkov angle for $\beta \sim 1$
particles in mineral oil, and the 
sum is over all hit phototubes. 
With the track direction determined,
the initial event vertex is found by using the measured track energy
(which is proportional to the total number of photoelectrons in
the event) to extrapolate
back to the vertex.

\begin{figure}
\centerline{\psfig{figure=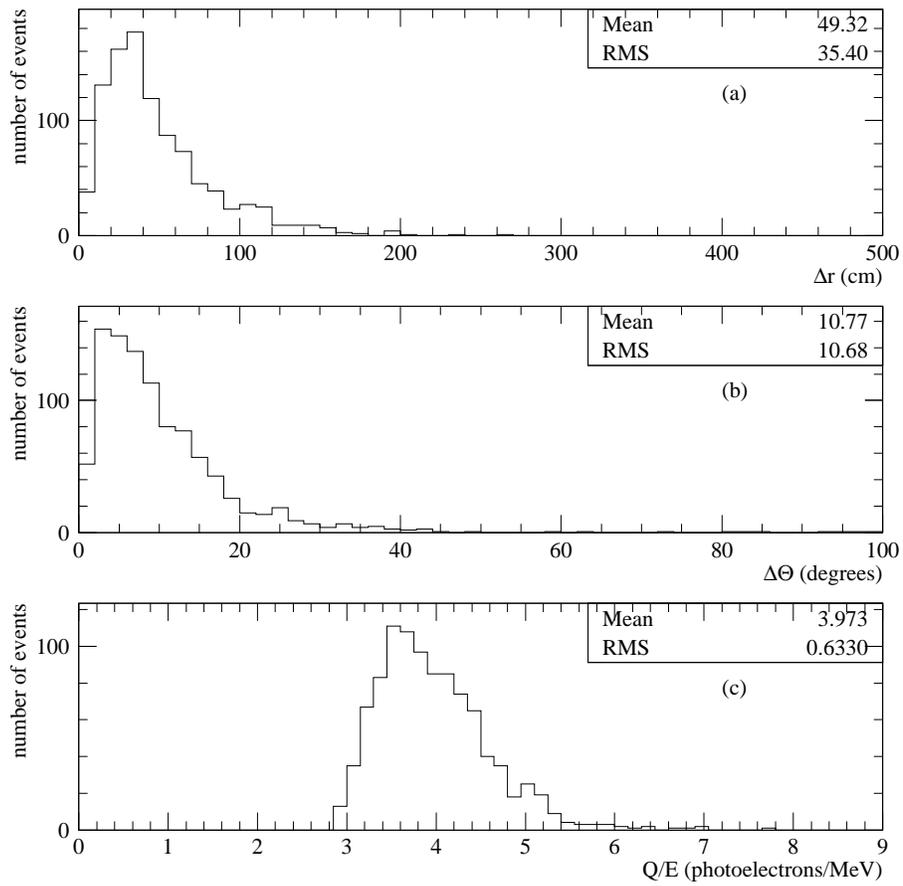,width=4.7in,silent=}}
\caption{The (a) position, (b) angular,
and (c) energy resolutions for a large sample of electrons generated
in the tank by the detector simulation.}
\label{fig:resolutions}
\end{figure}

\begin{figure}
\centerline{\psfig{figure=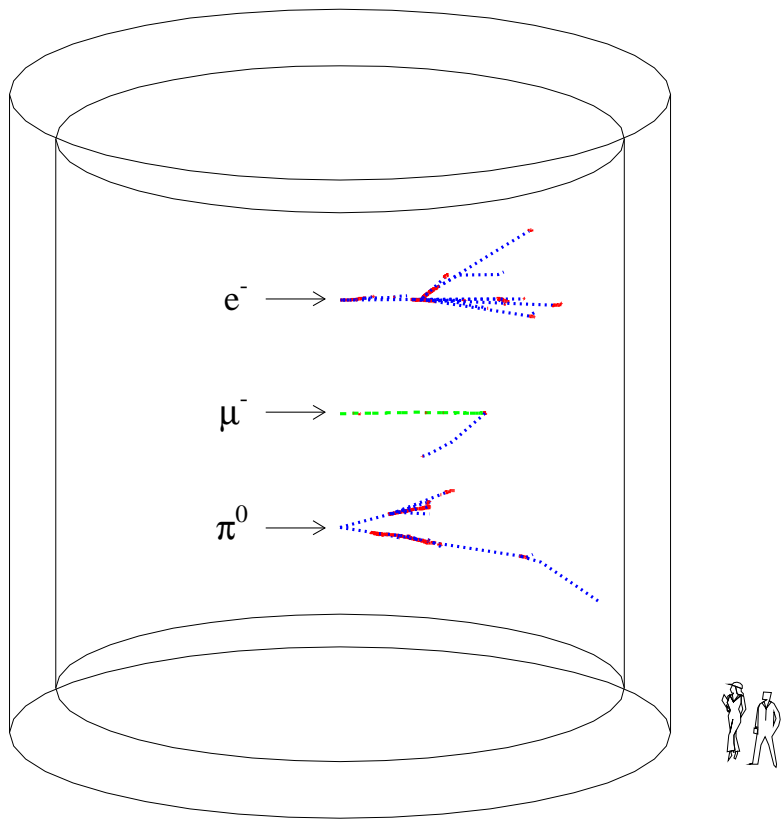,width=4.7in,silent=}}
\caption{Typical topologies of 500 MeV electrons, muons, and neutral pions
in the detector tank.} 
\label{fig:piddisplay}
\end{figure}

\begin{figure}
\centerline{\psfig{figure=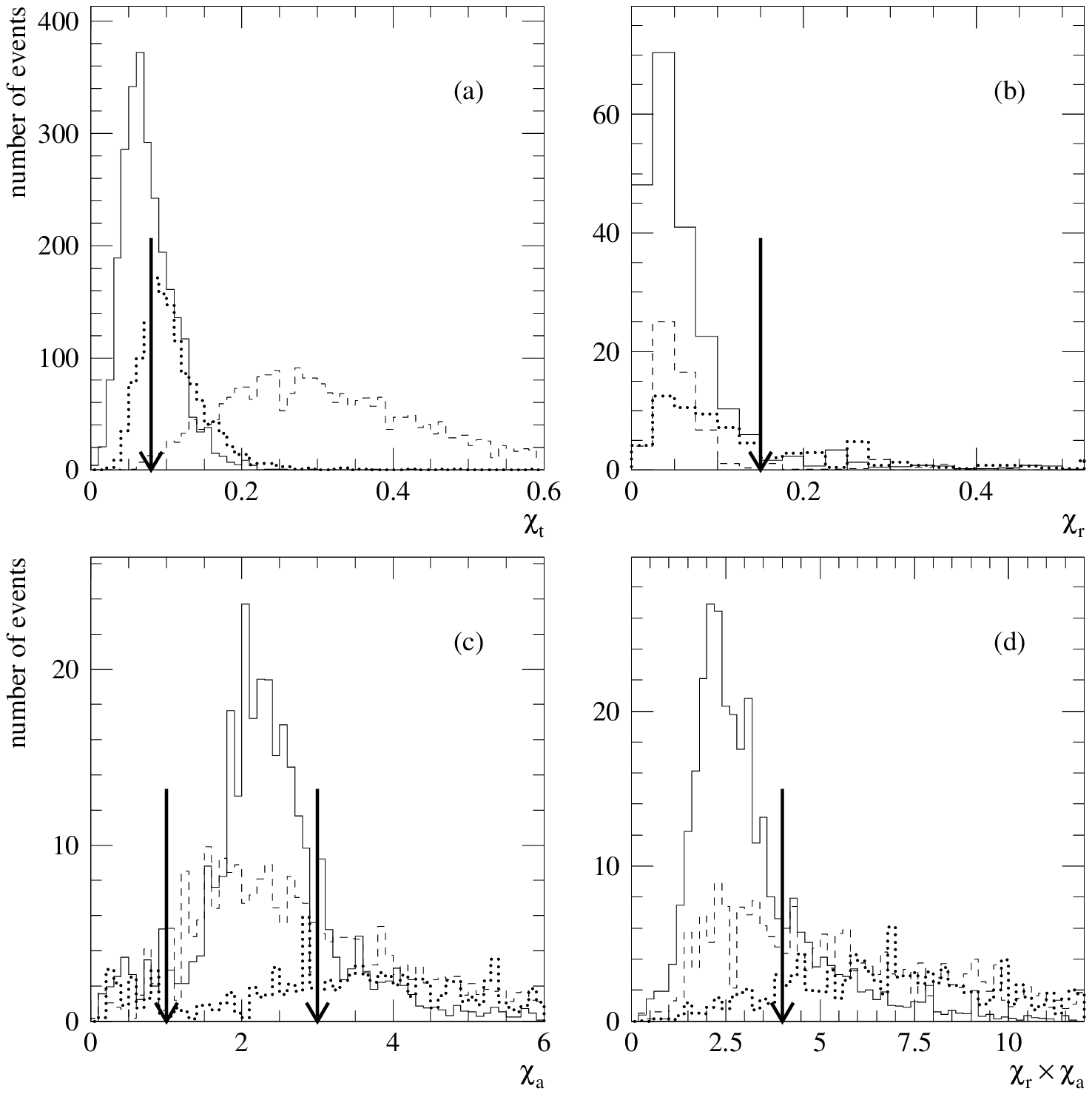,width=4.7in,silent=}}
\caption{The (a) $\chi_t$, (b) $\chi_{r}$, (c) $\chi_{a}$, 
and (d) $\chi_r \times \chi_a$ 
chi square
distributions for electrons (solid curve), muons (dashed curve),
and neutral pions (dotted curve) in the 50 to 1000 MeV energy range.
The numbers of events are normalized and 
many of the muons are in the overflow bin. The arrows show cuts that
reduce the $\mu^-$ and $\pi^0$ by factors of 1000 and 100 relative
to electrons.}
\label{fig:pid}
\end{figure}

This event reconstruction algorithm has been tested by our detector
simulation discussed above. A large sample of electrons were generated
uniformly in the tank and with random directions in the energy range from
50 to 1200 MeV. Fig. \ref{fig:resolutions} 
shows the resulting position, angular,
and energy resolutions. As can be seen in the figure, the position
resolution is about 50 cm, the angular resolution is about $10^o$,
and the energy resolution is about $15\%$. Note that for electrons there are
about 4 photoelectrons per MeV. Furthermore, electrons can easily be
distinguished from muons and neutral pions by fitting the
\v Cerenkov ring and event vertex and by measuring the fraction of PMTs
with a late hit. As seen in Fig. \ref{fig:piddisplay}, electrons, muons,
and neutral 
pions have quite different topologies in the tank. Fig. \ref{fig:pid}
shows $\chi_t$, the fraction of PMTs hit after 10 ns from the start
of the event, and the $\chi_{r}$, 
$\chi_{a}$, $\chi_r \times \chi_a$
chi square distributions for electrons (solid curve), muons (dashed curve),
and neutral pions (dotted curve) in the 50 to 1000 MeV energy range. 
The numbers of events are normalized and 
many of the muons are in the overflow bin.
A clear
separation is observed between electrons and muons and pions.

\section{Signal and Backgrounds for $\nu_\mu \rightarrow \nu_e$}

{\it The signature for $\nu_\mu \rightarrow \nu_e$
oscillations is $\nu_e$ quasi elastic
scattering off carbon nuclei. The main backgrounds come from intrinsic
$\nu_e$ contamination in the beam, mis-identified 
$\nu_\mu$ quasi elastic scattering,
and neutral current $\pi^0$ production.} \\

The quasi-elastic cross sections in this chapter are obtained from a
modified Fermi-Gas calculation that agrees approximately with a more
sophisticated continuum RPA calculation.\cite{kolbe} Also, the
single pion cross sections are taken from Rein and Sehgal.\cite{pi0}

\subsection{Signal Reactions}

\begin{figure}
\centerline{\psfig{figure=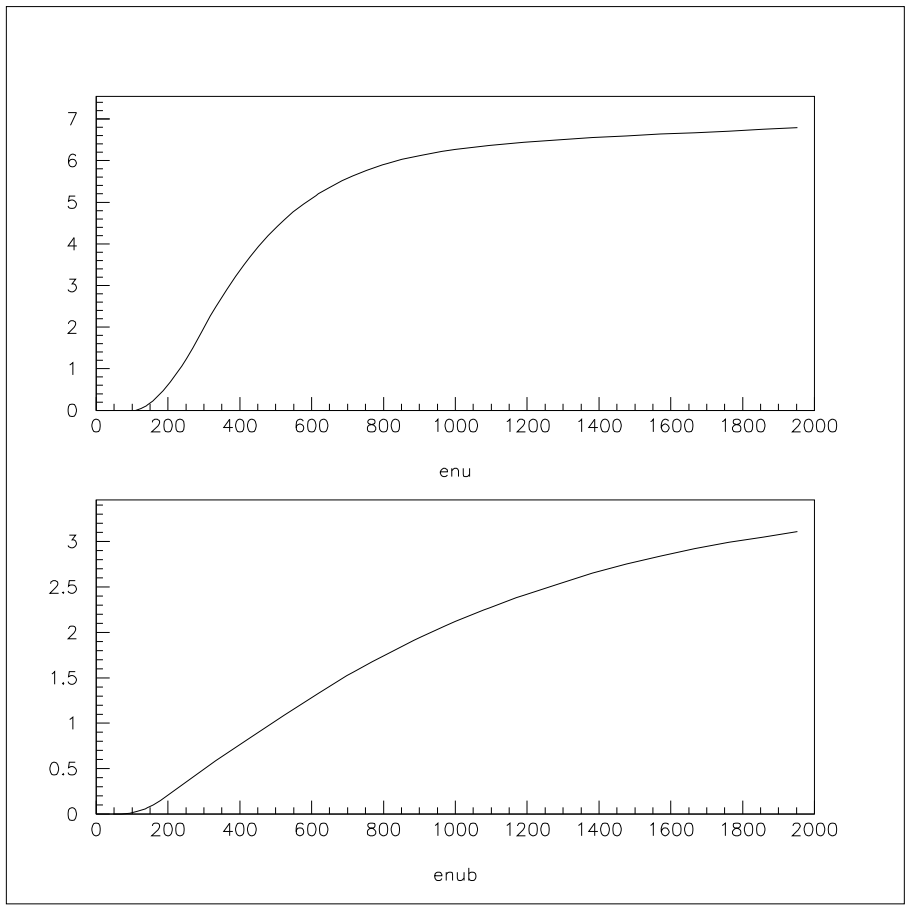,width=4.7in,silent=}}
\caption{The cross sections (in units of $10^{-38}$ cm$^2$)
for $\nu_e C \rightarrow e^- N$ and
$\bar \nu_e C \rightarrow e^+ B$ quasi elastic scattering 
as a function of incident neutrino energy in MeV.}
\label{fig:sig_cross}
\end{figure}

\begin{figure}
\centerline{\psfig{figure=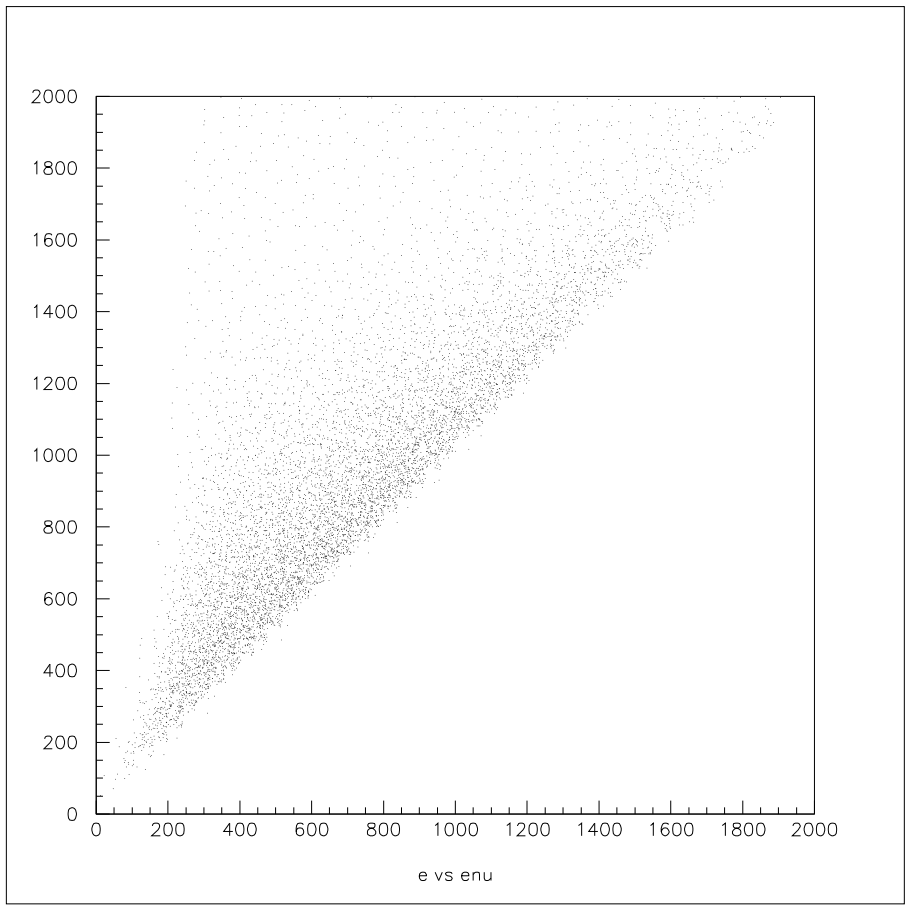,width=4.7in,silent=}}
\caption{The visible energy versus
the neutrino energy in MeV
from quasi elastic scattering
after integrating over the incident
neutrino energy spectrum.}
\label{fig:evsenu}
\end{figure}

\begin{figure}
\centerline{\psfig{figure=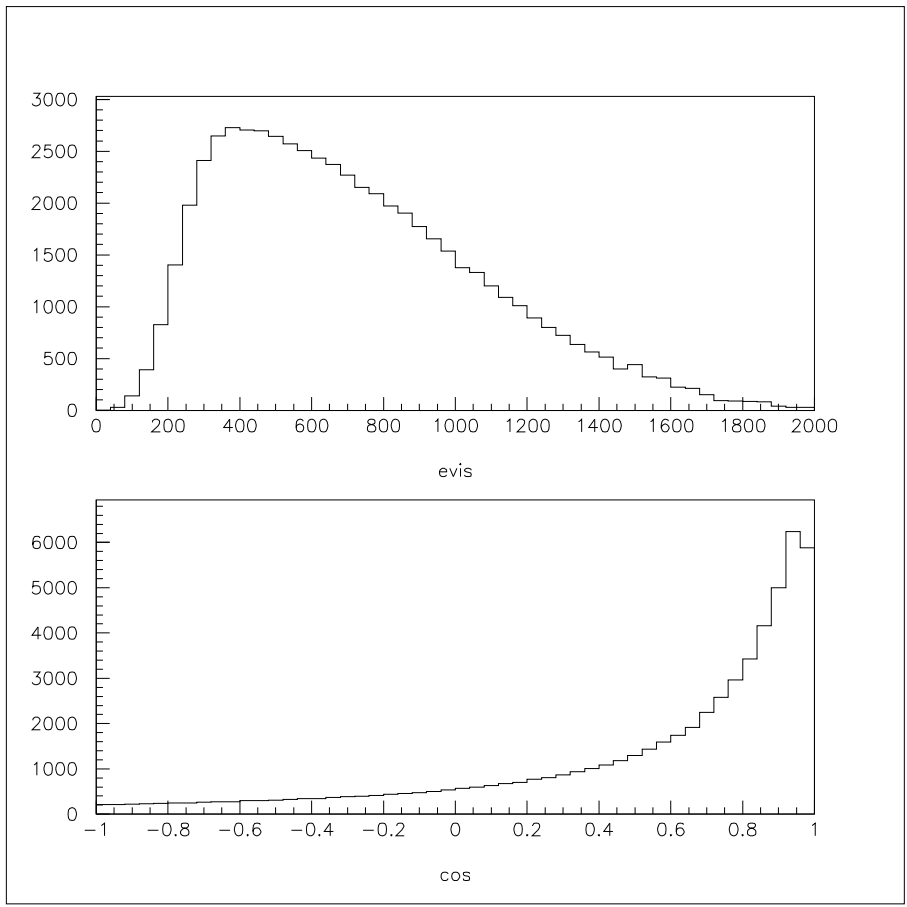,width=4.7in,silent=}}
\caption{The recoil $e^{\pm}$ (a) energy in MeV and
(b) $\cos \theta$ distributions from quasi elastic scattering
after integrating over the incident
neutrino energy spectrum.}
\label{fig:ea_cross}
\end{figure}

The neutrino oscillations ($\nu_{\mu} \rightarrow \nu_e$ and
$\bar \nu_{\mu} \rightarrow \bar \nu_e$)
are observed by $\nu_e C \rightarrow e^- N$ and 
$\bar \nu_e C \rightarrow e^+ B$ quasi elastic scattering. 
The cross sections for these
reactions\cite{kolbe} are shown
in Fig. \ref{fig:sig_cross} and Tables \ref{tab:cross} and \ref{tab:crossa}
as a function of incident neutrino energy. Note that
the cross sections rise rapidly with energy near threshold and
then taper off to a flat energy dependence well above threshold.
Also, Fig. \ref{fig:evsenu} shows the visible energy versus
the neutrino energy, and
Fig. \ref{fig:ea_cross} shows the recoil $e^{\pm}$ energy and 
$\cos \theta$ distributions after integrating over the incident
neutrino energy spectrum. 
The angle $\theta$ is the reconstructed $e^{\pm}$ direction
relative to the incident neutrino direction. Note
that the $e^{\pm}$ has an energy on average that is about 2/3 the
neutrino energy and a direction that becomes more forward peaked
with energy.

\begin{table}[t]

\caption{Signal and background neutrino cross sections as a function of
neutrino energy in MeV. The cross sections are for each C atom and are in
units of $10^{-40}$ cm$^2$.}

\label{tab:cross}
\vspace{0.4cm}
\begin{center}
\begin{tabular}{|c|c|c|c|c|c|c|}
\hline
{\footnotesize  $E_{\nu}$(MeV)}
&{\footnotesize $ \nu_e C \rightarrow e^-N$}
&{\footnotesize $ \nu_\mu C \rightarrow \mu^-N$}
&{\footnotesize $ \nu_\mu e^- \rightarrow \nu_\mu e^-$}
&{\footnotesize $ \nu_\mu C \rightarrow \nu_\mu \pi^0 X$}
&{\footnotesize $ \nu_\mu C \rightarrow \pi^{\pm} X$}
&{\footnotesize $ \nu_\mu C \rightarrow \mu^- \pi X$}\\
\hline
100&13&0&0.01&0&0&0\\
200&106&63&0.03&0&0&0\\
300&249&213&0.04&1&0&0\\
400&369&341&0.05&8&1&10\\
500&458&435&0.06&20&8&71\\
600&520&501&0.08&34&17&135\\
700&563&546&0.09&50&24&203\\
800&593&575&0.10&65&32&267\\
900&613&596&0.11&77&39&325\\
1000&627&609&0.13&87&49&364\\
1100&636&619&0.14&92&52&394\\
1200&644&625&0.16&99&56&415\\
1300&650&629&0.17&101&57&442\\
1400&655&632&0.18&101&57&444\\
1500&659&634&0.20&102&58&450\\
1600&664&635&0.21&105&59&453\\
1700&667&636&0.22&105&59&460\\
1800&671&637&0.23&105&59&460\\
1900&675&638&0.25&105&59&460\\
2000&679&640&0.26&105&59&460\\
\hline
\end{tabular}
\end{center}
\end{table}

\begin{table}[t]

\caption{Signal and background antineutrino
cross sections as a function of
antineutrino energy. The cross sections are for each C atom (the numbers in
parentheses are the cross sections for free protons) and are in
units of $10^{-40}$ cm$^2$.}

\label{tab:crossa}
\vspace{0.4cm}
\begin{center}
\begin{tabular}{|c|c|c|c|c|c|c|}
\hline
{\footnotesize $E_{\bar \nu}$(MeV)}
&{\footnotesize $\bar \nu_e C \rightarrow e^+B$}
&{\footnotesize $\bar \nu_\mu C \rightarrow \mu^+B$}
&{\footnotesize $\bar \nu_\mu e^- \rightarrow \bar \nu_\mu e^-$}
&{\footnotesize $\bar \nu_\mu C \rightarrow \bar \nu_\mu \pi^0 X$}
&{\footnotesize $\bar \nu_\mu C \rightarrow \pi^{\pm} X$}
&{\footnotesize $\bar \nu_\mu C \rightarrow \mu^+ \pi X$}\\
\hline
100&7 (4)&0 (0)&0.01&0&0&0\\
200&32 (10)&22 (8)&0.02&0&0&0\\
300&59 (15)&53 (14)&0.03&1&0&0\\
400&84 (19)&80 (18)&0.04&5&0&4\\
500&109 (24)&106 (23)&0.05&11&4&31\\
600&133 (28)&130 (27)&0.06&18&8&59\\
700&156 (32)&153 (31)&0.07&26&11&89\\
800&176 (35)&174 (34)&0.08&34&14&117\\
900&195 (39)&193 (38)&0.09&40&18&142\\
1000&212 (42)&210 (41)&0.10&45&22&160\\
1100&227 (44)&226 (43)&0.11&48&23&177\\
1200&241 (47)&240 (46)&0.12&54&26&191\\
1300&253 (49)&253 (48)&0.13&56&27&213\\
1400&265 (51)&265 (50)&0.14&57&27&221\\
1500&275 (53)&275 (52)&0.15&59&29&224\\
1600&284 (55)&285 (54)&0.16&63&31&231\\
1700&292 (57)&293 (56)&0.17&65&32&246\\
1800&299 (58)&301 (57)&0.18&66&32&250\\
1900&305 (60)&307 (59)&0.19&67&33&256\\
2000&311 (61)&313 (60)&0.20&68&33&261\\
\hline
\end{tabular}
\end{center}
\end{table}

\subsection{Background Reactions}

A principal beam-related background is due to the intrinsic
$\nu_e$ and $\bar \nu_e$ fluxes from kaon and muon decays
discussed in chapter 5. This intrinsic background has the
same reactions as the signal reactions described in section 8.1. 
The other main beam-related 
backgrounds are due to neutrino reactions that produce a
$\mu^{\pm}$ or $\pi^0$ in the final state. The BooNE design goal 
is to have particle identification (PID) sufficient to suppress
muon misidentification as an electron by a factor of 1000 and
$\pi^0$ misidentification as an electron by a factor of 100. As
shown in section 8.3, these suppression factors are achieved. Other
backgrounds that are considered are $\pi^{\pm}$ production and
$\nu e \rightarrow \nu e$ elastic scattering. 
A list of all of these background reactions is shown in Table
\ref{tab:list}, and the cross sections
are given in 
Tables \ref{tab:cross} and \ref{tab:crossa}
as a function of incident neutrino energy. Note that
beam-off backgrounds are expected to be very small due to the
very low duty factor and can be easily subtracted by a beam-on minus 
beam-off subtraction.

\begin{table}[t]

\caption{A list of the background neutrino reactions for
$\nu_\mu \rightarrow \nu_e$ appearance. Also shown is the
factor that these backgrounds can be suppressed relative
to $\nu_e C \rightarrow e^- X$ scattering and the approximate 
background level for the appearance oscillation search.}

\label{tab:list}
\vspace{0.4cm}
\begin{center}
\begin{tabular}{|c|c|c|}
\hline
Reaction&Suppression Factor&Background Level\\
\hline
$\nu_\mu C \rightarrow \mu^- X$&$10^{-3}$&$10^{-3}$\\
$\nu_\mu C \rightarrow \nu_\mu \pi^0 X$&$10^{-2}$&$10^{-3}$\\
$\nu_\mu C \rightarrow \mu^-\pi X$&$10^{-4}$&$10^{-4}$\\
$\nu_\mu C \rightarrow \nu_\mu \pi^{\pm} X$&$10^{-3}$&$10^{-3}$\\
$\nu_\mu e^- \rightarrow \nu_\mu e^- $&$10^{-1}$&$10^{-4}$\\
\hline
\end{tabular}
\end{center}
\end{table}

\smallskip

\begin{figure}
\centerline{\psfig{figure=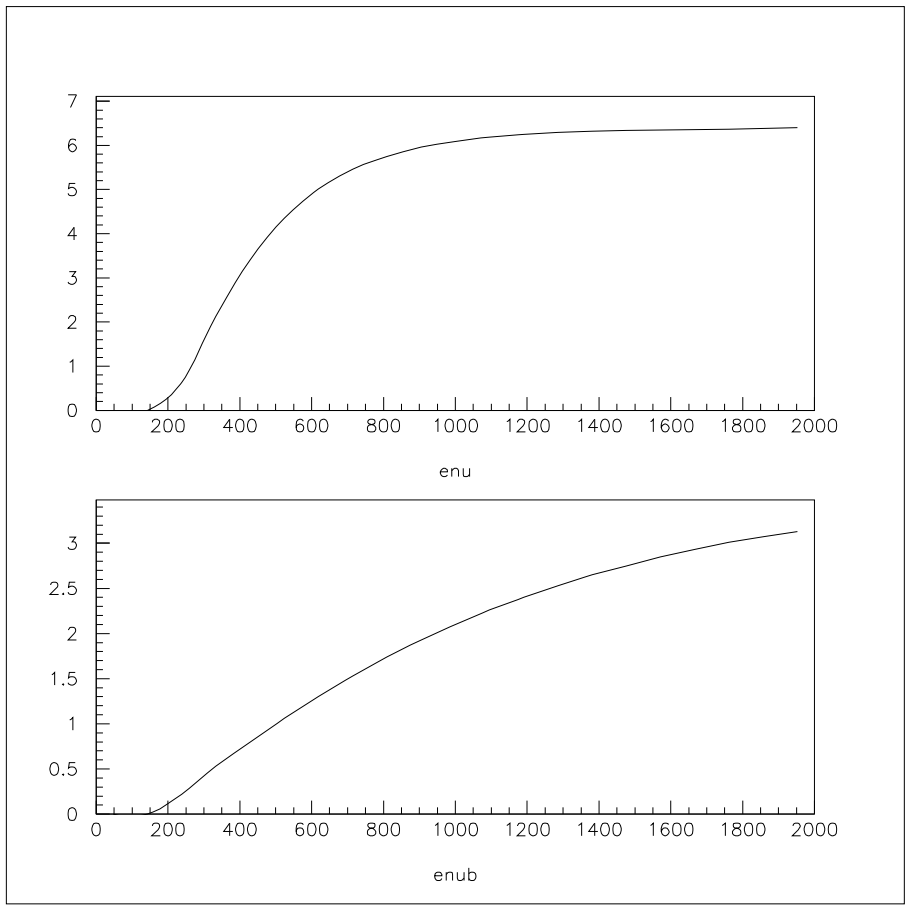,width=4.7in,silent=}}
\caption{The cross sections (in units of $10^{-38}$ cm$^2$)
for $\nu_{\mu} C$ 
and $\bar \nu_{\mu} C$ quasi elastic
scattering as a function of incident neutrino energy in MeV.}
\label{fig:mu_cross}
\end{figure}

\noindent 1. $\nu_{\mu} C \rightarrow \mu^- N$
\smallskip
The first background is due to $\nu_{\mu} C \rightarrow \mu^- N$
and $\bar \nu_{\mu} C \rightarrow \mu^+ B$ scattering, where the $\mu^{\pm}$
is misidentified as an electron. The
cross sections for these reactions as a function of incident
neutrino energy are shown in Fig. \ref{fig:mu_cross}. Because these cross
sections are comparable to the cross sections of the neutrino
oscillation signal reactions discussed in section 7.1 above, we must
be able to reject these events by a factor of $ \sim 1000$ in order
to achieve a neutrino oscillation sensitivity of $\sim 10^{-3}$. 
\smallskip

\begin{figure}
\centerline{\psfig{figure=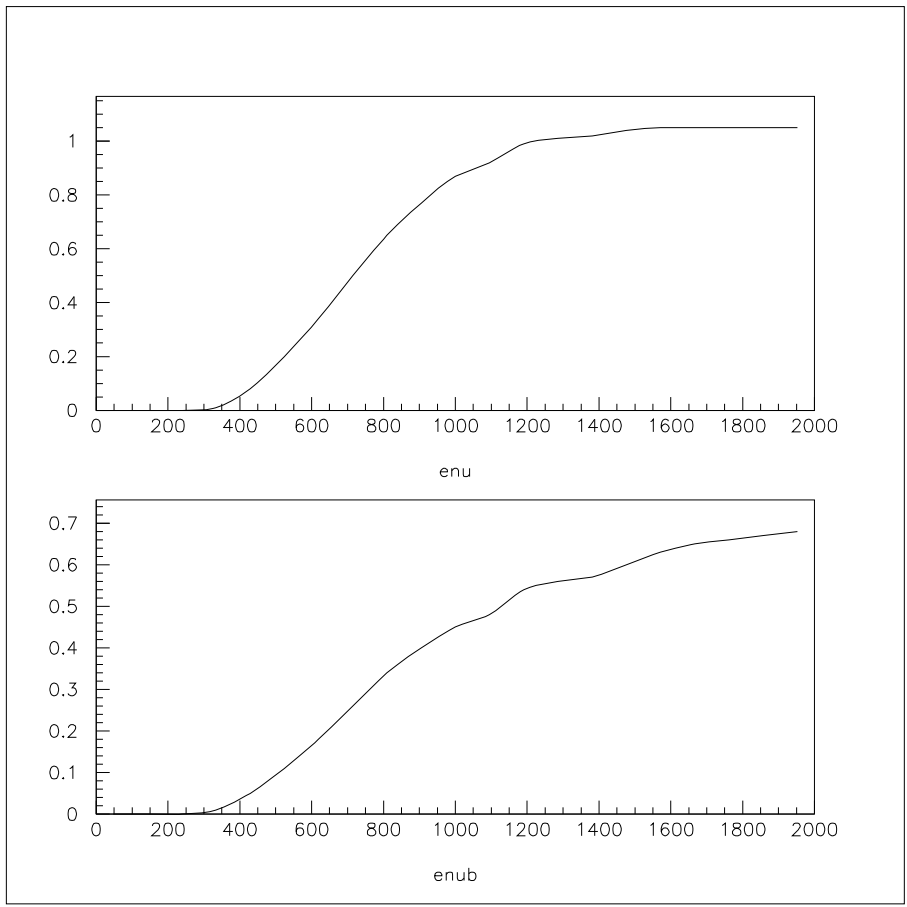,width=4.7in,silent=}}
\caption{The cross sections (in units of $10^{-38}$ cm$^2$)
for $\nu_{\mu} C$ 
and $\bar \nu_{\mu} C$ neutral current $\pi^0$ production
as a function of incident neutrino energy in MeV.}
\label{fig:pi0_cross}
\end{figure}

\noindent 2. $\nu_{\mu} C \rightarrow \nu_{\mu} \pi^0 X$
\smallskip
The second background is due to $\nu_{\mu} C \rightarrow \nu_{\mu}
\pi^0 X$ 
and $\bar \nu_{\mu} C \rightarrow \bar \nu_{\mu} \pi^0 X$ scattering. 
Fig. \ref{fig:pi0_cross} 
shows the cross sections for these reactions as a function of incident
neutrino energy.
By comparing with the signal cross sections,
it is clear that we must
be able to reject these events by a factor of $ \sim 100$ in order
to achieve a neutrino oscillation sensitivity of $\sim 10^{-3}$.
\smallskip

\begin{figure}
\centerline{\psfig{figure=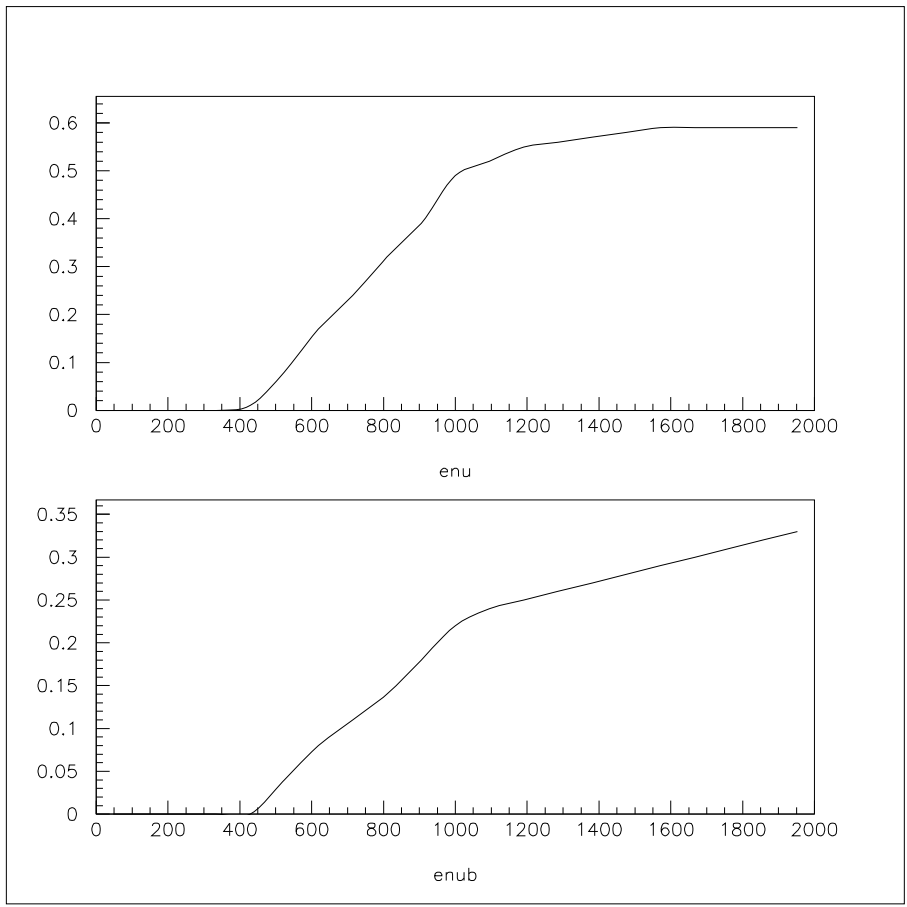,width=4.7in,silent=}}
\caption{The cross sections (in units of $10^{-38}$ cm$^2$)
for $\nu_{\mu} C$ 
and $\bar \nu_{\mu} C$ neutral current $\pi^{\pm}$ production
as a function of incident neutrino energy in MeV.}
\label{fig:pipm_cross}
\end{figure}

\noindent 3. $\nu_{\mu} C \rightarrow \nu_{\mu} \pi^{\pm} X$
\smallskip
The next background is due to $\nu_{\mu} C \rightarrow \nu_{\mu}
\pi^{\pm} X$ 
and $\bar \nu_{\mu} C \rightarrow \bar \nu_{\mu} \pi^{\pm} X$ scattering. 
Fig. \ref{fig:pipm_cross} 
shows the cross sections for these reactions as a function of incident
neutrino energy.
By comparing with the signal cross sections,
it is clear that we must
be able to reject these events by a factor of $ \sim 100$ in order
to achieve a neutrino oscillation sensitivity of $\sim  10^{-3}$.
\smallskip

\begin{figure}
\centerline{\psfig{figure=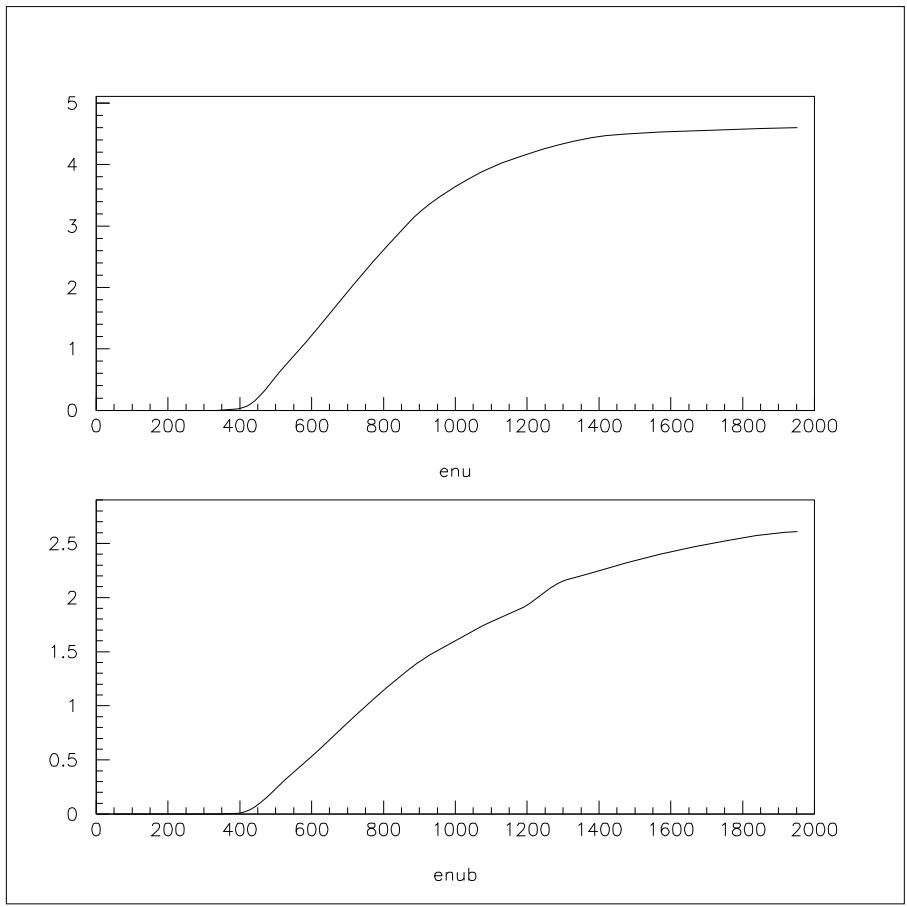,width=4.7in,silent=}}
\caption{The cross sections (in units of $10^{-38}$ cm$^2$)
for $\nu_{\mu} C$ 
and $\bar \nu_{\mu} C$ charged current pion production
as a function of incident neutrino energy in MeV.}
\label{fig:mupi_cross}
\end{figure}

\noindent 4. $\nu_{\mu} C \rightarrow  \mu^- \pi X$
\smallskip
Another background is due to $\nu_{\mu} C \rightarrow \mu^-
\pi X$
and $\bar \nu_{\mu} C \rightarrow \mu^+ \pi X$ scattering, where the recoil
$\pi$ is a $\pi^0$ or $\pi^{\pm}$.
Fig. \ref{fig:mupi_cross} 
shows the cross sections for these reactions as a function of incident
neutrino energy. 
By comparing with the signal cross sections,
it is clear that we must
be able to reject these events by a factor of $ \sim 1000$ in order
to achieve a neutrino oscillation sensitivity of $\sim 10^{-3}$.
\smallskip

\noindent 5. $\nu_{\mu} e^- \rightarrow \nu_{\mu} e^-$
\smallskip
The final background that we consider is $\nu_{\mu} e^- \rightarrow \nu_{\mu}
e^-$
and $\bar \nu_{\mu} e^- \rightarrow \bar \nu_{\mu} e^-$ elastic scattering.
The cross section for this reaction is proportional to the neutrino
energy and can be expressed as $\sigma(\nu_{\mu} e^- \rightarrow \nu_{\mu}
e^-) = 1.6 \times E_{\nu} 
\times 10^{-42}$ cm$^2$, where the neutrino energy is in GeV. 
This background is small and can be 
identified by the recoil electron direction.

\subsection{Rejection of Events with Muons and Pions}

\begin{figure}
\centerline{\psfig{figure=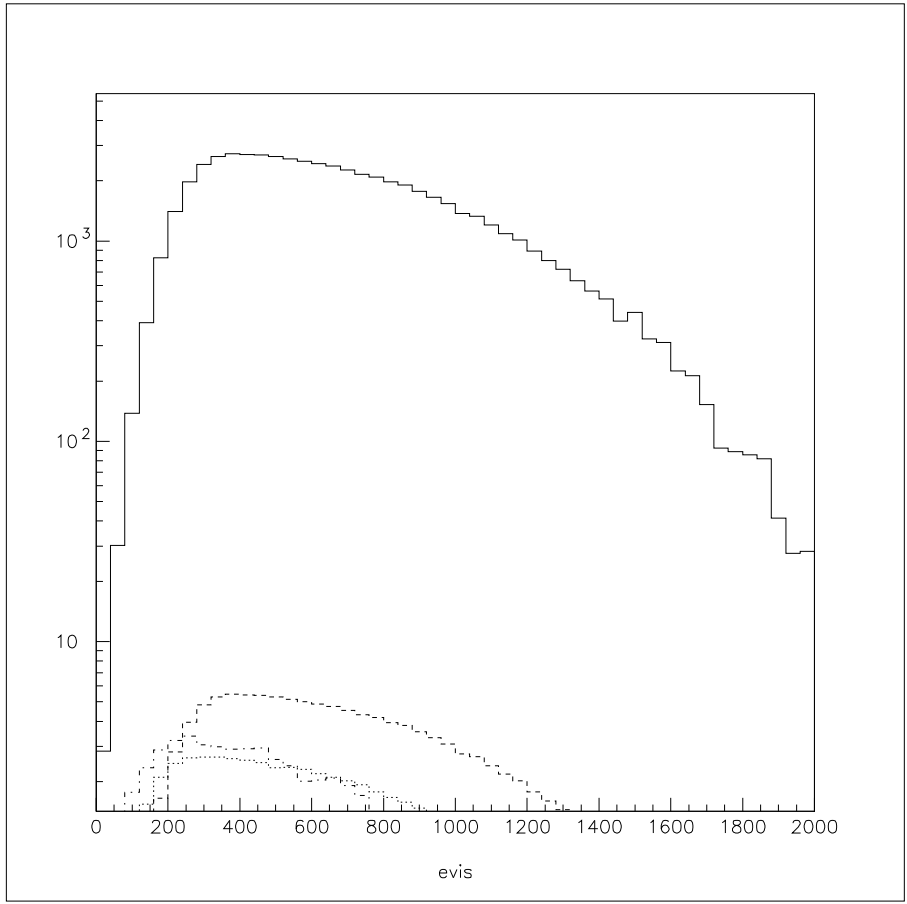,width=4.7in,silent=}}
\caption{The expected visible energy (electron equivalent)
distributions in MeV from the intrinsic $\nu_e$ background (dashed curve),
the $\nu_{\mu} C \rightarrow \mu^- X$ background (dotted curve),
and from the $\nu_{\mu} C \rightarrow \nu_{\mu} \pi^0
X$ background (dot-dashed curve). Also shown is the expected distribution from
$\nu_{\mu} \rightarrow \nu_e$ oscillations for
100\% transmutation (solid curve).}
\label{fig:energy}
\end{figure}

Events with a $\mu^{\pm}$, $\pi^0$, or $\pi^{\pm}$ in the final state
are rejected by using the $\chi^2$ of the position fit ($\chi_r$) and the
$\chi^2$ of the angle fit ($\chi_a$) as described in chapter 7, by vetoing
events with a signal in the veto shield, and by using $\chi_t$, the fraction
of PMTs with a late hit ($>10$ ns).
By requiring, for example, $1<\chi_{a}<4$,
$0< \chi_{r} <0.15$,
$0< \chi_r \times \chi_a<4$, and $\chi_t < 0.08$, we obtain a factor
of $\sim 1000$ rejection of $\mu^{\pm}$ and $\pi^{\pm}$ and a factor of
$\sim 100$ $\pi^0$ rejection relative to electrons (see Fig. \ref{fig:pid}). 
The $e^{\pm}$ efficiency is $\sim 50\%$.
Fig. \ref{fig:energy} shows the expected visible energy (electron equivalent)
distributions from the intrinsic $\nu_e$ background (dashed curve),
the $\nu_{\mu} C \rightarrow \mu^- X$ background (dotted curve),
and from the $\nu_{\mu} C \rightarrow \nu_{\mu} \pi^0
X$ background (dot-dashed curve). Also shown is the expected distribution from
$\nu_{\mu} \rightarrow \nu_e$ oscillations for
100\% transmutation.
The signal to background ratio is
greater than one for oscillation probabilities greater than $\sim 0.5\%$.

\subsection{Systematic Errors due to Background}

From the present studies, 250 background events are expected in 
$1 \times 10^7$ seconds of running due both to misidentification of
 $\mu^{\pm}$, $\pi^0$, or $\pi^{\pm}$ and to the intrinsic $\nu_e$
content of the beam.   We are in the process of studying the design of
our beam line and detector to improve this rate of background events.   
However, even with the present background expectation, a significant signal 
will be isolated over the full LSND region because the systematics 
associated with the background subtraction is expected to be less 
than 10\%.  This systematic 
error has been included in the sensitivities presented in this
letter of intent.

For this document we have assumed a 10\% error 
on the $\nu_e$ content of the beam, which is a conservative estimate
based on the experience of previous analyses.
Experiment 776 at Brookhaven, which used 
a horn focusing system, achieved a systematic error
of 11\% on the determination of the $\nu_e$ content.\cite{Borodovsky}
Contributions to this uncertainty came from the $K/\pi$ ratio (10\%), the 
$\nu_\mu$ statistical error (1\%) and the normalization uncertainty
(3\%).   The CCFR experiment, which used an 800 GeV
proton beam to produce neutrinos and antineutrinos 
with a Quad Triplet system, obtained a total
uncertainty in the $\nu_e$ content of the beam of 4.2\%.\cite{King}
Noting that the kaon production from an 8 GeV beam is significantly
less than for the 28 GeV proton beam at Brookhaven and 
using techniques developed by the CCFR and NuTeV experiment for 
analysis and monitoring of the flux,\cite{TM} 
we believe that we can obtain in the final analysis
an uncertainty in the 
$\nu_e$ content of approximately 5\%,
but choose to use a conservative 10\% error for this document.

Presently, we also assume a systematic error of 10\% on the number of 
background events due to particle misidentification.  Again, we believe that
this is conservative.   The Kamioka experiment, which faced similar 
issues of particle identification in their multi-GeV data, was able to achieve 
a 4\% systematic error.\cite{WIN}
The LSND decay-in-flight analysis, which has an average neutrino
energy of $\sim 150$ MeV, determines the systematic error on the 
muon identification to be less than 5\%.\cite{DIF}
In MiniBooNE, the systematic error on fake quasi elastic $\nu_e$ events 
due to a misidentified muon or pion is expected to be 
less than 5\% in the final analysis.   Note that this systematic error
can be determined directly from
the study of events that are not misidentified and from 
Monte Carlo studies.  Using the data to constrain the Monte Carlo 
in determining this background is a
very powerful method which significantly reduces the error.   
 
\section{Projected Measurements and Sensitivity}

{\it The MiniBooNE experiment will be able to clearly observe
neutrino oscillations in the 0.1 - 0.5 eV$^2$ mass range for
values of $\sin^22\theta > 10^{-3}$ ($\nu_\mu 
\rightarrow \nu_e$ appearance) and $\sin^22\theta \sim 0.5$ ($\nu_\mu$
disappearance).} \\

\subsection{Event Rates}

For the estimation of event rates we make the following assumptions.
First, we assume that the Booster operates at an energy of 8 GeV 
and at an average 
rate of 5 Hz ($2.5 \times 10^{13}$ protons/s) for one year
($1 \times 10^7$ s) at each horn-focusing polarity. Also, we assume
that the fiducial volume of the detector is 382 t ($1.7 \times 10^{31}$
$CH_2$ molecules) and that the total electron and muon
efficiencies, including PID, are 50\%. The
resulting numbers of quasi elastic events are shown in Table \ref{tab:numbers}
for both neutrino and antineutrino scattering and for the detector at 
a distance of 1000 m from the neutrino source.
The muon-neutrino 
quasi elastic scattering estimates assume no oscillations, while the
electron-neutrino quasi elastic scattering estimates assume 100\%
$\nu_{\mu} \rightarrow \nu_e$ transmutation.

\begin{table}[t]
\caption{The estimated numbers of quasi elastic events 
for both neutrino and antineutrino scattering and for the detector at a
distance of 1000 m from the neutrino source.
The muon-neutrino
quasi elastic scattering estimates assume no oscillations, while the
electron-neutrino quasi elastic scattering estimates assume 100\%
$\nu_{\mu} \rightarrow \nu_e$ transmutation.}
\label{tab:numbers}
\vspace{0.4cm}
\begin{center}
\begin{tabular}{|c|c|}
\hline 
Reaction&Number of Events\\
\hline
$\nu_{\mu} C \rightarrow \mu^- X$&50,800\\
$\bar \nu_{\mu} C \rightarrow \mu^+ X$&9,500\\
$\nu_e C \rightarrow e^- X$&52,500\\
$\bar \nu_e C \rightarrow e^+ X$&9,600\\
\hline
\end{tabular}
\end{center}
\end{table}

\subsection{$\nu_{\mu} \rightarrow \nu_e$ Appearance}

\begin{figure}
\centerline{\psfig{figure=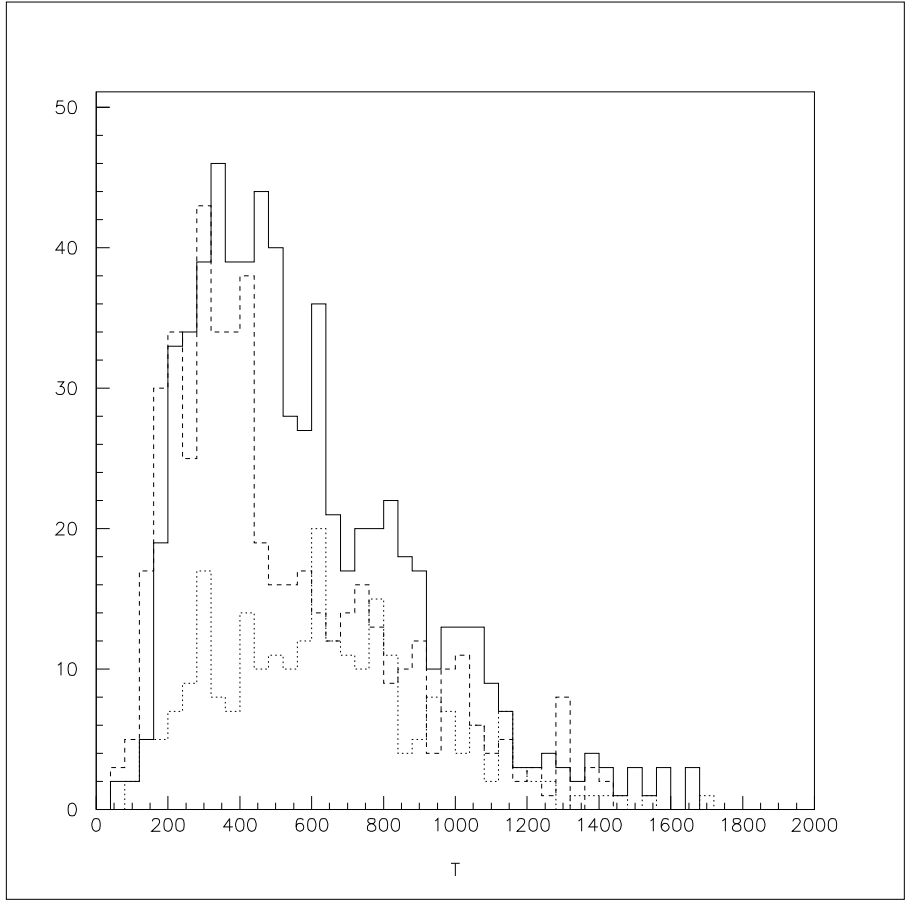,width=4.7in,silent=}}
\caption{The $\nu_e$
quasi elastic visible energy distribution in MeV, including
oscillations and background sources,
for two different possible oscillation parameters (motivated by LSND):
$\Delta m^2 = 0.2$ eV$^2$ and $\sin^22\theta = 0.04$ (dashed) and
$\Delta m^2 = 0.4$ eV$^2$ and $\sin^22\theta = 0.02$ (solid).
Also shown in the figure is a dotted line giving
the expectations for no oscillations. The two oscillation histograms
include the background of the no oscillation histogram.}
\label{fig:app}
\end{figure}

The detector located at 1000 m from the neutrino source
will measure the $\nu_e$ energy spectrum through quasi elastic scattering 
as described in chapters 8 and 9. The event energy distribution in the
detector will provide proof that neutrino oscillations
are occurring and will allow the determination of the neutrino oscillation
parameters (assuming that the LSND signal is indeed due to neutrino
oscillations). Fig. \ref{fig:app} shows the $\nu_e$
quasi elastic event energy distribution
for two different possible oscillation parameters (motivated by LSND): 
$\Delta m^2 = 0.2$ eV$^2$ and $\sin^22\theta = 0.04$ (dashed) and
$\Delta m^2 = 0.4$ eV$^2$ and $\sin^22\theta = 0.02$ (solid).
Also shown in the figure is a dotted line giving
the expectations for no oscillations. Note that, as discussed in
chapter 5, the neutrino flux varies as $r^{-2}$ to an excellent
approximation for $r>250$ m.
As is shown in the figure,
neutrino oscillations can be clearly observed and measured. We 
estimate that, for the oscillation parameters above, $\Delta m^2$
can be determined with an uncertainty of $<0.2$ eV$^2$ and $\sin^22\theta$
with an uncertainty of $<50\%$.

\subsection{Search for CP Violation in the Lepton Sector}

Assuming that $\nu_{\mu} \rightarrow \nu_e$ and
$\bar \nu_{\mu} \rightarrow \bar \nu_e$ oscillations are observed,
then, by comparing the neutrino and antineutrino oscillation parameters,
it will be possible to make a test for CP violation in the
lepton sector. CP violation may appear as a difference in the
measured values of $\sin^22\theta$ (or even $\Delta m^2$, because
in general all three different $\Delta m^2$ values can contribute
to an oscillation signal) for neutrinos
compared to antineutrinos. CP violation will be observed
easily if the violation is large ($>50\%$).

\subsection{$\nu_{\mu}$ Disappearance}

\begin{figure}
\centerline{\psfig{figure=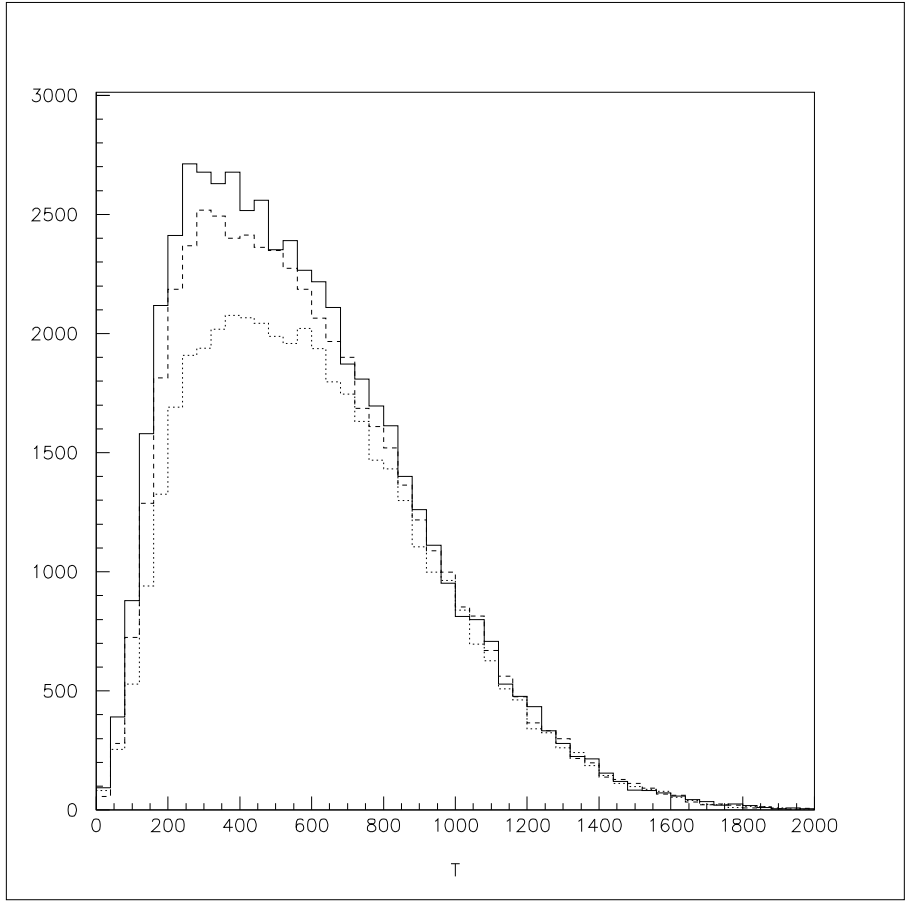,width=4.7in,silent=}}
\caption{The $\nu_{\mu}$
quasi elastic visible energy distribution
in MeV for two
different possible oscillation parameters (motivated by the
atmospheric neutrino problem):
$\Delta m^2 = 0.2$ eV$^2$ and $\sin^22\theta = 0.5$ (dashed) and
$\Delta m^2 = 0.4$ eV$^2$ and $\sin^22\theta = 0.5$ (dotted).
Also shown in the figure is a solid line giving
the expectations for no oscillations.}
\label{fig:dis}
\end{figure}

The detectors located at 1000 m from the neutrino source also
will measure the $\nu_{\mu}$ energy spectrum through quasi elastic scattering 
as described in chapters 8 and 9. The event energy distribution
can provide proof for $\nu_{\mu}$ disappearance oscillations
and determine the neutrino oscillation
parameters if the atmospheric neutrino problem is due to $\nu_{\mu}$
disappearance with $\Delta m^2 > 0.1$ eV$^2$.
Fig. \ref{fig:dis} shows the $\nu_{\mu}$
quasi elastic energy distribution
for two
different possible oscillation parameters (motivated by the atmospheric
neutrino problem): 
$\Delta m^2 = 0.2$ eV$^2$ and $\sin^22\theta = 0.5$ (dashed) and
$\Delta m^2 = 0.4$ eV$^2$ and $\sin^22\theta = 0.5$ (dotted).
Also shown in the figure is a solid line giving
the expectations for no oscillations. Note that, as discussed in
chapter 5, the neutrino flux varies as $r^{-2}$ to an excellent
approximation for $r>250$ m. As is shown in the figures,
neutrino oscillations can be clearly observed and measured. We 
estimate that, for the oscillation parameters above, $\Delta m^2$
can be determined with an uncertainty of $<0.2$ eV$^2$ and $\sin^22\theta$
with an uncertainty of $<50\%$. 

\subsection{A Method for Determining the Neutrino Flux at the Detector 
Location}

The determination of the expected neutrino event sample has been a limitation
in the systematic precision of neutrino oscillation experiments,
particularly disappearance measurements. Quasi-elastic
scattering off free protons has a well known cross section
and can be used to reduce the systematic errors associated with
scattering off neutrons and protons bound in $~^{12}C$.
The method is described in detail 
elsewhere\cite{tn114} and only the salient features are covered here.

There are two main points on which the method depends. First, the
antineutrino cross section and its $Q^2$ dependence on free protons 
is known to about 2\%
in this energy and momentum transfer range, mostly
because of well measured neutron decay and relatively little form
factor dependence. Second, $~^{12}C$ is a member of a highly
symmetrical isotriplet, so nuclear physics effects can be made to
cancel to high accuracy. In other words, scattering from bound protons
and bound neutrons are closely the same in $^{12}C$.

\begin{figure}
\centerline{\psfig{figure=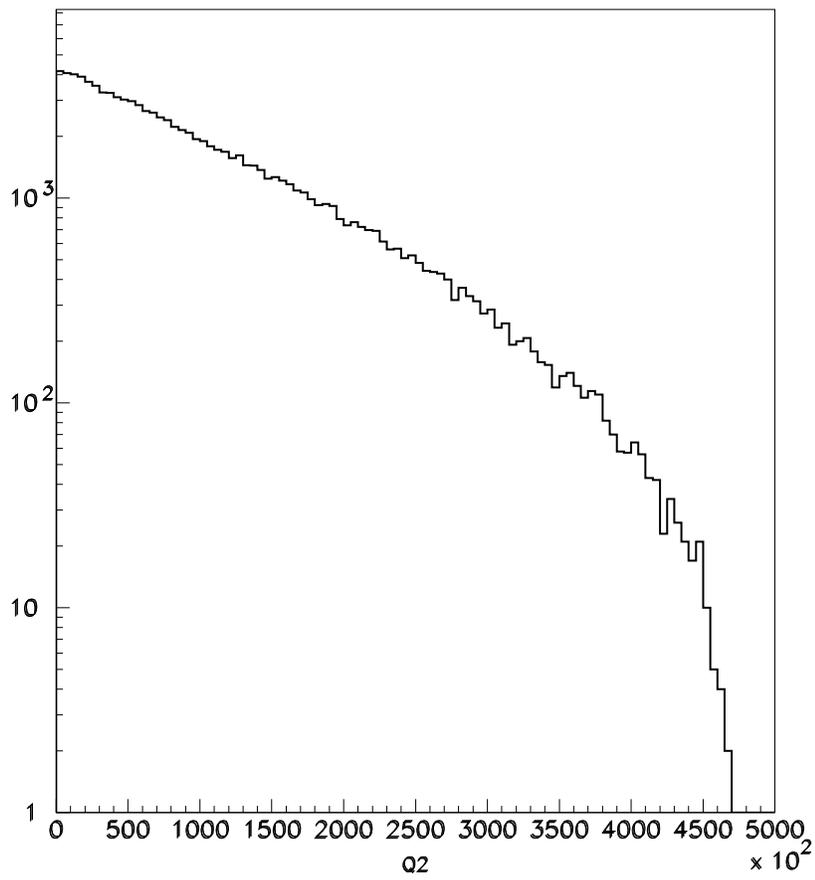,width=4.7in,silent=}}
\caption{The $Q^2$ distribution for $\bar \nu_\mu p \rightarrow \mu^+ n$
scattering in units of MeV$^2$/c$^2$.}
\label{fig:q2_nbarp}
\end{figure}

\begin{figure}
\centerline{\psfig{figure=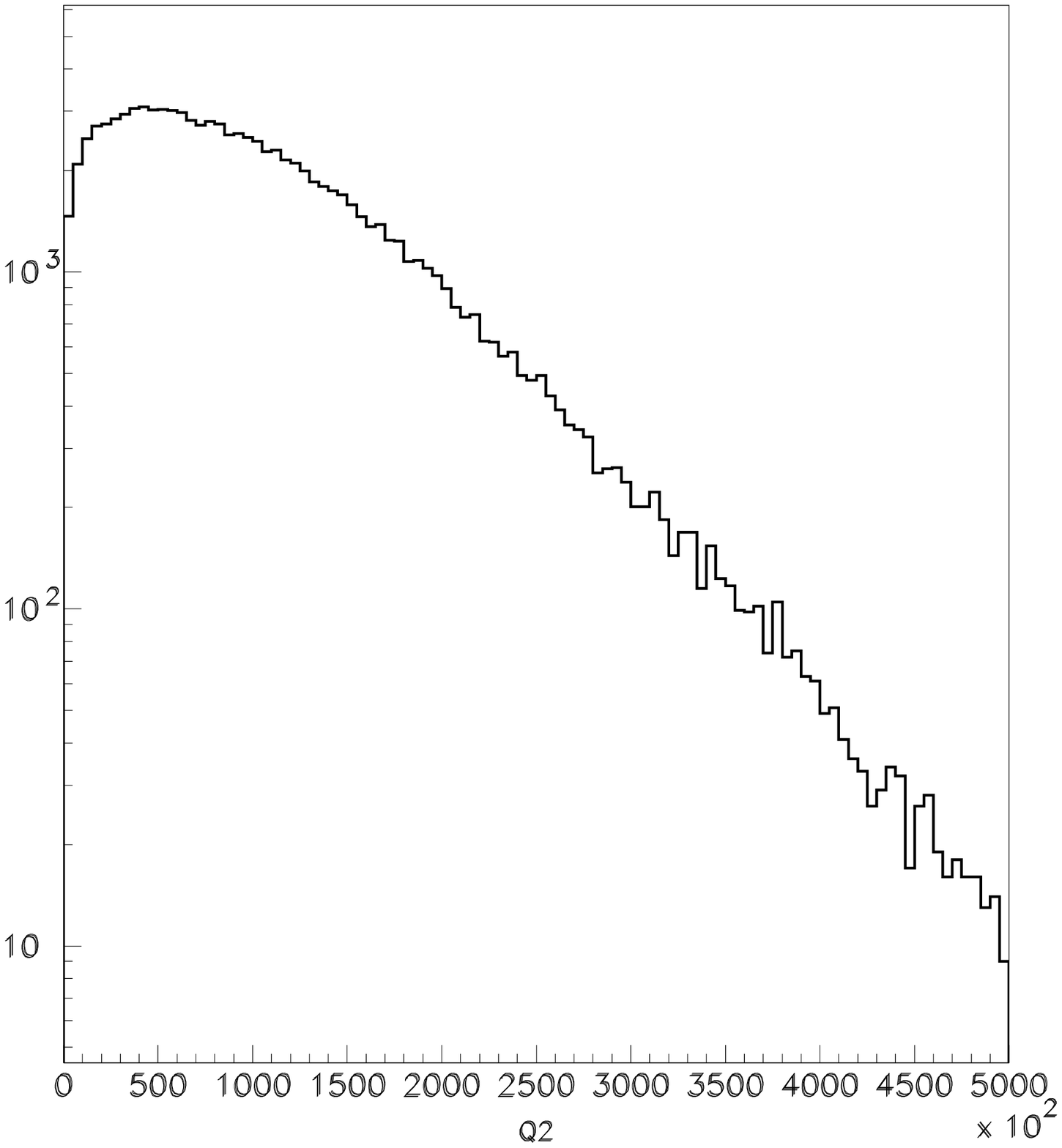,width=4.7in,silent=}}
\caption{The $Q^2$ distribution for $\bar \nu_\mu C \rightarrow \mu^+ X$
scattering in units of MeV$^2$/c$^2$,
using a Fermi gas model for the nuclear effects.}
\label{fig:q2_nbarc}
\end{figure}

The $\bar \nu_\mu$ on $CH_2$ scattering events can be separated into scattering
from bound and free protons using the different shapes of the respective $Q^2$
distributions. Since the $\bar \nu_\mu$ flux for these two samples is the same,
one can then scale the bound proton cross section from the known free proton
cross section. The $Q^2$ shape of free protons is well known,\cite{vogel} 
and the
$Q^2$ shape of bound protons in $^{12}C$ is deduced from $\nu_\mu$ scattering
on bound neutrons. Figs. \ref{fig:q2_nbarp} and \ref{fig:q2_nbarc} show the
expected $Q^2$ distributions for free and bound protons, using a Fermi gas
model for the nuclear effects. (The Fermi gas model is just being used to
indicate the expected difference in the two distributions. In the actual
measurement, the $Q^2$ distributions will be determined from the data as
discussed above.) The $\bar \nu_\mu - CH_2$ data is fit to the appropriate sum
of the two distributions, giving the amplitudes for bound and free scattering.
The amplitude of the free scattering gives an absolute value for the 
$\bar\nu_\mu$ flux. The amplitude of the bound scattering combined with this
measured $\bar \nu_\mu$ flux gives a measurement of an effective 
bound proton cross
section. This cross section on bound protons can then be used to infer the
related effective cross section for bound neutrons. 

Although there are other details discussed in reference 39, 
the method seems capable of yielding both neutrino and
antineutrino fluxes at a level much better than previously obtained ($<15\%$).
Thus, this method can be used to determine the observed flux at the
detector position.   A comparison of this observed flux with
predictions from the beam simulation can then be interpreted in terms
of neutrino oscillations.

\subsection{An Oscillation Analysis with Reduced Systematic Errors}

\begin{figure}
\centerline{\psfig{figure=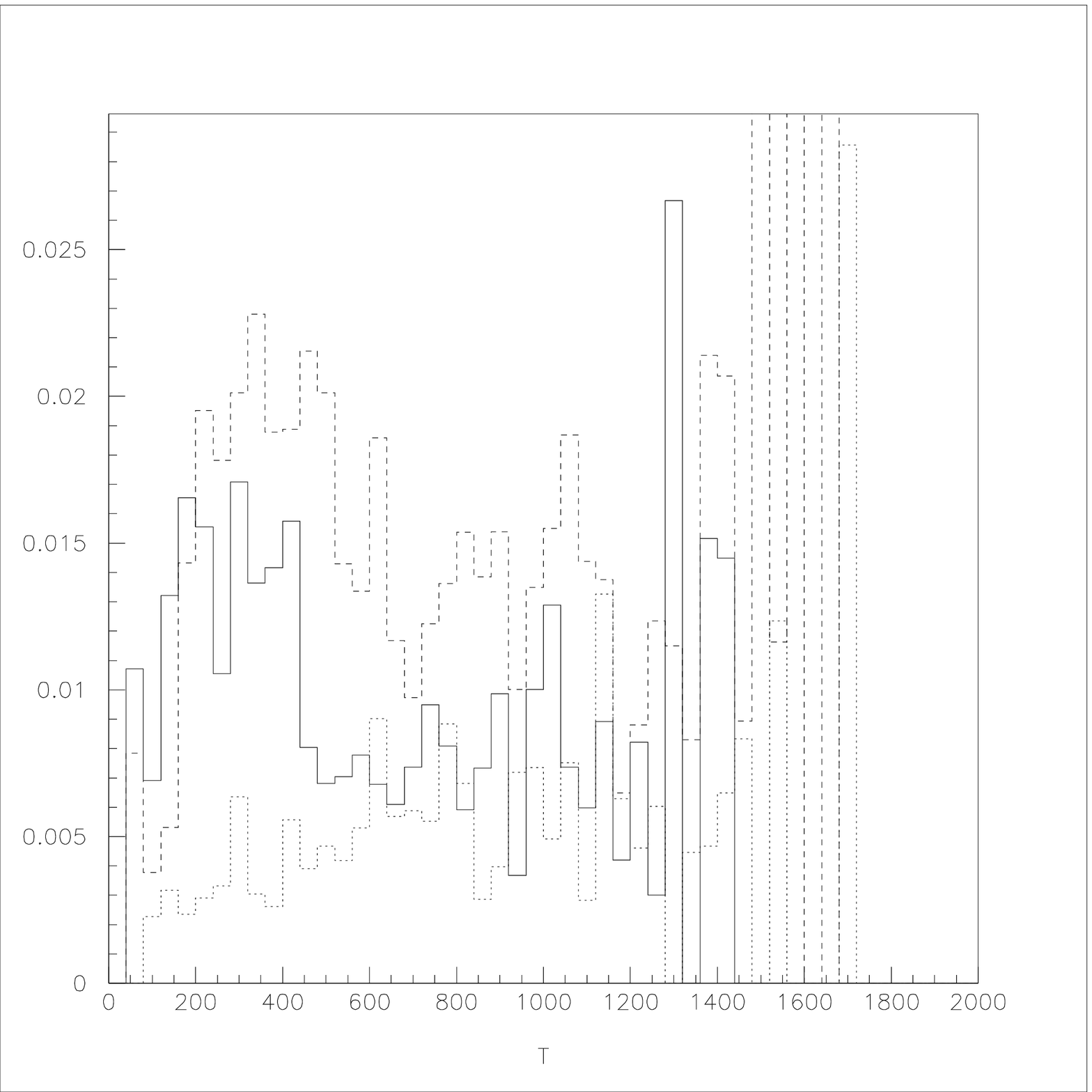,width=4.7in,silent=}}
\caption{The ratio of the
$\nu_e C \rightarrow e^- X$ and $\nu_\mu C \rightarrow \mu^- X$ visible energy
distributions in MeV for:
(a) $\Delta m^2 =0.2$ eV$^2$, appearance $\sin^22\theta = 0.04$,
and disappearance $\sin^22\theta = 0.5$ (solid curve); 
(b) $\Delta m^2 =0.4$ eV$^2$, appearance $\sin^22\theta = 0.02$,
and disappearance $\sin^22\theta = 0.5$ (dashed curve);  and
(c) no oscillations (dotted curve).}
\label{fig:eappdis}
\end{figure}

Another strategy for an oscillation analysis with reduced  
systematic errors is to 
divide the $\nu_e C \rightarrow e^- X$ visible energy
distribution by the $\nu_\mu C \rightarrow \mu^- X$ visible energy
distribution. This division removes uncertainties associated with the
neutrino flux and cross sections, and, furthermore, the effect of
oscillations is amplified if both $\nu_\mu \rightarrow \nu_e$
appearance and $\nu_\mu$ disappearance are occurring at the same
$\Delta m^2$. Fig. \ref{fig:eappdis} shows the ratio of the 
$\nu_e C \rightarrow e^- X$ and $\nu_\mu C \rightarrow \mu^- X$ visible energy
distributions (see Figs. \ref{fig:app} and \ref{fig:dis}) for:
(a) $\Delta m^2 =0.2$ eV$^2$, appearance $\sin^22\theta = 0.04$,
and disappearance $\sin^22\theta = 0.5$ (solid curve); 
(b) $\Delta m^2 =0.4$ eV$^2$, appearance $\sin^22\theta = 0.02$,
and disappearance $\sin^22\theta = 0.5$ (dashed curve);  and
(c) no oscillations (dotted curve). Neutrino oscillations with the
above parameters can be clearly distinguished from no oscillations.

\subsection{Non-oscillation Neutrino Physics with BooNE}

With the BooNE detector and FNAL booster neutrino source, a
plethora of nuclear and particle physics using the neutrino as a probe
could be investigated. These topics include the role of strangeness in
the proton, the behavior of the axial vector mass and coupling constant
in nuclear matter, the helicity structure of the weak neutral current,
and the neutrino magnetic moment.  The copious flux of intermediate
energy (100 MeV - 2 GeV) neutrinos from the FNAL booster source
would provide a new opportunity to pursue this physics.

The list below outlines some of the interesting physics that may be
investigated
with BooNE.  Several of the channels (namely,
$\nu_{\mu} C \rightarrow \mu^- N$ and $\nu_{\mu} C \rightarrow \nu_{\mu} \pi^0
X$)
will need to be understood thoroughly for neutrino oscillation background
estimates.
The others, while not potential backgrounds, have the possibility of yielding
exciting
physics.  More detailed feasibility studies are currently underway.

\begin{itemize}
\item Neutrino-Nucleon Elastic Scattering and a Measurement of $G_s$

The $\nu p \rightarrow \nu p$ and $\nu n \rightarrow \nu n$ reactions (where
$\nu$ is a $\nu_\mu$ or a $\bar{\nu}_\mu$) offer the possibility of extracting
$G_s$, the strange quark axial form factor of the nucleon.  The ratio
of neutrino elastic scattering events producing a proton to those producing a
neutron
on the isoscalar carbon nucleus is a sensitive measure of $G_s$
and dependent only weakly upon the $F_2^s$ form factor\cite{gar92}.
More exactly, if this ratio is measured for both antineutrinos
and neutrinos, $G_s$ and $F_2^s$ are separable\cite{gar93}.  The precision
of this measurement depends upon how well the BooNE detector can
distinguish neutrons from protons and is a topic that is currently
under study.

\item Neutrino Charged-Current Scattering

The $\nu_\mu\mbox{}^{12}C \rightarrow \mu^-\mbox{}^{12}N$ and
$\bar{\nu}_\mu\mbox{}^{12}C \rightarrow \mu^+\mbox{}^{12}B$
reactions will be measured to high precision with BooNE.  By comparing
the two as a function of $Q^2$, which allows a separation of the free
protons from the bound protons
in the $\bar{\nu}_\mu$ channel (see section 9.5),
$M_A$, the axial-vector dipole mass, in the nuclear medium will be measured.

\item Neutral-Current $\pi^0$ Production

A measure of $\nu_{\mu} C \rightarrow \nu_{\mu} \pi^0 X$ is a
sensitive probe of the structure of the weak neutral-current. Significant
gains in precision will be achieved over previous experiments in this
energy region.  This will enable a test of the standard model prediction
of the helicity structure of the weak neutral-current.

\item Neutrino-Electron Neutral-Current Scattering

By measuring the $\nu_{\mu} e^- \rightarrow \nu_{\mu} e^-$ cross section
and its behavior at low-$Q^2$, evidence for (or a limit on)
a magnetic moment of the muon-neutrino will be obtained.  If the neutrino
is a Majorana particle (and CPT holds) the neutrino must have no
magnetic moment. Thus, a measurement of a non-vanishing magnetic moment
is proof that the neutrino is a Dirac particle.  This is
also relevant to the solar neutrino problem.

\end{itemize}

\section{BooNE: A Future Upgrade to Two Detectors}

{\it Given an oscillation signal in the MiniBooNE experiment, a second
detector will be added to determine the oscillation
parameters.   This also permits a sensitive search for CP violation in the
lepton sector.} \\

Given that an oscillation signal is observed in MiniBooNE, then the 
next goals would be:
\begin{itemize}
\item to determine the oscillation parameters from both 
$\nu_\mu \rightarrow \nu_e$ and $\nu_\mu$ disappearance.
\item to search for CP violation in the Lepton Sector. 
\end{itemize}
A natural upgrade to the MiniBooNE experiment that addresses
these goals is to add a second
detector. The  energy dependence of the ratio of neutrino events
in the two detectors determines the oscillation parameters.
Comparison of results from running in neutrino mode to antineutrino mode
investigates CP violation. 
A summary of the expectations 
of this two-detector experiment is
provided here.

The second detector will be a double-wall cylindrical tank
of the same design as the MiniBooNE detector.
The outer volume serves as a veto shield for uncontained
events and is filled with high light-output liquid scintillator.
The inner (main detector) volume is a right cylinder, 9.6 m in height and
9.6 m in diameter,
that is filled with mineral oil.   The inner detector has
1220 eight-inch photomultiplier tubes of 
the type used 
for the MILAGRO experiment (Hammamatsu R5912).
The electronics and data 
acquisition would be similar to the LSND design.
The tank would be placed partially underground, with 
overburden 
to reduce the rate of cosmic ray muons and eliminate the
cosmic ray hadronic component.
An approximate cost estimate for this detector is given in
Table \ref{tab:det_cost2}.

The new, near detector will be placed approximately
500 meters from the neutrino source.   The MiniBooNE detector 
would be used as the far detector, located at 1000 meters from the 
beam line target.
These distances are chosen to provide 
the optimum comparison between the rate of events in the near and far
detector, as discussed below.
The MiniBooNE beam
line will be used to provide the neutrino and antineutrino beams
for both detectors.

\subsection{Event Rates for BooNE}

\begin{table}[t]
\caption{The estimated numbers of quasi elastic events 
for both neutrino and antineutrino scattering and for both the near (500 m)
and far (1000 m) detectors. The muon-neutrino
quasi elastic scattering estimates assume no oscillations, while the
electron-neutrino quasi elastic scattering estimates assume 100\%
$\nu_{\mu} \rightarrow \nu_e$ transmutation.}
\label{tab:numbers2}
\vspace{0.4cm}
\begin{center}
\begin{tabular}{|c|c|c|}
\hline 
Reaction&Near Detector&Far Detector\\
\hline
$\nu_{\mu} C \rightarrow \mu^- X$&      609,600 & 152,400\\
$\bar \nu_{\mu} C \rightarrow \mu^+ X$& 114,000 & 28,500\\
$\nu_e C \rightarrow e^- X$&            630,000 & 157,500\\
$\bar \nu_e C \rightarrow e^+ X$&       115,200 & 28,800\\
\hline
\end{tabular}
\end{center}
\end{table}

For the estimation of event rates we make the following assumptions.
First, we assume that the Booster operates at an energy of 8 GeV 
and at an average 
rate of 5 Hz ($2.5 \times 10^{13}$ protons/s) for 3 years of operation
($3 \times 10^7$ s) at each focusing polarity. Also, we assume
that the fiducial volume of the detector is 382 t ($1.7 \times 10^{31}$
$CH_2$ molecules) and that the total electron and muon 
efficiencies, including PID,  are 50\%. The
resulting numbers of quasi elastic events are shown in Table \ref{tab:numbers2}
for both neutrino and antineutrino scattering and for both the near (500 m)
and far (1000 m) detectors. The muon-neutrino 
quasi elastic scattering estimates assume no oscillations, while the
electron-neutrino quasi elastic scattering estimates assume 100\%
$\nu_{\mu} \rightarrow \nu_e$ transmutation.

\subsection{$\nu_{\mu} \rightarrow \nu_e$ Appearance}

\begin{figure}
\centerline{\psfig{figure=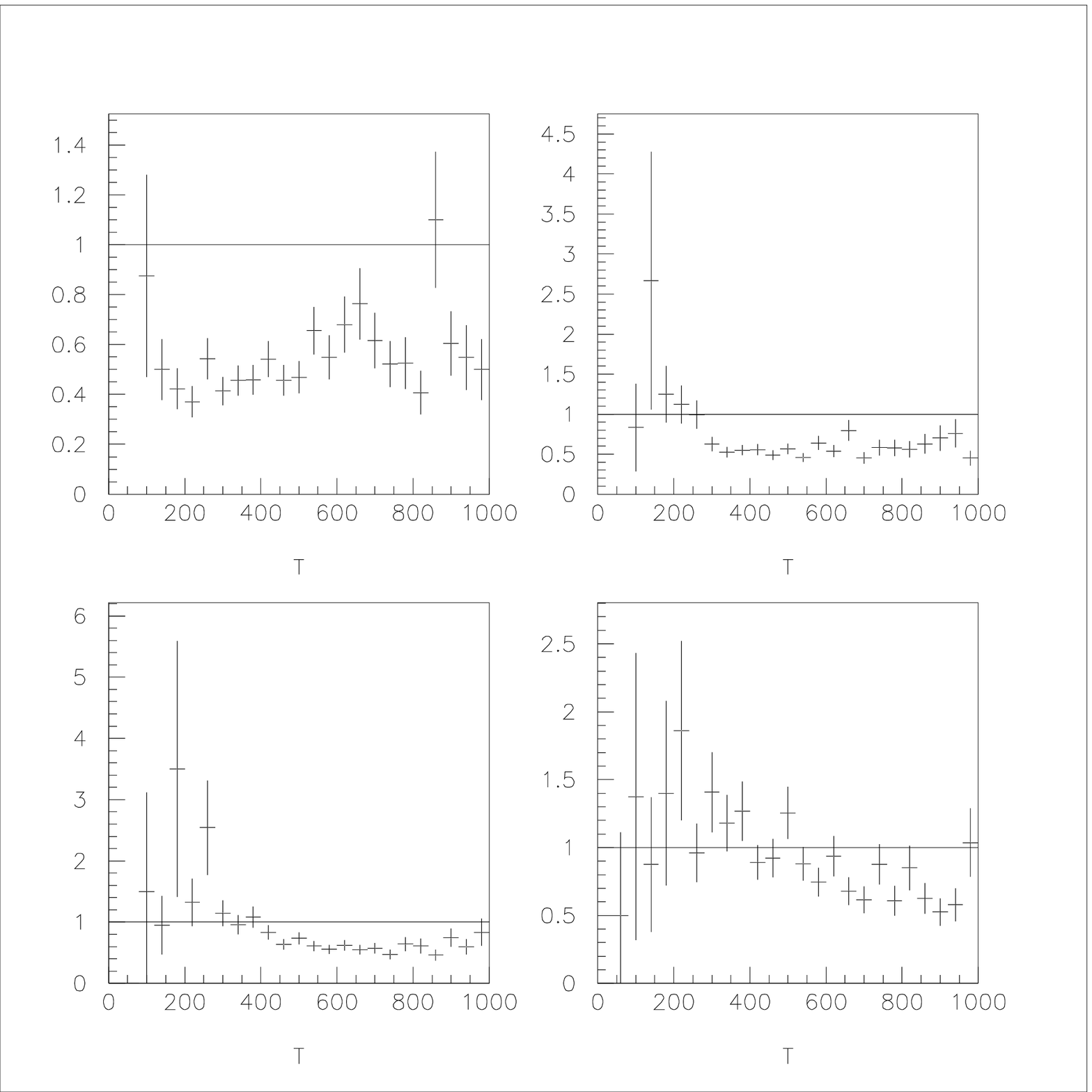,width=4.7in,silent=}}
\caption{The ratio of $\nu_e$
quasi elastic events from the detectors at 500 m and 1000 m as a
function of visible energy in MeV for four
different possible oscillation parameters (motivated by LSND):
(a) $\Delta m^2 = 0.2$ eV$^2$ and $\sin^22\theta = 0.04$;
(b) $\Delta m^2 = 0.4$ eV$^2$ and $\sin^22\theta = 0.02$;
(c) $\Delta m^2 = 0.6$ eV$^2$ and $\sin^22\theta = 0.01$;
(d) $\Delta m^2 = 0.8$ eV$^2$ and $\sin^22\theta = 0.008$.
Also shown in the figures (at a ratio of 1) are solid lines giving
the expectations for no oscillations.}
\label{fig:app_ratio}
\end{figure}

The two detectors located at 500 m and 1000 m from the neutrino source
will measure the $\nu_e$ energy spectrum through quasi elastic scattering 
as described in chapters 8 and 9. The energy-dependent ratio of event rates 
from the two detectors will provide proof that neutrino oscillations
are occurring and will allow the determination of the neutrino oscillation
parameters (assuming that the LSND signal is indeed due to neutrino
oscillations). Fig. \ref{fig:app_ratio} shows the ratio of $\nu_e$
quasi elastic events from the detectors at 500 m and 1000 m as a
function of visible energy for four
different possible oscillation parameters (motivated by LSND): 
(a) $\Delta m^2 = 0.2$ eV$^2$ and $\sin^22\theta = 0.04$;
(b) $\Delta m^2 = 0.4$ eV$^2$ and $\sin^22\theta = 0.02$;
(c) $\Delta m^2 = 0.6$ eV$^2$ and $\sin^22\theta = 0.01$;
(d) $\Delta m^2 = 0.8$ eV$^2$ and $\sin^22\theta = 0.008$.
For each set of oscillation parameters it is conservatively assumed
that the total background rate from all sources (intrinsic $\nu_e$
component in the beam, $\pi^0$ background, $\mu$ background, etc.)
is 0.5\% of the $\nu_{\mu} \rightarrow \nu_e$ 100\% transmutation rate.
Also shown in the figures (at a ratio of 1) are solid lines giving
the expectations for no oscillations. Note that, as discussed in
chapter 5, the neutrino flux varies as $r^{-2}$ to an excellent
approximation for $r>250$ m.
As is shown in the figures,
neutrino oscillations can be clearly observed and measured. We 
estimate that, for the oscillation parameters above, $\Delta m^2$
can be determined with an uncertainty of $<0.1$ eV$^2$ and $\sin^22\theta$
with an uncertainty of $<25\%$.

\begin{figure}
\centerline{\psfig{figure=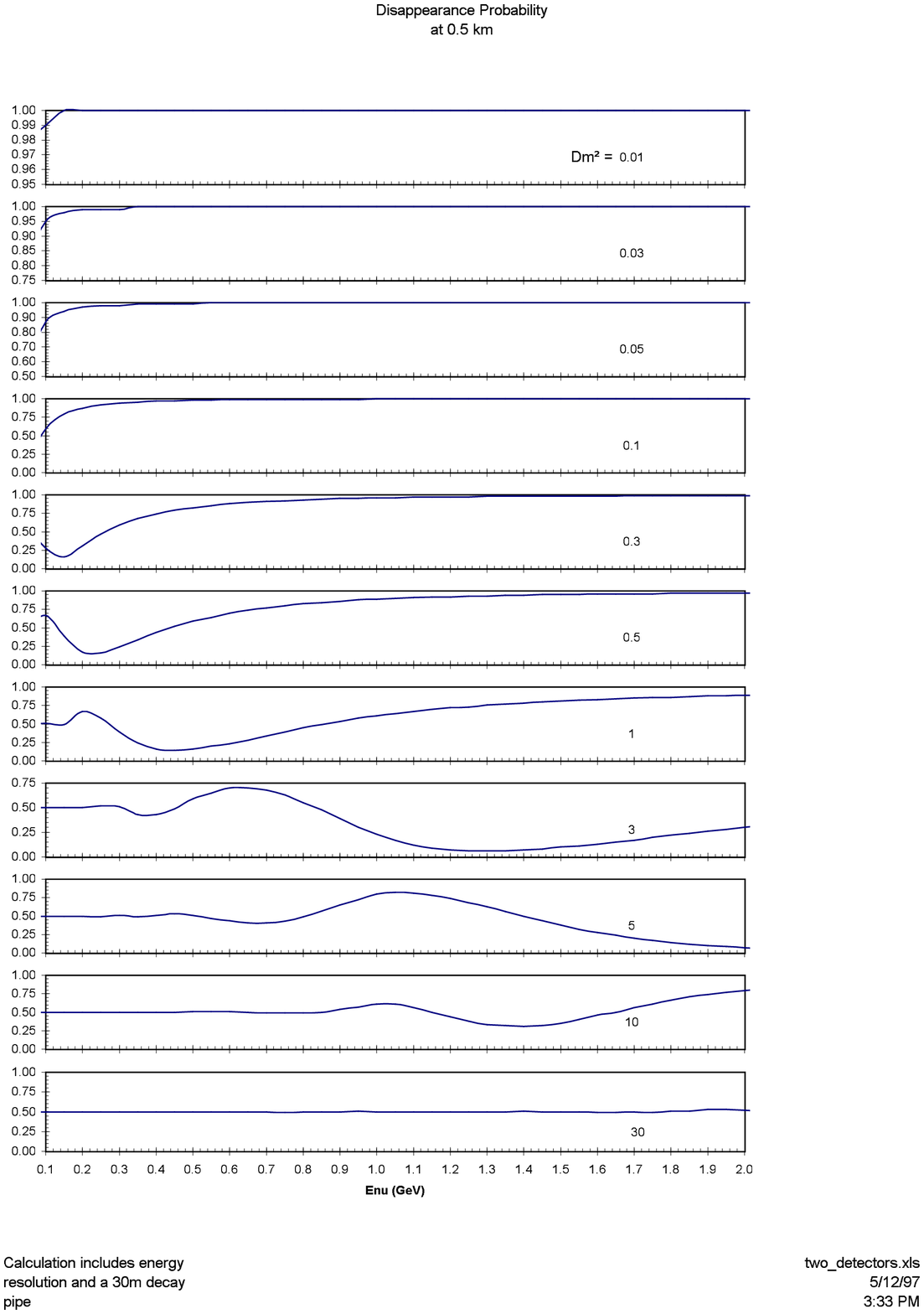,bbllx=0bp,bblly=75bp,bburx=600bp,bbury=700bp,height=6.in,width=5.5in,clip=T}}
\caption{The dependence on energy of events at the 
near (500 m) detector for full mixing and various $\Delta m^2$
values.}
\label{fig:near_disapp}
\end{figure}

\begin{figure}
\centerline{\psfig{figure=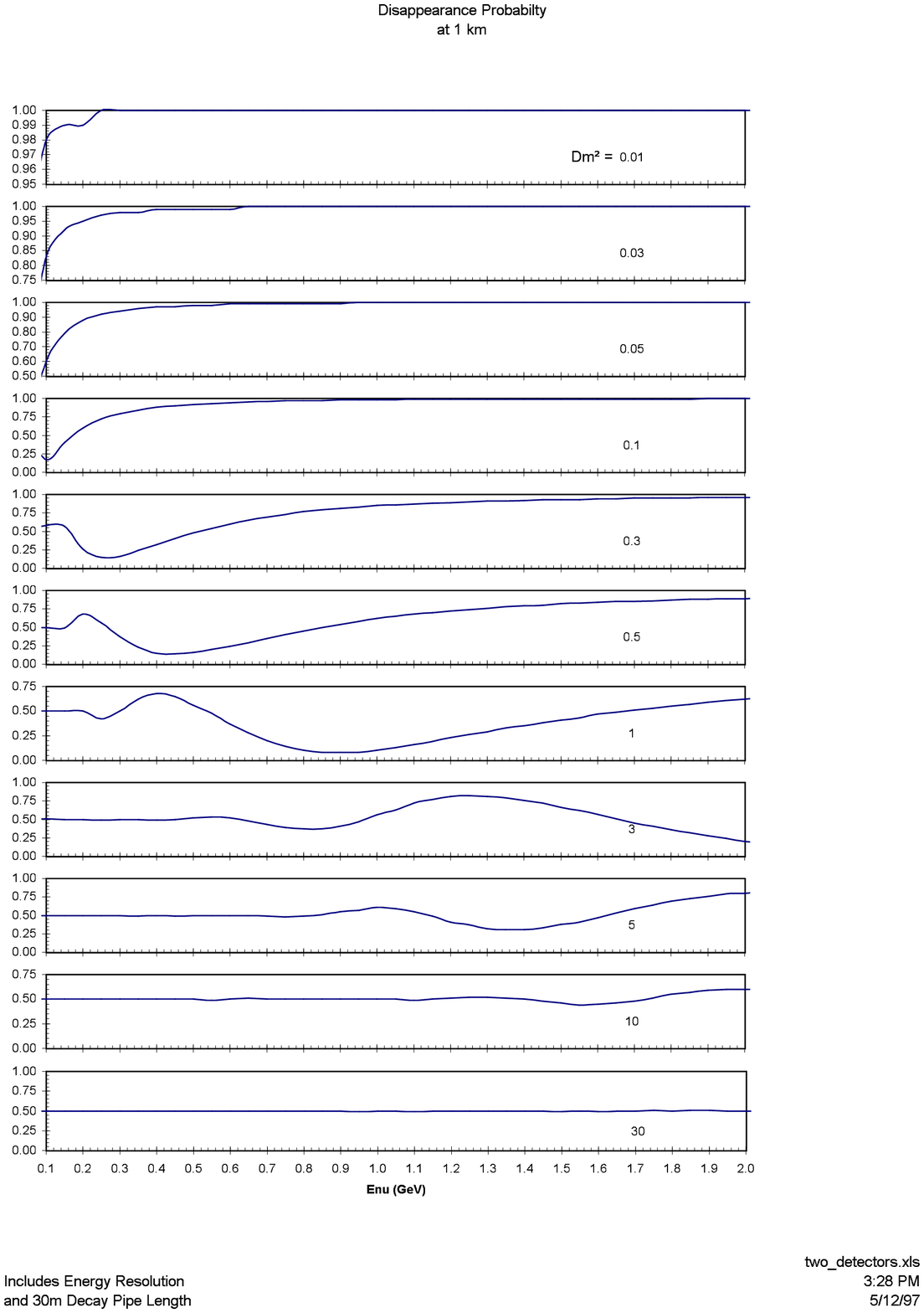,bbllx=0bp,bblly=75bp,bburx=600bp,bbury=700bp,height=6.in,width=5.5in,clip=T}}
\caption{The dependence on energy of events at the 
far (1000 m) detector for full mixing and various $\Delta m^2$
values.}
\label{fig:far_disapp}
\end{figure}

\begin{figure}
\centerline{\psfig{figure=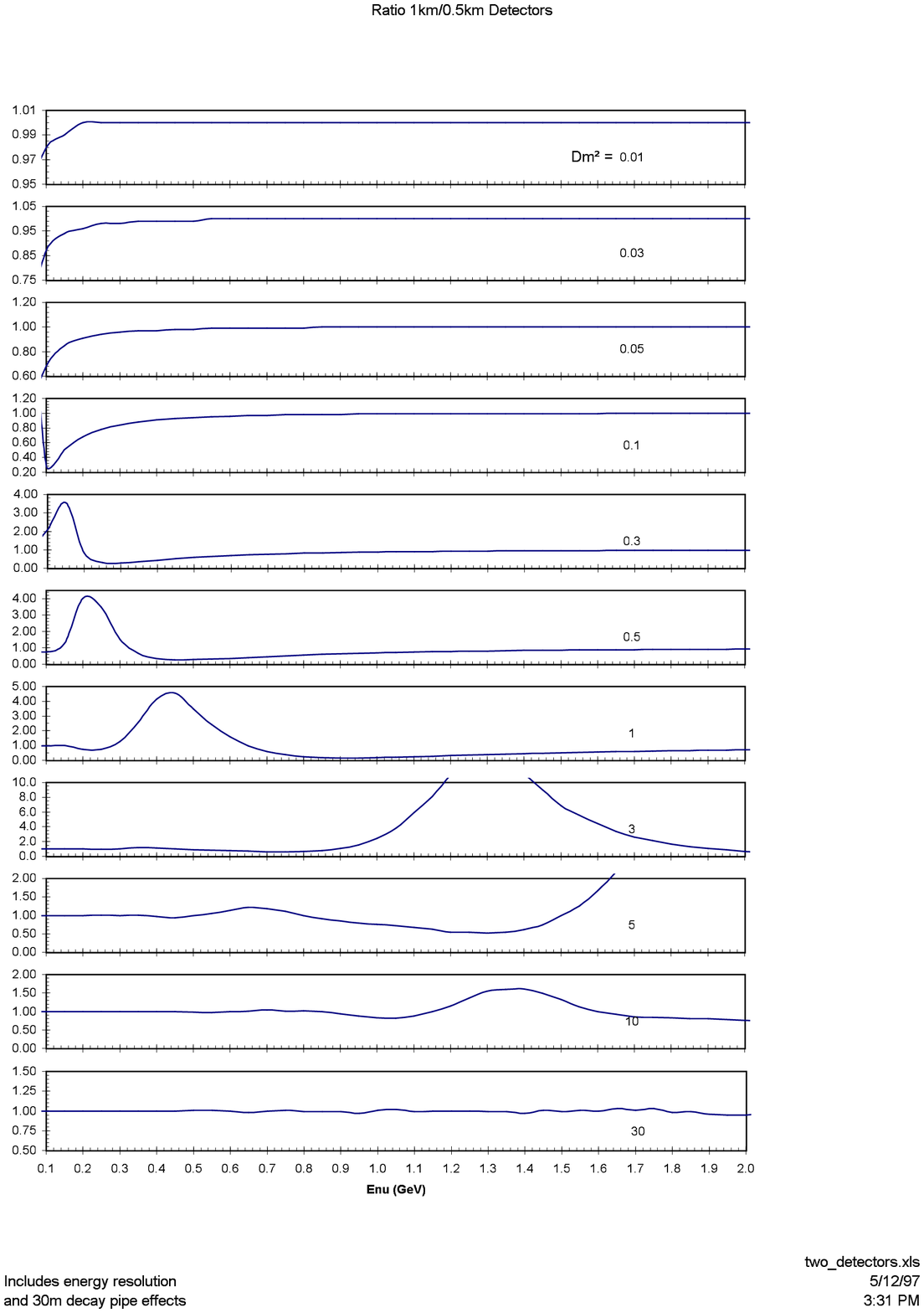,bbllx=0bp,bblly=75bp,bburx=600bp,bbury=700bp,height=6.in,width=5.5in,clip=T}}
\caption{The ratio of events at the 
near (500 m) and far (1000 m) detectors as a function of energy
for full mixing and various $\Delta m^2$
values.}
\label{fig:ratio_disapp}
\end{figure}

\begin{figure}
\centerline{\psfig{figure=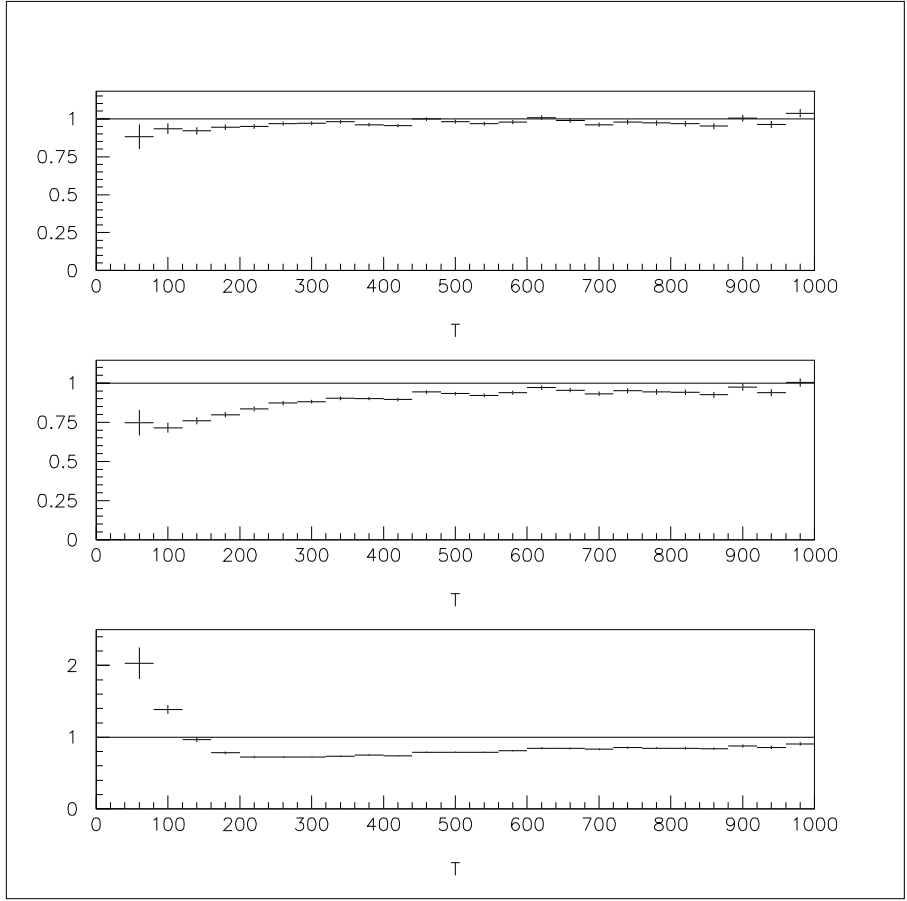,width=4.7in,silent=}}
\caption{The ratio of $\nu_{\mu}$
quasi elastic events from the detectors at 1000 m and 500 m as a
function of visible energy in MeV for three
different possible oscillation parameters (motivated by the
atmospheric neutrino problem):
(a) $\Delta m^2 = 0.1$ eV$^2$ and $\sin^22\theta = 0.5$;
(b) $\Delta m^2 = 0.2$ eV$^2$ and $\sin^22\theta = 0.5$;
(c) $\Delta m^2 = 0.4$ eV$^2$ and $\sin^22\theta = 0.5$.
Also shown in the figures (at a ratio of 1) are solid lines giving
the expectations for no oscillations.}
\label{fig:dis_ratio}
\end{figure}

\subsection{$\nu_{\mu}$ Disappearance}

\begin{figure}
\centerline{
\psfig{figure=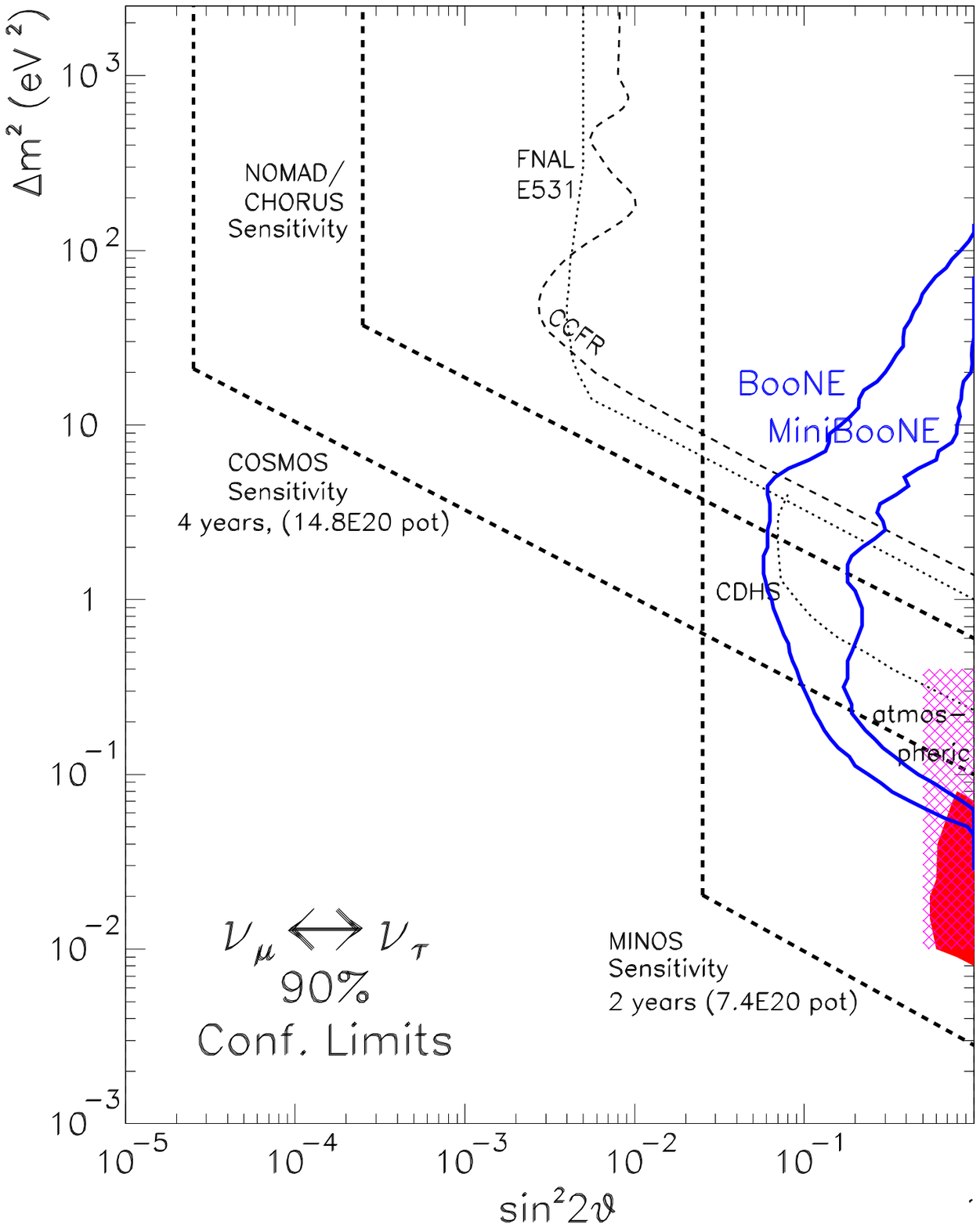,bbllx=100bp,bblly=150bp,bburx=600bp,bbury=700bp,height=5.5in,width=5.5in,clip=T}
}
\caption{Summary of results from past experiments (narrow, dashed and dotted), 
future approved experiments (wide, dashed) and 
90\% C.L. limit expected for the two-detector configuration, 
BooNE, (solid) for $\nu_{\mu}$ 
disappearance after one year of running at 1 km.   
Solid region indicates the favored region for the atmospheric neutrino
deficit from the Kamioka experiment.  
A result from Kamiokanda
indicating no zenith angle dependence extends the favored region to
higher $\Delta m^2$ as indicated by the hatched region.}
\label{fig:boonedisap}
\end{figure}

The two detectors located at 500 m and 1000 m from the neutrino source also
will measure the $\nu_{\mu}$ energy spectrum through quasi elastic scattering 
as described in chapters 8 and 9. The energy-dependent ratio of event rates 
from the two detectors can be used to 
observe $\nu_{\mu}$ disappearance oscillations
and determine the neutrino oscillation
parameters.
Figures \ref{fig:near_disapp} and \ref{fig:far_disapp}
shows how the visible energy will change with $\Delta m^2$ for
full mixing for the 500 m and 1000 m detectors, respectively.   
Figure \ref{fig:ratio_disapp}, shows the ratio for full mixing.
Motivated by the atmospheric neutrino problem,
Fig. \ref{fig:dis_ratio} shows the ratio of $\nu_{\mu}$
quasi elastic events from the detectors at 1000 m and 500 m as a function
of visible energy for: 
(a) $\Delta m^2 = 0.1$ eV$^2$ and $\sin^22\theta = 0.5$;
(b) $\Delta m^2 = 0.2$ eV$^2$ and $\sin^22\theta = 0.5$;
(c) $\Delta m^2 = 0.4$ eV$^2$ and $\sin^22\theta = 0.5$.
Also shown in these figures (at a ratio of 1) are solid lines giving
the expectations for no oscillations. Note that, as discussed in
chapter 5, the neutrino flux varies as $r^{-2}$ to an excellent
approximation for $r>250$ m. As is shown in the figures,
neutrino oscillations can be clearly observed and measured. We 
estimate that, for the oscillation parameters above, $\Delta m^2$
can be determined with an uncertainty of $<0.1$ eV$^2$ and $\sin^22\theta$
with an uncertainty of $<25\%$.    The expected limits if no signal 
is observed is shown in Fig.~\ref{fig:boonedisap}.

\subsection{Search for CP Violation in the Lepton Sector}

Assuming that $\nu_{\mu} \rightarrow \nu_e$ and 
$\bar \nu_{\mu} \rightarrow \bar \nu_e$ oscillations are observed,
then, by comparing the neutrino and antineutrino oscillation parameters,
it will be possible to make a test for CP violation in the
lepton sector. CP violation may appear as a difference in the
measured values of $\sin^22\theta$ (or even $\Delta m^2$, because
in general all three different $\Delta m^2$ values can contribute
to an oscillation signal) for neutrinos
compared to antineutrinos. We estimate that CP violation can be observed
easily if the violation is large ($>25\%$).

\section{Cost and Schedule}

{\it The cost of the MiniBooNE detector and neutrino beam line are
estimated to be \$1.6M and \$3M, respectively. A reasonable time scale
is the commencement of data taking in 2001.} \\

\begin{table}[t]

\caption{Existing materials for the MiniBooNE Experiment. a) Materials
committed by Los Alamos National Laboratory and universities. b) 
Request to Fermilab for use of materials existing on site.}

\label{tab:reuse}

\vspace{0.4cm}

\begin{center}

\begin{tabular}{|c|c|r|}
\hline 
\multicolumn{3}{|c|}{A. Materials to be provided by LANL and universities
for Detector}
\\ \hline
Item & Source & Value in thousands\\ \hline
1512 Phototubes & LANL &  \$1210 \\
Cabling and connectors & LANL & \$50 \\
Channel Electronics & LANL &    \$422 \\
Electronics Hut & LANL    & \$100 \\
Remote DAQ & LANL          & \$250 \\
Dilute Scintillator Oil & LANL & \$185 \\
Scintillator Oil & LANL & \$110 \\
Solid Scintillator & Univ, LANL & \$400 \\ 
Phototubes for solid scintillator & Univ & \$25 \\ \hline
\multicolumn{2}{|c|}{ Total value of goods:} & { \$2535} \\ \hline \hline
\multicolumn{3}{|c|}{B. Materials requested from Fermilab}
\\ \hline
Item & Present Location & Value in thousands \\ \hline
Liquid Scintillator (5500 gallons) & NuTeV Detector & \$56 \\ 
Iron (2200 ft$^3$) & NuTeV Detector  & \$110 \\
Solid Scintillator & NuTeV Detector  & \$100 \\
Beam Monitors \& Electronics & NuTeV Beam Line & \$100 \\
Portakamp & Not yet identified & \$100 \\ \hline
\multicolumn{2}{|c|}{ Total value of goods:} & { \$366} \\ \hline
\hline
\end{tabular}

\end{center}

\end{table}

This chapter outlines the costs of MiniBooNE.  The detector and the
beam line are considered separately.   The cost of a second detector
for the Phase II upgrade is also presented.   The costs described here 
include equipment and construction.   

The cost of engineering is not
included.   LANL is prepared to provide some engineering, with 
\$100K in funds earmarked for engineering design of the MiniBooNE
detector in fiscal year 1998.   

The cost of MiniBooNE is held to a modest level by re-using equipment.
Table \ref{tab:reuse}a lists the
materials which will be contributed by LANL and the universities on
the experiment.   The total of equipment which is already committed to
the project, mainly by LANL, is over \$2.5 million, where 
the listed value is based on the present purchase 
cost of similar equipment.  

Fermilab has solicited requests to reuse equipment from detectors
which are to be dismantled.\cite{workshop}
Table \ref{tab:reuse}b lists existing 
materials at Fermilab which we are specifically requesting.   We 
request to be notified if any further liquid scintillator or 
solid scintillator becomes available from Fermilab experiments which
will be dismantled.  The portakamp will be for temporary use during the
construction phase in order to maintain a clean entrance to the
detector and provide space for testing equipment.   Thus the portakamp
will become available to the lab again after 2000.   We plan to make
use of some NIM electronics from PREP which is presently in use within
the NuTeV trigger system.   We request that all of the NuTeV trigger 
system NIM logic be reassigned to MiniBooNE.  We request that the
lemo cables and terminators from NuTeV be reserved for MiniBooNE.

\begin{table}[t]
\caption{Cost Estimate for the 8 GeV proton-generated neutrino beam.}
\label{tab:beam_cost}
\vspace{0.4cm}
\begin{center}
\begin{tabular}{|c|r|}
\hline 
Item&Cost in thousands \\
\hline \hline
Permanent Quadrupole Magnets (@ \$2K per 20 m) & \$10 \\
Permanent Dipole Magnet (@ \$10K per magnet)    & \$10 \\
Vacuum system (@ \$300K per km) & \$30 \\ 
Focusing apparatus & \$100 \\
Rapid cycling power supply for focusing & \$200 \\ 
Beam monitoring equipment & \$50 \\ \hline
Total of beam line elements  & \$400 \\ \hline \hline
Shielding surrounding target and decay pipe  & \$300 \\
Dump, including cooling and shielding & \$150 \\ \hline 
Total for shielding and dumps & \$450 \\ \hline \hline
Conventional construction & \$ 2000 \\
Utilities & \$ 150 \\ \hline
Total enclosure costs & \$2150 \\ \hline \hline
Total without contingency&  \$3000 \\ \hline
\hline
\end{tabular}
\end{center}
\end{table}

A cost estimate for the beam is 
summarized in Table \ref{tab:beam_cost}.   
The cost of the beam line is mainly due to the enclosures which must
be centered 25 ft below surface level.
The low energy neutrino beam, which is used for both phases of the
experiment, is of conventional design and therefore of relatively low cost.
The 8 GeV transport system using permanent magnets is particularly 
economical.   The cost of the focusing 
apparatus is estimated by the cost of a Lithium Lens.   Note that the 
rapid cycling power supply for the focusing system appears as a separate item.
Beam monitoring equipment from the NuTeV experiment is expected to be
re-used,  so only \$50K is assigned to this item.

\begin{table}[t]
\caption{Unit cost of the 0.6 kton MiniBooNE detector with mineral oil.
Costs reflect the the fact that some material are already available to
members of the BooNE Collaboration.}
\label{tab:det_cost}
\vspace{0.4cm}
\begin{center}
\begin{tabular}{|c|r|}
\hline 
Item &Cost in thousands\\
\hline
Two-tank system  & \$350 \\
Tank installation (Conventional construction) & \$300 \\
Temporary inflatable domed roof & \$25 \\
Phototubes &\$0 \\
Phototube Mounts & \$15 \\
Mineral Oil &\$555\\
Veto Scintillator Oil & \$236+\$0\\
Veto Solid Scintillator & \$0 \\
Cabling and connectors & \$0 \\
Channel Electronics &\$0\\
Electronics Hut&\$0\\
Remote DAQ&\$0\\
Liquid Handling & \$50\\
Other Shipments & \$50\\
Support Utilities & \$50\\ \hline
Total without contingency& \$1631\\
\hline
\end{tabular}
\end{center}
\end{table}

Approximate costs for the MiniBooNE detector are listed in 
Table \ref{tab:det_cost}.   These represent the actual expected 
expenditures rather than the net worth of the equipment.  
Therefore, the cost has been adjusted for materials which are 
already owned or are available to members of the MiniBooNE
collaboration.   The cost of the tank system and 
construction comes from discussions 
with the Chicago Bridge and Iron Company or Plainfield Illinois.
These represent only cost estimates.\cite{Ned}
We assume that the detector will contain a dilute scintillator mixture
formed by combining the LSND scintillator (1/4 of the total volume)
with mineral oil.   Computing for analysis will be supplied by LANL
and the universities.

\begin{table}[t]
\caption{Milestones for MiniBooNE}
\label{tab:miles}
\vspace{0.4cm}
\begin{center}
\begin{tabular}{|rr|}
\hline 
1997  & Finalize Design \\
      & Complete Approvals Process \\
1998  & Detector engineering complete \\
      & Engineering of rapid-cycling power supply \\
      & Engineering of the beam line enclosures \\
1999  & Tank installation \\
      & Construction of the beam line enclosures \\
      & Construction of horn and power supply system \\
      & Engineering of the remaining beam line elements \\
      & Recommissioning/ building of beam line monitors \\
2000  & Electronics and Phototube installation \\
      & Installation of the beam line elements \\
2001  & Begin taking data \\  \hline
\end{tabular}
\end{center}
\end{table}

Once funding has been secured, it is estimated that the detector 
construction will take two years. A reasonable time scale has 
construction starting in 1999 and the commencement of data taking
in 2001.   Table~\ref{tab:miles} describes the milestones in the 
process of commissioning MiniBooNE.    A substantial amount of work 
is already scheduled for the summer, including: finalizing the 
horn design (G. Mills, H. White), addressing whether the detector
should contain dilute mineral oil (W. Louis), writing the triggering
software for MiniBooNE (R. Tayloe, D. Matzner), testing phototubes 
(T. Ochs, B. Tamminga),  completing CASIM studies (J. Conrad, M.
Petravick), and other projects.   
The expectation is that a formal proposal will be provided
to the PAC in the autumn.   

As discussed in chapter 10, a second detector is needed if a signal is
observed in MiniBooNE in order to accurately measure the mixing and
mass parameters.  The costs associated with this second detector,
assuming that it is identical in design to the MiniBooNE detector,  
are listed  
in table ~\ref{tab:det_cost2}.  Located at 500m from the beam dump,
this detector will be centered 12ft below ground level. Thus the cost 
of the conventional construction is higher than for MiniBooNE. 
The MiniBooNE domed roof will be re-used.

\begin{table}[t]
\caption{Unit cost of a second 0.6 kton second detector with mineral
oil for Phase II (BooNE).}
\label{tab:det_cost2}
\vspace{0.4cm}
\begin{center}
\begin{tabular}{|c|r|}
\hline
Item & Cost in thousands\\
\hline
Two-tank system  & \$450 \\
Tank installation (Conventional construction) & \$600 \\
Temporary inflatable domed roof & \$0 \\
Phototubes &\$1210 \\
Phototube Mounts & \$15 \\
Dilute Scintillator &\$740\\
Veto Scintillator Oil & \$402\\
Veto Solid Scintillator & \$280 \\
Cabling and connectors & \$50 \\
Channel Electronics &\$422\\
Electronics Hut&\$100\\
Remote DAQ&\$250\\
Liquid Handling & \$50\\
Support Utilities & \$100\\ \hline
Total without contingency& \$4669\\
\hline
\end{tabular}
\end{center}
\end{table}

\section{Conclusions}

{\it The MiniBooNE experiment will have the capability of observing
both $\nu_\mu \rightarrow \nu_e$ appearance and $\nu_\mu$
disappearance. It will also search for CP violation in the
lepton sector.} \\

We propose to construct and operate a neutrino detector system and
broad band neutrino beam generated from the Fermilab Booster that has
the capability of observing and measuring $\nu_{\mu} \rightarrow
\nu_e$ oscillations and $\nu_{\mu}$ disappearance over a
wide range of $\Delta m^2$. The motivation for this experiment stems
from the LSND neutrino oscillation result\cite{bigpaper2} and the
atmospheric neutrino problem. The relatively low energy (0.1 - 2.0
GeV) neutrino beam combined with a detector distance of 1km will cover
the interesting $\Delta m^2$ region between $0.01$ and $1.0$ eV$^2$
for modest cost and effort. The proposed MiniBooNE experiment would
start with a single detector with the goal of probing this mass region
and establishing definitive indications of neutrino oscillations. If a
positive signal is observed, a second stage, BooNE, would use two
detectors in the same neutrino beam to accurately determine the
oscillation parameters and investigate any CP violating
$\nu_{\mu}/\bar \nu_{\mu}$ effects.

The MiniBooNE experiment will use one detector at a distance of 1 km
from the neutrino beam source. The high intensity available from the
Booster combined with an efficient horn-focused secondary beam will
provide over 50,000 events per year in the 400 ton detector. With this
setup, the intrinsic $\nu_e$ component of the beam is less than $3
\times 10^{-3}$ due to the little kaon production by 8 GeV primary protons
and reduced muon decay in the short 30 m decay pipe. The totally active
MiniBooNE detector also has good particle identification capabilities
and the expected mis-identification fraction is at the $2
\times 10^{-3}$ level.  Comparisons of the observed $\nu_e$ and
$\nu_\mu$ energy distributions with expectations will allow a
sensitive search for $\nu _\mu \rightarrow \nu _e$ oscillations and
$\nu _\mu$ disappearance.

If oscillations exist at the LSND level, MiniBooNE should see several
hundred anomalous $\nu_e$ events over background and  establish the
signal at the $>5\;\sigma $ level. From the energy dependence of the
excess, the oscillation parameters can be determined in the LSND
region with a $\Delta m^2$ $(\sin^22\theta)$ uncertainty of $<0.2$
eV$^2$ ($<50\%$) respectively. Examining the signal with both incident
$\nu _\mu$ and $\bar \nu _\mu$ will allow a first look at possible
CP violation effects in the lepton mixing matrix. On the other hand,
if no oscillation signal is observed, the experiment will
significantly extend the region probed for oscillations and exclude
$\nu _\mu \rightarrow\nu _e$ oscillations with
$\sin ^22\theta >6\times 10^{-4}$ for large $\Delta m^2$ and
$\Delta m^2>0.01$ eV$^2$ for $\sin ^22\theta = 1$.

In summary, the MiniBooNE experiment will make a sensitive search for
$\nu _e$ appearance and $\nu _\mu$ disappearance from neutrino
oscillations in the $\Delta m^2$ region between 0.01 and 1 eV$^2$. The
experiment has unique capabilities and sensitivity to measure
oscillations in this region in a cost effective and timely manner.

\end{document}